\documentclass[12pt]{article}

\usepackage{bm}
\usepackage{subfigure}
\usepackage{amssymb,amsmath,graphicx,color}
\usepackage[numbers,compress]{natbib}

\newcommand{\be}{\begin{equation}}
\newcommand{\ee}{\end{equation}}
\newcommand{\bea}{\begin{eqnarray}}
\newcommand{\eea}{\end{eqnarray}}
\newcommand{\nn}{\nonumber}
\def\si{^1 \hskip -0.03in S _0}
\def\siii{^3 \hskip -0.025in S _1}

\def\L1Abar{\tilde{l}_{1,A}}

\def\e{{\varepsilon}}
\def\g{{\gamma}}

\def\L{{\Lambda}}

\def\cC{{\mathcal C}}

\def\cO{{\mathcal O}}

\def\cR{{\mathcal R}}

\def\cW{{\mathcal W}}

\def\tnubb{$2\nu\beta\beta$}
\def\znubb{$0\nu\beta\beta$}
\def\eqref#1{{(\ref{#1})}}

\usepackage{lineno,hyperref}
\modulolinenumbers[5]

\newcommand{\lsim}{\raisebox{-0.7ex}{$\stackrel{\textstyle <}{\sim}$ }}

\def\si{^1 \hskip -0.03in S _0}
\def\siii{^3 \hskip -0.025in S _1}

\newcommand{\spacevec}[1]{{\mathbf #1}} 
\newcommand{\NLDBD}{$0 \nu \beta \beta$}

\newcommand{\vL}{\ensuremath{\mathcal{L}}}    
\newcommand{\al}{\alpha}
\newcommand{\bt}{\beta}
\newcommand{\boldsigma}{\mbox{\boldmath $\sigma$}}
\newcommand{\sq}{^{2}}
\renewcommand{\vec}[1]{{\mathbf #1}} 
\newcommand{\slashpi}{\protect{\slash\hspace{-0.5em}\pi}}
\newcommand{\xpt}{$\chi$PT}

\begin{document}

\title{ \vspace{1cm} Lattice QCD Inputs for Nuclear Double Beta Decay}
\author{Vincenzo Cirigliano,$^{1}$ William Detmold,$^2$ \\Amy Nicholson,$^3$ Phiala Shanahan$^2$\\
\\
$^1$Theoretical Division, Los Alamos National Laboratory, \\ Los Alamos, NM 87545, USA\\
$^2$Center for Theoretical Physics, Massachusetts Institute of Technology, \\ Cambridge, MA 02139, USA\\
$^3$Department of Physics and Astronomy, University of North Carolina, \\
Chapel Hill, NC 27516, USA}
\maketitle

\begin{abstract}
Second order $\beta$-decay processes with and without neutrinos in the final state are key probes of nuclear physics and of the nature of neutrinos. Neutrinoful double-$\beta$ decay is the rarest Standard Model process that has been observed and provides a unique test of the  understanding of weak nuclear interactions. Observation  of neutrinoless double-$\beta$ decay would reveal that neutrinos are Majorana fermions and that lepton number conservation is violated in nature. While significant progress has been made in phenomenological approaches to understanding these processes, establishing a  connection between these processes and  the physics of the Standard Model and beyond is a critical task as it will provide input into the design and interpretation of future experiments. The strong-interaction contributions to double-$\beta$ decay processes are non-perturbative and can only be addressed systematically through a combination of lattice Quantum Chromoodynamics (LQCD) and nuclear many-body
calculations. In this review, current efforts to establish the LQCD connection are discussed  for both neutrinoful and neutrinoless double-$\beta$ decay. LQCD calculations of the hadronic contributions to the neutrinoful process $nn\to pp e^- e^- \bar\nu_e\bar\nu_e$ and to various neutrinoless pionic transitions are reviewed, and the connections of these calculations to the phenomenology of double-$\beta$ decay through the use of effective field theory (EFTs) is highlighted. At present,  LQCD calculations are limited to small nuclear systems, and to pionic subsystems, and require matching to appropriate EFTs to have direct phenomenological impact. However, these calculations have already revealed qualitatively that there are terms in the EFTs that can only be constrained from double-$\beta$ decay processes themselves or using inputs from LQCD. Future prospects for direct calculations in larger nuclei are also discussed.

\end{abstract}

\date{\today}

\newpage

\tableofcontents

\newpage

\section{Introduction }

The study of nuclear beta decays has been instrumental in the development of the modern theory of electroweak interactions 
encoded in the Standard Model (SM)~\cite{Weinberg:2009zz}.
 {\it Single} beta decay of the neutron and nuclei with sufficiently high precision has been used to test the SM and probe  new physics in 
charged-current electroweak interactions~\cite{Gonzalez-Alonso:2018omy}.  
{\it Double}  beta decay~\cite{GoeppertMayer:1935qp} (DBD) is a rare nuclear process, observable only in certain nuclei with even numbers of protons and neutrons (even-even nuclei) for which single beta decay is energetically forbidden. In such decays, two neutrons decay into two protons with emission of two electrons and two anti-neutrinos.   
Neutrinoful double-$\beta$ (\tnubb) decay is the rarest SM process whose rate has been measured~\cite{Tanabashi:2018oca} 
and it therefore   offers a non-trivial test of  our understanding of weak interactions in nuclei.

It was realized as early as 1939 \cite{Furry:1939qr}  that a neutrinoless variant of  double beta decay could occur  if neutrinos are Majorana fermions~\cite{Majorana:1937vz}.  
In the neutrinoless double beta decay mode (\znubb),  two neutrons convert into two protons with emission of two electrons and no neutrinos ($nn \to p p e^- e^-$),  thus changing the number of leptons by two units. Since lepton number 
 (more precisely at the quantum level the difference of baryon and lepton number,  $B-L$)  is conserved in the SM,   observation of \znubb\
  would be direct evidence of new physics, with far reaching implications: it would demonstrate that neutrinos are Majorana fermions~\cite{Schechter:1981bd}, shed light on the mechanism of neutrino mass generation~\cite{Minkowski:1977sc,Mohapatra:1979ia,GellMann:1980vs}, and probe a key ingredient, lepton number violation (LNV),  
needed to generate the matter-antimatter asymmetry in the universe via  ``leptogenesis''~\cite{Davidson:2008bu}. 
Because of these outstanding scientific motivations,  a vigorous worldwide experimental program exists  searching for \znubb. 
Current experimental limits are very stringent~\cite{Gando:2012zm,Agostini:2013mzu,Albert:2014awa,Andringa:2015tza,KamLAND-Zen:2016pfg,Elliott:2016ble,Agostini:2017iyd,Aalseth:2017btx, Albert:2017owj,Alduino:2017ehq,Agostini:2018tnm, Azzolini:2018dyb,Anton:2019wmi};
for example, the \znubb\ lifetime of ${}^{136}$Xe  is $T^{0\nu}_{1/2}>1.07\times 10^{26}$ yr
\cite{KamLAND-Zen:2016pfg}. Next-generation, ton-scale, experiments 
aim to improve sensitivity by one or two orders of magnitude.

\begin{figure}[t]
\centering
\vspace{-.1cm}
\includegraphics[width=.5\textwidth]{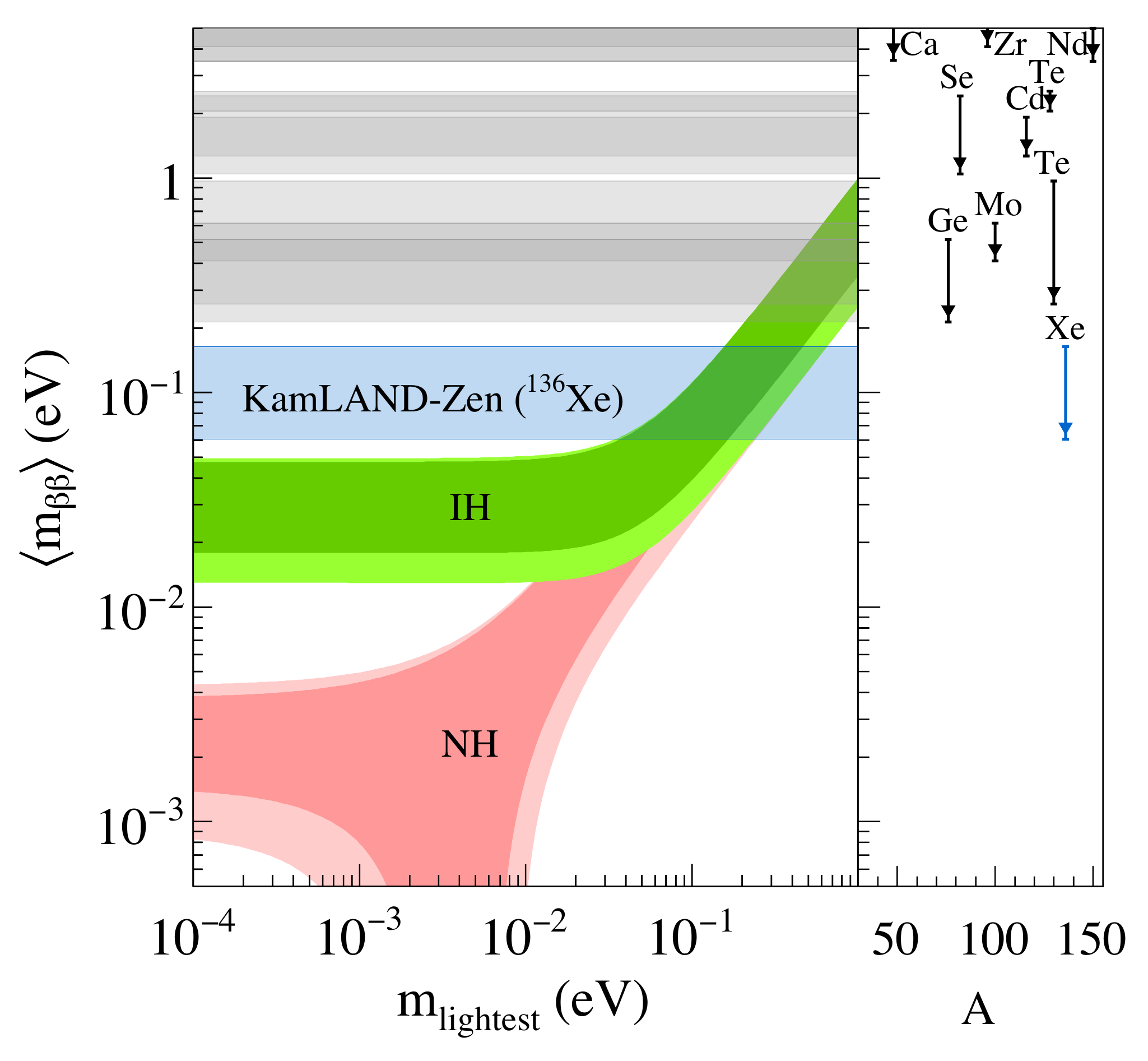}
\vspace{-.1cm}
\caption{$ |m_{\beta \beta}|$ versus lightest neutrino mass for  both normal (pink band)  and inverted neutrino mass hierarchy  (green band). 
Current experimental sensitivities are reported as gray and blue horizontal bands, whose width is determined by 
order-of-magnitude theoretical uncertainties on the nuclear transition matrix elements. 
(Figure from Ref.~\cite{KamLAND-Zen:2016pfg})}
\label{Fig:Kamland}
\end{figure}

By itself, the observation of \znubb\   would not  immediately resolve the underlying mechanism of  LNV. 
In fact, ton-scale \znubb\ searches will constrain  LNV from a variety of mechanisms at unprecedented precision~\cite{Rodejohann:2011mu,DellOro:2016tmg}.
For example, the standard see-saw mechanism for neutrino mass generation originates at very high scale~\cite{Minkowski:1977sc,GellMann:1980vs}, 
 through the exchange of heavy right-handed (RH) neutrinos which leave behind 
a  single  dimension-5 operator  at low-energy~\cite{Weinberg:1979sa} written in terms of lepton and Higgs fields, 
suppressed by the scale $\Lambda$ associated with LNV, which in this case coincides with the mass of RH neutrinos ($\Lambda \sim M_R$). 
Below the electroweak scale  the dimension-5 operator provides a Majorana mass term  for the light neutrinos. 
In this case $0 \nu \beta \beta$  is a direct probe of the neutrino mass matrix.  The decay rate scales as   $\Gamma \propto  |M_{0\nu}|^2  m_{\beta \beta}^2$,  
where $M_{0\nu}$ is the nuclear matrix element and  $m_{\beta \beta} = \left|\sum_i  U_{ei}^2 m_i\right|$ is   the ``$ee$'' component of the 
neutrino Majorana mass matrix 
in the flavor basis with $U$ being the Pontecorvo-Maki-Nakagawa-Sakata  (PMNS) matrix \cite{Pontecorvo:1967fh,Maki:1962mu}.   
From neutrino oscillation experiments, some inputs for  $m_{\beta \beta}$ are constrained \cite{Tanabashi:2018oca}.  
However, the two Majorana phases,  the ordering of the spectrum ($m_{\rm lightest} = m_1$ or $m_{\rm lightest} = m_3$), 
and the value of $m_{\rm lightest}$ remain unknown.  This implies that in the 
$m_{\beta \beta}$ vs  $m_{\rm lightest}$ plane one has two bands, whose width is due to the unknown Majorana phases.  
The current understanding is summarized  in Fig.~\ref{Fig:Kamland}, including  the  experimental constraint on $m_{\beta \beta}$ from 
Ref.~\cite{KamLAND-Zen:2016pfg}. 
Experimental sensitivities appear as horizontal bands rather than lines, due to estimates of the uncertainties in the nuclear matrix elements 
$M_{0\nu}$,  which vary by a factor of 3 depending on the nucleus and on which model is adopted in the calculation~\cite{Engel:2016xgb}. 
Next-generation experiments aim to explore  parameter space, covering the entire ``inverted hierarchy'' band in $m_{\beta \beta}$  (green region), assuming nuclear matrix elements are given by the minimum of available calculations.    A discovery will be possible if nature realises an inverted spectrum, or if $m_{\rm lightest} > 50$~meV, irrespective of the ordering.  
Note that the interplay  of $0 \nu \beta \beta$ with other neutrino mass probes, 
namely constraints on $m_\beta \equiv ( \sum_i |U_{ei}|^2 m_i^2 )^{1/2}$  from tritium beta decay~\cite{Aker:2019uuj,Esfahani:2017dmu} and $\Sigma \equiv \sum_{i} m_i$ from cosmology~\cite{Aghanim:2018eyx}, 
can test the high-scale see-saw model and possibly unravel new sources of LNV or physics beyond the standard 
$\Lambda$CDM + $m_\nu$ cosmological paradigm.   

Alternatively, LNV could originate at an intermediate scale, close to the TeV-scale,  as in for example  the 
Left-Right Symmetric Model~\cite{Mohapatra:1979ia}.  TeV sources of LNV (such as a TeV-mass  right-handed neutrino) may lead to sizeable contributions to $0\nu \beta \beta$ not directly related to the exchange of light neutrinos, 
provided the  scale  $\Lambda$ is not too high compared to the weak scale.   
In such cases, the Large Hadron Collider (LHC) can compete with $0\nu \beta \beta$ to constrain the parameter space of  certain  models, see for example Refs.~\cite{Tello:2010am,Peng:2015haa}.
Note that the new contributions from TeV scale LNV can interfere with $m_{\beta \beta}$  or add incoherently, significantly affecting the interpretation of experimental  results.  
In these scenarios  the exchange of  heavy particles  leaves behind  operators of odd dimension ($d=7, 9, 11, ...$)~\cite{Kobach:2016ami} written in terms of lepton, Higgs, and quark fields, and suppressed by $\Lambda^{d-4}$.  
Importantly, these TeV scale mechanisms can lead to different transition operators  at the hadronic and nuclear scale, 
which in most cases probe the structure of nuclei at very short distances. 
Finally,   LNV could be lurking at very low-scale, through mass terms of light sterile neutrinos. 
Implications of light sterile neutrino exchange for \NLDBD \ have been studied in the context of  current
short-baseline neutrino oscillation anomalies~\cite{Gariazzo:2015rra}.

To summarize, 
ton-scale $0\nu \beta \beta$ searches, which will reach sensitivities $T_{1/2} >10^{27-28}$ yr, will probe uncharted territory 
and thus have significant discovery potential.   
In combination with oscillation experiments, direct mass measurements, and  cosmology, 
$0\nu \beta \beta$  can effectively probe the high-scale see-saw paradigm. 
At the same time, $0\nu \beta \beta$  is quite sensitive to LNV originating at scales lower than the GUT scale, 
and by itself, or  in combination with the LHC experiments, can discover  LNV at the multi-TeV scale.

For any of the possible underlying mechanisms, 
the interpretation of \znubb\ experiments and the constraints on 
fundamental LNV parameters,
such as the Majorana masses of left-handed neutrinos, rely on having a 
theoretical framework 
that provides reliable 
predictions of transition rates with controlled uncertainties. 
As shown by the horizontal bands in Fig.~\ref{Fig:Kamland}, 
current knowledge of the relevant hadronic and nuclear matrix elements  is  unsatisfactory as it is a key source of uncertainty~\cite{Engel:2016xgb}. 
For reviews of the standard approach to nuclear matrix elements, we refer the reader to Refs.~\cite{Ejiri:2019ezh,Engel:2016xgb,Bilenky:2014uka,Vergados:2012xy,Faessler:1999zg}.
Here we note that various approaches  to the many-body problem  lead to estimates of the matrix elements that differ by a factor of two to three, and, more importantly, lack a systematic way in which to assess the uncertainties. In fact, few  of the current calculations of nuclear matrix elements are  based on ab-initio many-body methods and effective field theory (EFT) analysis of the transition operators, where lattice QCD can provide key input. 
To improve upon this situation, recent  efforts have advocated an ``end-to-end''  EFT analysis of \znubb\ 
to link the scale $\Lambda$ of LNV to nuclear scales. 
This multi-prong approach includes various steps:  
\begin{enumerate}

\item   The use of  the Standard Model  EFT to link the scale $\Lambda$ of LNV to the hadronic scale $\Lambda_\chi \sim O(1)$~GeV,  where non-perturbative QCD effects arise.  
This step is by now mature:  the operator 
basis (to which any underlying model can be matched) is known up to dimension-nine and the renormalization 
group evolution of these operators under strong 
interactions is known.   
Light new degrees of freedom (such as sterile neutrinos) can also be included in this framework.

\item  The  matching of the quark-gluon level EFT to hadronic  EFTs such as Chiral Perturbation Theory ($\chi$PT) in the meson and single nucleon sector, 
and chiral EFT and pionless EFT in the multi-nucleon sector. 
This step can be performed consistently in the strong and  weak sectors of the theory, which in the case of interest here involves $\Delta L=2$ transition operators.
The form of the transition operators is known 
to leading order in the hadronic EFT expansion  for all underlying LNV mechanisms, and sub-leading corrections are also known for most mechanisms. 
%
%
The matching procedure typically requires the introduction of hadronic interactions that are short-range compared to the typical nuclear scale 
and have effective couplings encoding non-perturbative strong-interaction physics.  In what 
follows, we refer to these effective couplings as low-energy constants (LECs). 

\item  The use of  lattice QCD (LQCD)  to determine the LECs relevant to double beta decay, including the ones  controlling the  $\Delta L=2$ transition operators needed 
to predict neutrinoless double beta decay.  This step involves  matching a given hadronic or few-body amplitude computed in LQCD to the corresponding expression in the hadronic EFT.
This is a relatively new area of research.
Recent activity  has focused 
on the calculation of polarizability effects for 
neutrinoful double beta decay, as well as mesonic LECs 
of relevance for both TeV LNV mechanisms and light 
Majorana neutrino exchange for neutrinoless double beta decay.  
Addressing the challenges associated with two-nucleon  (such as $nn \to pp ee$) and  multi-nucleon matrix 
elements  in LQCD is  an active area of  research.  
In concert,  EFT calculations for LNV transitions 
will need to be extended to finite-volume to 
optimize the matching procedure.

\item  The solution of the nuclear many-body problem 
for nuclei of experimental interest, through  {\it ab initio} nuclear structure methods, 
relying on QCD-rooted  chiral potentials and weak transition operators (including those with  $\Delta L=2$). 
These calculations are in their infancy for nuclei of experimental interest such as $^{76}$Ge, and  
can be    benchmarked in lighter nuclei where  ab initio many-body methods are available~\cite{Pastore:2017ofx,Basili:2019gvn}. 
\end{enumerate} 
With this framework in mind, this review focuses on recent progress  and future prospects in the  use of LQCD to compute the LECs relevant for neutrinoful and neutrinoless double beta decay (step 3 above), 
which is  intertwined with the  EFT description of the these processes (step 2 above) and serves as input for many-body calculations (step 4). Whenever appropriate, we discuss the correspondence between the EFT and  traditional approaches to nuclear matrix elements~\cite{Ejiri:2019ezh,Engel:2016xgb,Bilenky:2014uka,Vergados:2012xy,Faessler:1999zg}.
For a broader overview of the role of LQCD and EFT in searches for violations of fundamental symmetries, the reader is referred to Ref.~\cite{Cirigliano:2019jig}.

The manuscript is organized as follows:   the EFT framework for double-$\beta$ decay 
and the basics of Lattice QCD are reviewed in 
Sections \ref{sec:EFTforDBD} and \ref{sec:LQCD}, 
respectively. 
In Section~\ref{sec:tnubb} neutrinoful double-$\beta$ decay is discussed.
Section~\ref{sec:znubb} is devoted to 
neutrinoless double beta decay, 
focusing on short-range  (TeV scale) 
mechanisms of LNV in Section~\ref{sec:znubb_short} 
and on the  light Majorana neutrino exchange mechanism  
in Section~\ref{sec:znubb_long}. 
The prospects for two- and multi-nucleon 
matrix elements in \znubb\ are discussed in Section~\ref{sec:znubb_nn} 
before this review is concluded with an outlook in Section~\ref{sec:outlook}.

\pagebreak

\section{Lattice QCD and effective field theory}

\subsection{Effective field theory for double-$\beta$ decay}
\label{sec:EFTforDBD}


In most beyond the Standard Model (BSM) scenarios, the LNV source responsible for $0\nu\beta\beta$ is induced at an energy scale $\Lambda$ well above the electroweak scale. This scale separation justifies an effective field theory approach. Such an approach has the advantage that $0\nu\beta\beta$ and its correlation with collider observables can be described in a model-independent fashion. The Standard Model can be seen as the renormalizable part of an EFT that includes higher-dimensional operators which are suppressed by powers of the scale of BSM dynamics~\cite{Weinberg:1979sa,Wilczek:1979hc}: 
\begin{equation}
 {\cal L}_{\rm SMEFT} = {\cal L}_{\rm SM}   +  \sum_{n,\ d \geq 5}  \frac{C_n^{(d)}}{ \Lambda^{d-4} } \, O_n^{(d)} ~.
 \label{eq:SMEFT}
 \end{equation}
The dimension-$d$ operators $O_n^{(d)}$, where $n$ indexes the allowed forms of operators of the given dimension,  are built out SM fields (or additional light degrees of freedom) and 
are invariant under the SM gauge group $SU(3)_c\times  SU(2)_{L} \times U(1)_{Y}$. 
If the underlying BSM model is known, the dimensionless Wilson coefficients $C_n^{(d)}$  can be calculated in terms of the model parameters. 
The  effective Lagrangian in Eq.~(\ref{eq:SMEFT}) describes the low-energy limit of any high-scale  extension of the SM, 
and defines the so-called SM Effective Field Theory (SMEFT).  

Within this EFT, the $\Delta L=2$ operators have odd dimension \cite{Kobach:2016ami}. The first  $\Delta L=2$ term therefore appears at dimension five~\cite{Weinberg:1979sa} and provides a contribution to the neutrino Majorana mass \cite{Weinberg:1979sa}. In the standard type-I see-saw mechanism, this dimension-five operator arises from integrating out heavy right-handed neutrinos typically at the  GUT-scale, $\Lambda \sim 10^{15}$ GeV. LNV operators with a dimension $d> 5$ are then suppressed by multiple powers of $v/\Lambda \simeq 10^{-13}$, where $v\simeq 246$ GeV is the Higgs vacuum expectation value, and can be safely neglected. In various models, however, the scale of  LNV new physics is much lower and the dimension-five operator  may be suppressed by loop factors and/or  small Yukawa couplings. For instance, in the above-mentioned left-right symmetric models the dimension-five operator  scales as $y^2 /\Lambda$, where $y$ is a Yukawa coupling  scaling as $y \sim m_e/v \sim 10^{-6}$. While 
dimension-seven~\cite{Lehman:2014jma} and -nine~\cite{Prezeau:2003xn,deGouvea:2007qla,Graesser:2016bpz} LNV operators  are suppressed by additional powers of $\Lambda$, they can be suppressed by only one power of $y$, or even by ${\cal O}(y^0)$. As such, the dimension-seven and -nine operators can have contributions at the same order as the  dimension-five operator, for $\Lambda$ in the $1-10$ TeV range.
Since for operators  at dimension 11 and larger, the  usual $v/\Lambda$ suppression holds (no factors of Yukawa couplings can compensate it), 
in order  to describe $0\nu\beta\beta$ in a model-independent way, one should include all 
$SU(3)_c\times  SU(2)_{L} \times U(1)_{Y}$-invariant   $\Delta L=2$  operators up to dimension-nine in the SMEFT.
Dimension-seven and -nine operators have been discussed in the literature  in the context of models of radiative neutrino mass generation~\cite{Zee:1980ai,Zee:1985id,Babu:1988ki, Babu:1988ig,Babu:1988wk,Cepedello:2017lyo},  R-Parity Violating Supersymmetry~\cite{Faessler:1996ph, Faessler:1998qv, Prezeau:2003xn,Faessler:2007nz}, and Left-Right Symmetric model~\cite{Tello:2010am,Nemevsek:2011aa,Dev:2014xea,Bambhaniya:2015ipg,Senjanovic:2015yea,Ge:2015yqa} (this is not an exhaustive list of references).

Below the electroweak scale, these high-scale operators induce an $SU(3)_c\times U(1)_{\rm em}$-invariant $\Delta L=2$ Lagrangian 
that includes operators of dimension three, six, seven, and nine~\cite{Pas:1999fc,Pas:2000vn,Cirigliano:2018yza}. 
The mismatch in dimensions is due to the fact that the Higgs field acquires a vacuum expectation value (vev) and the Higgs and $W$ boson are integrated out of the EFT. 
At the scale $\Lambda_\chi \sim 1$ GeV, characteristic of  nonperturbative QCD
effects, the effective Lagrangian is given by~\cite{Cirigliano:2018yza}
\begin{eqnarray}
{\cal L}_{\rm eff} &=& {\cal L}_{\rm QCD}  
+ {\cal L}_{\rm W}  
- \frac{m_{\beta \beta}}{2} \, {\nu}_{eL}^T C\nu_{eL} 
+ \mathcal L^{(6)}_{\Delta L = 2} + \mathcal L^{(7)}_{\Delta L = 2} 
+ \mathcal L^{(9)}_{\Delta L = 2} ~, 
\label{eq:intro.1}
\\
{\cal L}_{\rm W} &=& - {\cal H}_{\rm W} = 
-  2 \sqrt{2} G_F  V_{ud} \, \bar{u}_L \gamma^\mu d_L \,\bar{e}_L \gamma_\mu \nu_{eL} 
\label{eq:intro.1.5}
\end{eqnarray}
where $C$ is the charge conjugation matrix. Here, the first term denotes the strong interactions among quarks and gluons,
and the second term represents the charged-current weak interactions of up and down 
quarks with leptons, whose strength is determined by the Fermi constant $G_F$ and
the $V_{ud}$ element of the Cabibbo-Kobayashi-Maskawa (CKM) matrix.  
Analogous terms involving strange quarks can be included but are irrelevant in the context of this review so are omitted.
The remaining terms denote the $\Delta L =2$ contributions, 
with the dimension-three Majorana mass term displayed explicitly. 
Details of ${\cal L}^{(6,7,9)}_{\Delta L=2}$ can be found in Ref.~\cite{Cirigliano:2018yza}.

In order to calculate $0\nu\beta\beta$ transitions,  below the scale $\Lambda_\chi \sim 1$~GeV,
the Lagrangian in   Eq. \eqref{eq:intro.1} needs to be matched onto a theory of hadrons. 
Since the relevant hadronic and nuclear processes  involve momentum transfer $Q \ll \Lambda_\chi$,   
the tool of choice is  chiral effective field theory ($\chi$EFT) 
\cite{Weinberg:1978kz,Weinberg:1990rz,Weinberg:1991um}
(for reviews see \cite{Epelbaum:2008ga,Hammer:2019poc}), which organizes the effective Lagrangian 
according to  the scaling of operators in powers of 
the typical momentum in units of the breakdown scale,
\begin{equation}
\epsilon_\chi= Q/\Lambda_\chi~, 
\qquad   Q \sim m_\pi \sim k_F,
\qquad \Lambda_\chi \sim 4\pi F_\pi \sim 1\ {\rm GeV}\,,
\label{eq:scales}
\end{equation}
where $m_\pi\simeq 140$ MeV and $F_\pi \simeq 92.2$ MeV are the pion mass 
and decay constant, respectively, and $k_F$ represents a typical Fermi momentum inside a nucleus. 
The $\chi$EFT Lagrangian schematically reads 
\begin{eqnarray}
\mathcal L_{\chi {\rm EFT}}  &=&  \mathcal L_{\rm strong}(\pi,N,\Delta)   
- \frac{4G_F}{\sqrt{2}} V_{ud} \, \mathcal J_\mu(\pi,N,\Delta) \,   \bar{e}_L \gamma^\mu \nu_{eL}
 \  - 4 G_F^2 V_{ud}^2  {\cal T}_{\mu \nu}   (\pi,N,\Delta)   \,  \bar{e}_L \gamma^\mu \nu_{eL}  \,   \bar{e}_L \gamma^\nu \nu_{eL}     
\nonumber \\
& &- \frac{1}{2} m_{\beta \beta} \, {\nu}_{eL}^T C \nu_{eL}   
- \frac{4 G_F}{\sqrt{2}} V_{ud} \, \mathcal O(\pi,N,\Delta) \, 
\bar{e} \, \Gamma C \bar{\nu}_{e L }^T
- G_F^2 \, \mathcal O^\prime(\pi,N,\Delta) \, \bar{e} \, \Gamma^\prime C \bar e^T  
+ \textrm{H.c.}, 
\label{eq:intro.2}
\end{eqnarray}
where  the first line contains the strong,  first-  and second-order charged current weak interactions,   respectively, while 
the operators in the second line
violate $L$ by two units.
Here $\mathcal L_{\rm strong}$, $\mathcal J_\mu$,  
$\mathcal{T}_{\mu \nu}$, $\mathcal O$, and 
$\mathcal O^\prime$ are 
combinations of pion, nucleon, and Delta isobar fields, organized according to increasing powers of $\epsilon_\chi$. 
They encode the non-perturbative QCD effects arising from distances 
shorter than $\Lambda_\chi^{-1}$.   $\mathcal{J}_\mu$  is  the  hadronic realization of the weak 
current~\cite{Park:1993jf,Park:1995pn,Pastore:2008ui,Pastore:2009is,Kolling:2009iq,Pastore:2011ip,Kolling:2011mt,Hoferichter:2015ipa,Baroni:2015uza,Krebs:2016rqz}, 
while $\mathcal{T}_{\mu \nu}$  encodes the weak hadronic polarizability~\cite{Shanahan:2017bgi,Tiburzi:2017iux}. 
Finally $\mathcal O$ and  $\mathcal O^\prime$ parametrize the hadronic component of  $\Delta L=2$ interactions. 
$\Gamma$ and $\Gamma^\prime$  represents the possible Dirac structures of the leptonic bilinear,
and  for simplicity, possible Lorentz indices in
$\Gamma$, $\Gamma^\prime$,  $\mathcal O$, and $\mathcal O^\prime$ are suppressed.

For situations in which the momentum transfer $Q \to \aleph \ll m_\pi \sim \Lambda_\slashpi$, which can be realized in 
particular kinematic regions or in LQCD calculations at unphysically large values of the quark masses,  the appropriate EFT to use is the so-called 
pionless EFT~\cite{Kaplan:1998tg, Kaplan:1998we, Chen:1999tn, Beane:2000fi,Bedaque:2002mn} ($\slashpi$EFT), in which pion degrees of freedom are  integrated out.  
The structure of  Eq.~(\ref{eq:intro.2}) carries over in ${\cal L}_{\chi{\rm EFT}} \to {\cal L}_{\slashpi {\rm EFT}}$, 
with  $\mathcal L_{\rm strong}$, $\mathcal J_\mu$,  
$\mathcal{T}_{\mu \nu}$, $\mathcal O$, and 
$\mathcal O^\prime$  now combinations of the nucleon fields only. 
In $\slashpi$EFT, operators and amplitudes are expanded in powers of $\epsilon_{\slashpi} = \aleph / \Lambda_\slashpi$. 
Pionless EFT allows one to gain analytic insight into the structure of strong and electroweak amplitudes and 
is  particularly useful in matching to current LQCD calculations in the multi-nucleon sector 
(see Section~\ref{sec:tnubb} for a concrete example).
The lepton-number conserving terms in the first line of Eq.~(\ref{eq:intro.2}) are discussed in greater depth in Section~\ref{sec:pionless}.
In particular, the contributions to neutrinoful double beta decay  
using the dibaryon formulation~\cite{Kaplan:1996nv} of pionless EFT are presented, setting the stage for matching to 
LQCD in Section~\ref{sec:tnubb}.

Concerning $\Delta L=2$ effects, 
the Lagrangian in Eq.~\eqref{eq:intro.2} can be used to calculate  
few-body amplitudes, from which one can then 
obtain non-relativistic potentials and 
weak transitions operators to be used in 
nuclear many-body calculations 
(see for example Ref.~\cite{Hammer:2019poc}).
This step is equivalent to  integrating out pions (and Majorana neutrinos)  with  ``soft"  ($(k^0, |\spacevec{k}|) \sim (\epsilon_\chi, \epsilon_\chi) \Lambda_\chi$)  
and ``potential" ($(k^0, |\spacevec{k}|) \sim (\epsilon_\chi^2, \epsilon_\chi) \Lambda_\chi$)  
scaling of their four-momenta,  
while keeping in mind that the effects of 
``hard" Majorana neutrinos 
($k^0 \sim |\spacevec{k}| \gg \Lambda_\chi$) 
are already included in the local terms in Eq.~\eqref{eq:intro.2}.
Through this procedure, one arrives at 
nuclear-level weak currents 
that contribute to single beta decay 
and neutrinoful double beta decay,  as well as  
the $0\nu\beta\beta$ transition operators, 
often referred to as  ``neutrino potentials''. 
%
The part of the effective nuclear Hamiltonian 
controlling \NLDBD \ can be written as 
\begin{eqnarray}
H_{\Delta L = 2}^{\rm (Nucl)} =  2 G_F^2 V_{ud}^2 \,
\bar e_{L} C \bar e_{L}^T  \, 
\sum_{a\neq b} 
\left(  m_{\beta \beta} \,     V^{(a,b)}_\nu  \, 
 +  V^{(a,b)}_6  \,   +  V^{(a,b)}_7   \, 
  +  V^{(a,b)}_9 \right)~,   
\label{eq:HV}
\end{eqnarray}
where $a,b$ label nucleons in the system and 
$V^{(a,b)}_\nu$ and $V^{(a,b)}_d$ are 
the two-body transition operators induced by the dimension-3 operator (neutrino mass) and dimension-$d$ operators of Eq.~(\ref{eq:intro.1}), 
respectively. 
Within the current EFT framework,  the two-body transition operators admit an expansion in 
powers of $v/\Lambda$,  $\Lambda_\chi/v$, and  $\epsilon_\chi = Q/\Lambda_\chi$ or $\epsilon_\slashpi = \aleph/\Lambda_\slashpi$~\cite{Cirigliano:2018yza,Cirigliano:2019vdj}.  
In the case of Majorana neutrino exchange, three-body   transition operators (suppressed by $\epsilon_\chi^2$ compared to the  leading two-body ones) have been considered in Refs.~\cite{Menendez:2011qq,Wang:2018htk}.  

Full details of each of the matching steps outlined above can be found in Ref.~\cite{Cirigliano:2018yza}. 
The potentials  $V^{(a,b)}_{6,7}$ require the 
hadronic and nuclear  realization of  isovector quark bilinears  of the form $\bar{u} \Gamma d$ ($\Gamma \in 
\{1, \gamma_5,  \gamma_\mu, \gamma_\mu \gamma_5, \sigma_{\mu \nu}\}$), which also appear in the analysis of single beta decay of nuclei in the SM and beyond.  
The LQCD input needed here is the single-nucleon 
charges~\cite{Aoki:2019cca,Gupta:2018qil,Chang:2018uxx,Liang:2018pis,Capitani:2017qpc} 
as well as LECs associated with two-body 
currents~\cite{Beane:2015yha,Savage:2016kon,Chang:2017eiq}. 
In  Sections~\ref{sec:LNV5} and \ref{sec:LNV9}, only  $V^{(a,b)}_\nu$ and $V^{(a,b)}_9$ are discussed in detail, 
since at the leading order (LO),  these 
potentials involve genuinely non-factorizable contributions 
with new  LECs 
that cannot be extracted from data and 
whose first-principles determination 
requires input from LQCD.

\subsubsection{Lepton number conserving interactions}
\label{sec:pionless}

It is useful to begin with consideration of  the lepton number conserving  first and second order weak interactions in nuclei.
As is appropriate for the analysis of the LQCD calculations of \tnubb, presented in Section \ref{sec:tnubb}, in the dibaryon formulation~\cite{Kaplan:1996nv,Beane:2000fi, Phillips:1999hh} of pionless EFT~\cite{Kaplan:1998tg, Kaplan:1998we, Chen:1999tn, Beane:2000fi}  the lepton number conserving strong and weak interactions (the first line in Eq.~(\ref{eq:intro.2})) are expressed in terms of nucleon and dibaryon fields for each possible two-nucleon channel. The purely strong interaction Lagrangian, and the form of the dibaryon propagators that re-sum effects on multiple nucleon-nucleon scatterings, are given in Refs.~\cite{Beane:2000fi, Phillips:1999hh} and reproduced in the current context in Ref.~\cite{Tiburzi:2017iux}. The corresponding Lagrangians for the first-order~\cite{Butler:1999sv, Butler:2000zp, Kong:2000px} and second-order axial current interactions (implemented as an axial background field $W^a_i$ where $a$ ($i$) denotes the isovector (vector) indices of the field) are 
\begin{eqnarray}
{\cal L}^{(1)} & = & -\frac{g_{A,0}}{2} N^\dagger \sigma_3 \left[ W_3^- \tau^++W^3_3 \tau^3+W_3^+ \tau^- \right] N
\nonumber\\
&&- \frac{l_{1,A}} {2M\sqrt{r_s r_t}} \left[ W_3^- t_3^\dagger s^++W^3_3t_3^\dagger s^3+W_3^+ t_3^\dagger s^- +\text{h.c.} \right],
\label{eq:L-dibaryon-1}
\end{eqnarray}
and %
\begin{eqnarray}
{\cal L}^{(2)} & \supset & - \frac{h_{2,S}} {2Mr_s} {\cW}^{ab} {s^a}^{\dagger} s^b\,,
\label{eq:L-dibaryon-2}
\end{eqnarray}
respectively. 
Here,  $N=(p,n)^T$ is the  nucleon doublet field and $t_i$ and $s_a$ are the isosinglet ($\siii$) and isotriplet ($\si$) 
 dinucleon fields. 
 The nucleon mass is $M$, the chiral limit axial coupling of the nucleon is $g_{A,0}$, and the spin singlet and triplet effective ranges are $r_{s,t}$, respectively. 
 The two body counterterms at first and second order in the weak interaction are  $l_{1,A}$ and $h_{2,S}$. Finally, $\sigma_i$ and $\tau_i$ are Pauli matrices in spin and isospin space, respectively, with $\tau^{\pm}=(\tau_1 \pm i\ \tau_2)/\sqrt{2}$. 
 For simplicity, the background axial field is defined to be non-vanishing only for the $i=3$ component.
 It is useful to define a new coupling, $\tilde{l}_{1,A}$, that encapsulates solely two-body contributions to the amplitudes,
\begin{eqnarray}
\tilde{l}_{1,A}=l_{1,A}+2M\sqrt{r_s r_t}g_A.
\label{eq:l1Atilde}
\end{eqnarray}

For the second order coupling, ${\cW}^{ab}=W^{\{a}_3W^{b\}}_3$ is the symmetric traceless combination of two background field insertions at the same location, and  only terms relevant for the $nn$ to $pp$ isotensor transition are shown and involve only the isovector dibaryon field. As with $\tilde{l}_{1,A}$,
a new coupling $\tilde{h}_{2,S}$ can be defined to exclude the one-body contributions to the transition amplitudes from the interaction in Eq.~(\ref{eq:L-dibaryon-2}),
\begin{eqnarray}
\tilde{h}_{2,S}=h_{2,S}-\frac{M^2r_s}{2\gamma_s^2}g_A^2.
\label{eq:h2Stilde}
\end{eqnarray}

\subsubsection{Lepton number violation from light Majorana neutrino exchange }
\label{sec:LNV5}

\begin{figure}[t]
\includegraphics[width=\textwidth]{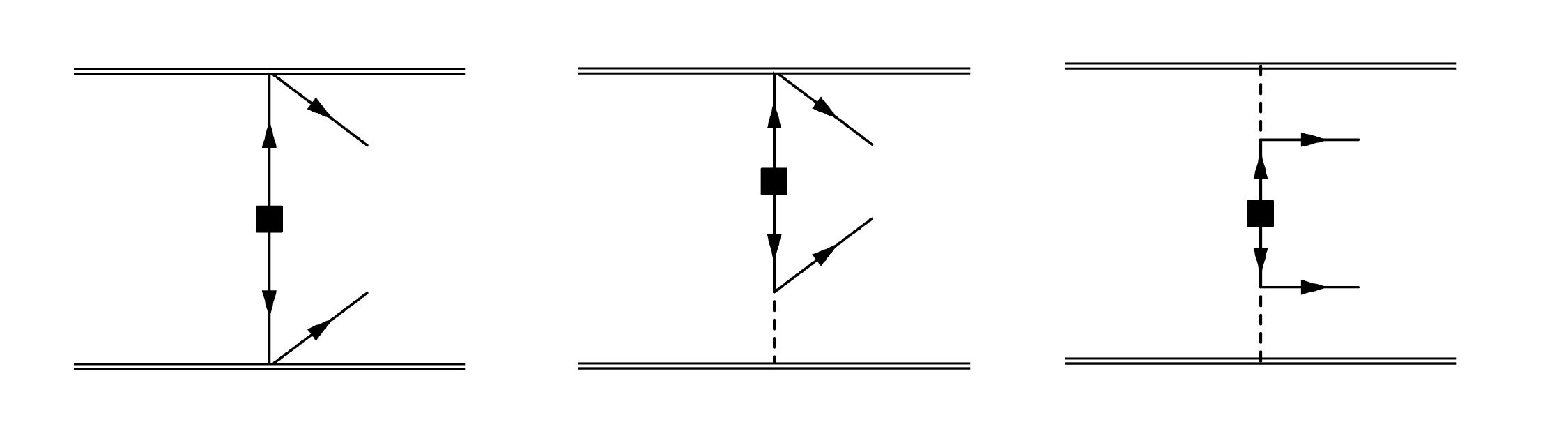}
\caption{Long-range contributions to the neutrino potential $V_\nu^{(a,b)}$. Double and dashed 
lines denote, respectively, nucleons and pions. Single lines denote electrons 
and neutrinos, and squares denote insertions of $m_{\beta\beta}$.}
\label{treelevel}
\end{figure}

In $\chi$EFT, the LO neutrino potential  $V^{(a,b)}_\nu$ 
induced by light Majorana neutrino exchange (see Eqs.~(\ref{eq:intro.2}) and (\ref{eq:HV}))
arises from double insertions of the weak current depicted in  Fig.~\ref{treelevel}, 
entailing long- and pion-range effects, 
as well as a short-range contact interaction~\cite{Cirigliano:2018hja,Cirigliano:2019vdj}: 
\begin{eqnarray}
V^{(1,2)}_{\nu} &=& \frac{\tau^{(1)+} \tau^{(2)+} }{\spacevec{q}^2} 
\left[
1- \frac{2g_{A,0}^2}{3} \boldsigma^{(1)} \cdot \boldsigma^{(2)} 
\left(1 + \frac{m_\pi^4}{2(\spacevec q^2 + m_\pi^2)^2}\right)  
- \frac{g_{A,0}^2 }{3} S^{(12)} \left(1 - \frac{m_\pi^4}{(\spacevec q^2 + m_\pi^2)^2}
\right)
\right]
\nonumber \\
&& -  2 g_\nu^{\rm NN} \ \tau^{(1)+} \tau^{(2)+}~.
 \label{eq:Vnu0}
\end{eqnarray}
Here 
$\spacevec{q}$ is the nucleon momentum-transfer
and
$S^{(12)} = \boldsigma^{(1)} \cdot \boldsigma^{(2)} - 3 \boldsigma^{(1)} \cdot 
\spacevec q \, \boldsigma^{(2)} \cdot \spacevec q/\spacevec q^2$ 
is the spin tensor operator.
Finally, $g_\nu^{\rm NN}$ is an a priori unknown  LO 
contact coupling, scaling as $g_\nu^{\rm NN} \sim O(F_\pi^{-2})$, which encodes the exchange of ``hard" neutrinos with virtualities much  greater than the nuclear scale.   
A term like this is  expected to arise from   
the interactions in Eq.~(\ref{eq:intro.1}), through  
non-factorizable terms induced by quark and gluon exchange in the T-ordered product of two weak currents.  

\begin{figure}[t]
\centering
\includegraphics[width=0.7\textwidth]{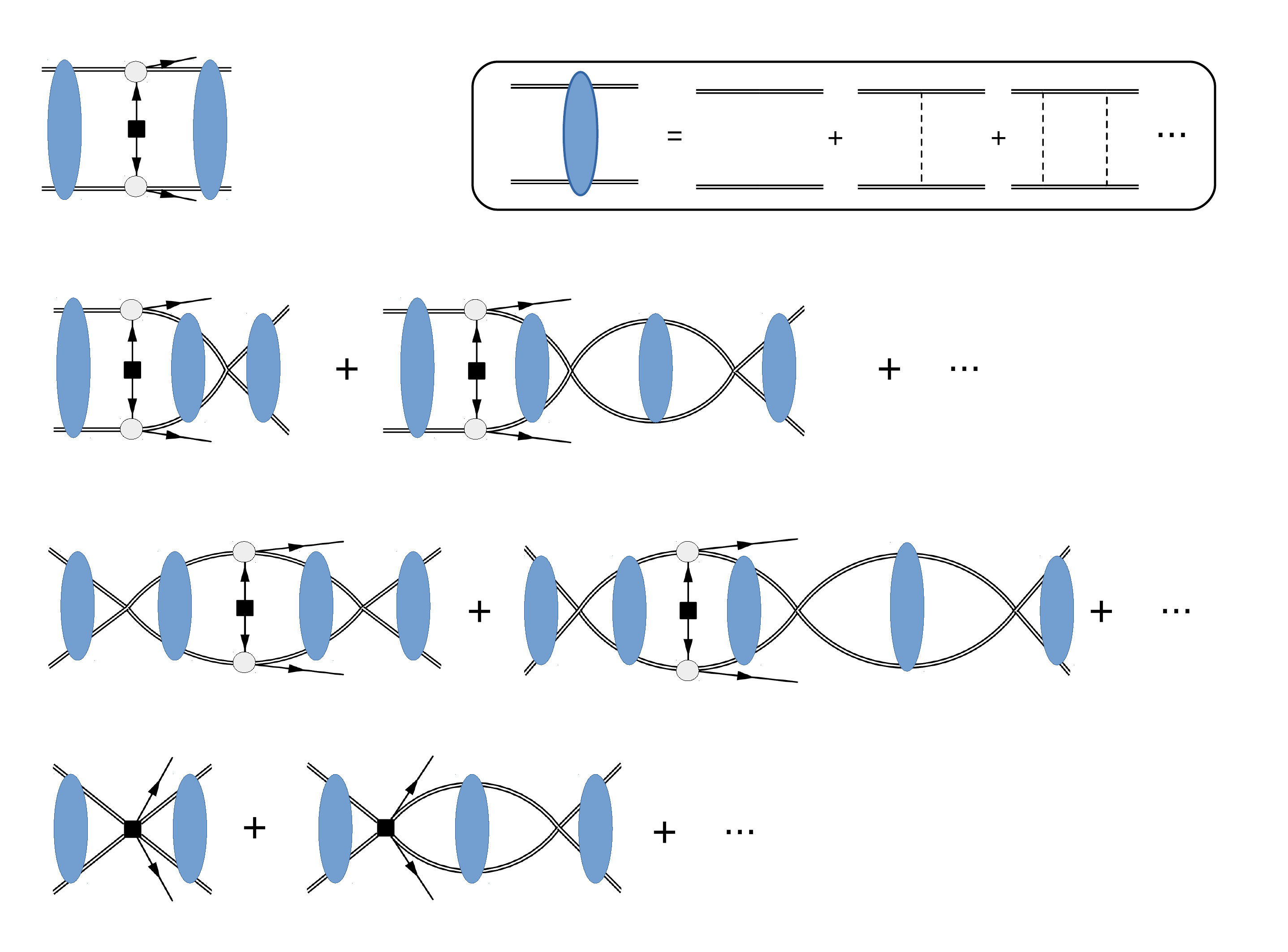}
\caption{Diagrammatic representation of LO contributions to 
$n n \rightarrow p p ee$. Double, dashed, and plain lines denote nucleons, 
pions, and leptons, respectively. 
Gray circles denote the nucleon axial and vector currents, and 
the black square represents an insertion of $m_{\beta \beta}$. 
The blue ellipse represents iteration of 
$V_\pi$. In the counterterm amplitude (fourth line) the black square represents
$g^{N\!N}_\nu$. The ellipses in the second, third, and fourth lines denote diagrams with 
arbitrary numbers of bubble insertions.  
}\label{Fig-ren}
\end{figure}

In the low-energy EFT, the presence of the contact interaction is required by renormalization of the $nn \to pp ee$ amplitude~\cite{Cirigliano:2018hja,Cirigliano:2019vdj}.
The argument is as follows: 
the first term in the  $0\nu\beta\beta$ transition operator in Eq.\eqref{eq:Vnu0} has 
Coulomb-like behavior at large $|\spacevec q|$, which induces ultraviolet (UV) 
divergences in LNV scattering amplitudes, such as 
$nn \to pp ee$, when both 
the two neutrons in the initial state and the two protons in the final state are in the $^1S_0$ channel.
The UV divergence arises when including 
strong re-scattering effects in the $^1S_0$ channel 
through the  LO $\chi$EFT 
potential in the ${}^1S_0$ channel
\begin{equation}
V_{N\!N}^{{}^1S_0}  = 
C_{\si} - \frac{g_A^2 }{4F_\pi^2}
\frac{m_\pi^2}{\spacevec q^2 + m_\pi^2}.
\label{1S0LOpot1}
\end{equation}
Here, $C_{\si} \sim \mathcal O(F_\pi^{-2}, m_\pi^2F_\pi^{-4})$
is a contact interaction that
accounts for short-range physics from pion exchange and other QCD effects.
This term is needed for renormalization and 
to generate the observed, shallow $^1S_0$ virtual state. 
It is expected at 
LO \cite{Weinberg:1990rz,Weinberg:1991um} 
on the basis of naive dimensional analysis (NDA), and it is present at LO 
in all formulations of chiral EFT.
The diagrammatic contributions to the $nn \to pp ee$ 
amplitude to LO in $\chi$EFT are given in Fig.~\ref{Fig-ren}.
The first  diagram in  the third row of Fig.~\ref{Fig-ren} 
has a logarithmic  divergence, which stems from an insertion of the most singular component 
of the neutrino potential, i.e. $1/\spacevec{q}^2$, and two insertions of $C_{\si}$. 
This divergence requires the introduction of the 
contact term $g_\nu^{\rm NN}$, and the associated  
renormalization group  equation determines its 
scaling: $g_\nu^{\rm NN} \sim O(F_\pi^{-2})$~\cite{Cirigliano:2018hja,Cirigliano:2019vdj}.

The contact term $g^{\rm NN}_\nu$ corresponds to a genuine new contribution due to 
the exchange of  neutrinos with momenta 
$|\spacevec{q}|  > \Lambda_\chi$. 
The new coupling encodes  a non-factorizable 
 two-nucleon effect,  beyond the factorizable one-nucleon corrections 
captured by  the radii of 
weak form factors, which also give a short-range neutrino
potential.  
Moreover, $g^{\rm NN}_\nu$ is not part of the so-called
``short-range correlations''~\cite{Miller:1975hu,Simkovic:2009pp,Engel:2011ss,Benhar:2014cka},
as it is  needed even when  one works with fully correlated wavefunctions, i.e.  exact solutions of 
the Schr\"odinger equation with the appropriate strong potential.  
The situation is somehow analogous to $\beta$ decay,
where two-nucleon weak currents and short-range correlations
are both present, and can both be viewed as an ``in-medium quenching''
of $g_A$, as recently discussed in  Refs.~\cite{Pastore:2017uwc,Gysbers:2019uyb}.

The neutrino potential $V_\nu^{(1,2)}$ in $\slashpi$EFT can be obtained by taking the 
$m_\pi \rightarrow \infty$ limit in Eq. \eqref{eq:Vnu0}. 
While the  tensor component of $V_\nu$ vanishes in this limit, 
the most singular part of $V_{\nu}$,  proportional to $1/\spacevec q^2$, survives. 
Therefore, re-scattering effects (in particular, the first diagram in the third row of Fig.~\ref{Fig-ren}, 
without the pion ladder) induced by the LO strong potential 
$V_{N\!N}^{{}^1S_0}  = C_\slashpi \sim 4 \pi/ (M_N \aleph)$ produce a UV divergence and 
require  a LO counterterm $g_\nu^{\rm NN}$ in $\slashpi$EFT as well~\cite{Cirigliano:2017tvr}.

The leading order LEC $g_\nu^{\rm NN}$ is currently unknown. 
Chiral and isospin symmetry arguments relate 
$g_\nu^{NN}$ to one of two $\Delta I=2$ $NN$ contact interactions of electromagnetic origin.
A fit to $NN$ scattering data confirms the 
LO scaling of such couplings, but does not 
allow  the two couplings to be  disentangled and 
hence  $g_\nu^{\rm NN}$ to be extracted~\cite{Cirigliano:2018hja,Cirigliano:2019vdj}. 
Therefore,  the first-principles determination of $g_\nu^{\rm NN}$ 
through LQCD is of the greatest importance. 
This calculation is quite challenging and will be most likely be first performed 
at unphysically large values of the quark  masses. 
This can still be useful for phenomenology, as it will 
allow one to extract $g_\nu^{\rm NN}$ in $\slashpi$EFT, 
which can be related through an EFT matching calculation to $g_\nu^{\rm NN}$ in $\chi$EFT, 
in the schemes suitable for implementation in many-body calculations. 
Subsequent LQCD calculations  closer to  the physical pion mass 
will allow for direct matching to $\chi$EFT.

In Ref.~\cite{Cirigliano:2019vdj}, 
it has been shown that the  short-range \NLDBD \ operator is  only needed in spin-singlet $S$-wave transitions, while leading-order 
transitions involving higher partial waves depend solely on long-range  currents. 
Moreover, Ref.~\cite{Cirigliano:2019vdj}  extended the 
calculation to include next-to-leading order (NLO) corrections 
finding that no additional undetermined parameters appear.

At next-to-next-to-leading order (N$^2$LO), 
a number of corrections to Eq.~(\ref{eq:Vnu0}) arise. 
These include corrections from the momentum dependence of the nucleon vector and axial form 
factors, as well as from weak magnetism. 
These are usually included in the neutrino 
potential, see for example Refs.~\cite{Engel:2016xgb,Ejiri:2019ezh}. 
However, at the same order (N$^{2}$LO) in 
$\chi$EFT,  there appear many other non-factorizable  contributions, 
for instance from pion loops that dress the neutrino exchange~\cite{Cirigliano:2017tvr}.
Due to  UV divergences,  at  N$^{2}$LO  
there appear three new 
$\mathcal O(1)$  LECs, namely  
$g_\nu^{\pi \pi}$, $g_\nu^{\pi N}$,  
and $g_\nu^{NN(2)}$~\cite{Cirigliano:2017tvr}:
\bea 
\label{eq:CTN2LO}
\vL^{(2)}_{\Delta L = 2} = \frac{2 G_F^2 V_{ud}^2  m_{\beta \beta}}{(4 \pi F_0)^2}
\left[    \frac{5}{6}  F_0^2   g_\nu^{\pi \pi} \,  \partial_\mu \pi^- \partial^\mu \pi^- 
+   \sqrt{2}g_{A,0} F_0    g_\nu^{\pi N}  \   \bar p S_\mu n   \,  \partial^\mu  \pi^- 
+ g_\nu^{\rm NN(2)}   \bar p n \,  \bar{p} n   \right]    \bar e_L C\bar e_L^T+\dots ,
\eea
where $S_\mu$ is the nucleon spin vector and $F_0$ is the pion decay constant in the chiral limit. In the nucleon rest frame  one has $S^\mu = (0,\, \boldsigma/2)$.
As in the case of the LO contact term, the couplings  $g_\nu^{\pi \pi}$, $g_\nu^{\pi N}$,  
and $g_\nu^{\rm NN(2)}$
are  related to LECs in the electromagnetic sector. 
Using large-$N_C$ based resonance saturation estimates for the electromagnetic LECs~\cite{Ananthanarayan:2004qk}, one finds  $g_\nu^{\pi \pi}   (\mu = m_\rho)= - 7.6$, 
to which one should attach a conservative 50\% uncertainty.
The LECs can also be computed by matching LQCD and $\chi$EFT  expressions for appropriate scattering amplitudes. So far LQCD efforts have focused on the determination of 
$g_\nu^{\pi \pi}$, see Section~\ref{sec:znubb_long}.

Before concluding this section, we compare the neutrino potential derived in the EFT with the standard approach. 
We begin by noting that in the EFT approach the 
neutrino potential $V_\nu^{(a,b)}$  depends only on the momentum scale $q \sim k_F$ 
and not  on infrared scales 
corresponding to the  excitation  energies of the intermediate 
odd-odd nucleus in \NLDBD. 
In the standard approach, the energy difference  
$E_n - 1/2(E_i + E_f)$ appearing in the denominator 
of second-order perturbation theory is 
often approximated by the average  $\bar{E} - 1/2(E_i + E_f)$, 
which is called the closure approximation. 
Now, up to the contact term $g_\nu^{NN}$, the standard neutrino potential~\cite{Bilenky:2014uka,Engel:2016xgb}   
reduces to the LO potential in Eq.~\eqref{eq:Vnu0} 
when $\bar{E} - 1/2(E_i + E_f)$ is set to zero 
and the form factors are evaluated at zero momentum. 
In the EFT approach, the form-factor effects 
appear to N$^2$LO. 
Similarly, the sensitivity to nuclear intermediate states 
appears in the EFT approach to N$^2$LO 
through the exchange of `ultrasoft' neutrinos ($q^0 \sim |\spacevec{q}| \ll k_F$), as discussed in detail in Ref.~\cite{Cirigliano:2017tvr}.

In summary, for the light Majorana neutrino exchange 
mechanism, LQCD input is needed 
for the LO coupling $g_{\nu}^{\rm NN}$ 
as well as for the N2LO couplings  
$g_\nu^{\pi \pi}$, $g_\nu^{\pi N}$,  
and $g_\nu^{\rm NN(2)}$.
Clearly a determination of $g_{\nu}^{\rm NN}$ 
is the most urgent task 
in order to address the  implications 
of next-generation experimental results 
on the LNV parameter $m_{\beta \beta}$.

\subsubsection{Lepton number violation from short range mechanisms}
\label{sec:LNV9}

The set of $SU(3)_c\times U(1)_{\rm em}$ invariant four-quark, two-lepton operators at dimension-9 can be written as \cite{Graesser:2016bpz,Prezeau:2003xn}
\bea \label{eq:Lag}
\vL^{(9)}_{\Delta L =2} = \frac{1}{v^5}\sum_i\bigg[\left( C^{(9)}_{i\, \rm R}\, \bar e_R C \bar e^T_{R} + C^{(9)}_{i\, \rm L}\, \bar e_L C \bar e^T_{L} \right)  \, O_i +  C^{(9)}_i\bar e\g_\mu\g_5  C \bar e^T\, O_i^\mu\bigg],
\eea
where $O_i$ and  $O_i^\mu$ are four-quark operators that are Lorentz scalars and vectors, respectively. 
The renormalization group evolution of the Wilson coefficients, $C^{(9)}_{i L,R}$ is known and summarized in Ref.~\cite{Cirigliano:2018yza}. 
The scalar operators have been discussed in Refs.~\cite{Graesser:2016bpz,Prezeau:2003xn} and can be written as
\begin{eqnarray}
O_ 1  &=&  \bar{q}_L^\alpha  \gamma_\mu \tau^+ q_L^\alpha \ \bar{q}_L^\beta  \gamma^\mu \tau^+ q_L^\beta\,, \qquad O^\prime_ 1  =  \bar{q}_R^\alpha  \gamma_\mu \tau^+ q_R^\alpha \ \bar{q}_R^\beta  \gamma^\mu \tau^+ q_R^\beta      
\,\,,\nonumber
\\
O_ 2  &=&  \bar{q}_R^\alpha  \tau^+ q_L^\alpha \  \bar{q}_R^\beta  \tau^+ q_L^\beta\,, \qquad \qquad     O^\prime_ 2  =  \bar{q}_L^\alpha  \tau^+ q_R^\alpha \  \bar{q}_L^\beta  \tau^+ q_R^\beta
\,\,,\nonumber
\\
O_ 3  &=&  \bar{q}_R^\alpha  \tau^+ q_L^\beta \  \bar{q}_R^\beta  \tau^+ q_L^\alpha\,, \qquad \qquad    O^\prime_ 3  =  \bar{q}_L^\alpha  \tau^+ q_R^\beta \  \bar{q}_L^\beta  \tau^+ q_R^\alpha
\,\,, 
\label{LagSca}
\\
O_ 4  &=&  \bar{q}_L^\alpha  \gamma_\mu \tau^+ q_L^\alpha \  \bar{q}_R^\beta  \gamma^\mu \tau^+ q_R^\beta    
\,\,,
\nonumber
\\
O_ 5  &=&  \bar{q}_L^\alpha  \gamma_\mu \tau^+ q_L^\beta \  \bar{q}_R^\beta  \gamma^\mu \tau^+ q_R^\alpha\,,
\nonumber
\end{eqnarray}
where $\al$, $\bt$ are color indices. The $O_i'$ operators are related to  $O_i$ by parity. 
Finally, there are four vector operators $O_{6,7,8,9}^\mu$, whose explicit form can 
be found in Ref.~\cite{Graesser:2016bpz}.

\begin{figure}[t]
\centering
\includegraphics[width=0.9\textwidth]{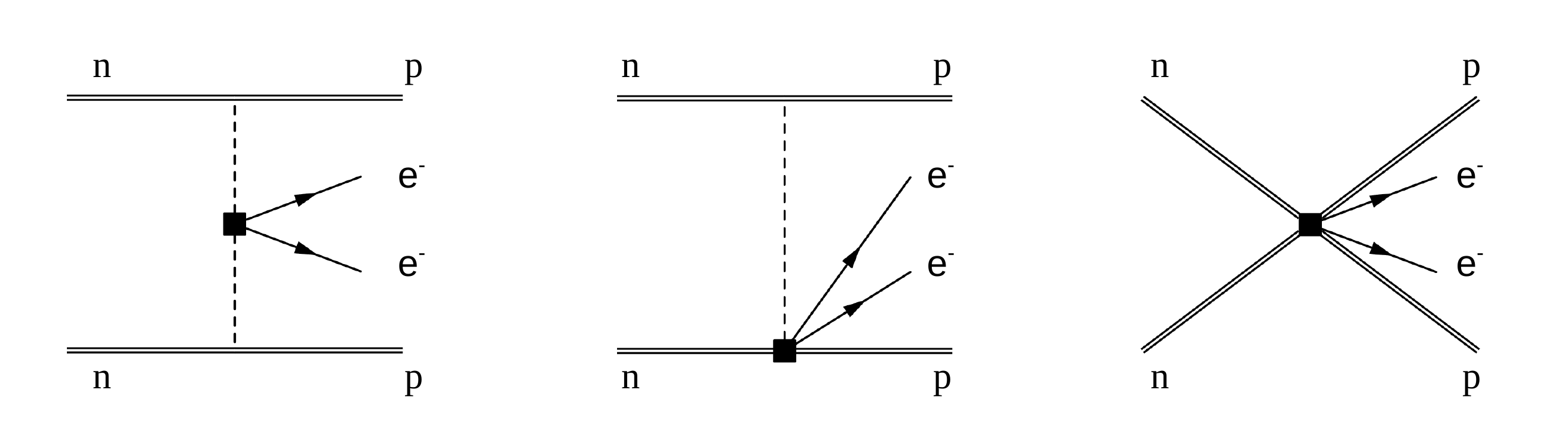}
\caption{The different contributions of dimension-9 LNV operators to the \NLDBD\  potential $V^{(a,b)}_9$, see  Refs.~\cite{Pontecorvo:1968wp, Vergados:1981bm,Faessler:1996ph,Prezeau:2003xn}.
Double, dashed, and single lines denote, respectively, nucleon, pion, and lepton fields. The black square denotes $\Delta L=2$ $\pi\pi$, $\pi N$, and $N\!N$ operators realizing the the dimension-9 quark-level operators at the hadronic level. 
The remaining vertices
are SM interactions between nucleons and pions.}\label{feyndiag}
\end{figure}

In $\chi$EFT, the scalar operators $\{O_{1},\ldots,O_5\}$ generate the $\pi\pi ee $, $\pi N\!N e e$, and $N\!N\, N\!N\,ee$ LNV vertices shown 
as black squares in Fig.~\ref{feyndiag}.
The operators $O_{2,3,4,5}$ induce non-derivative pionic operators~\cite{Cirigliano:2017ymo}, while the first pionic operators induced by $O_{1}$ contain two derivatives~\cite{Savage:1998yh}.
Based on this, within the Weinberg power counting 
of $\chi$EFT, one finds that for $O_{2,3,4,5}$, the dominant contribution to the $nn \to pp$ transition 
operator arises from the double pion exchange diagrams in Fig.~\ref{feyndiag},  while for $O_1$ all diagrams in  Fig.~\ref{feyndiag} are equally 
important~\cite{Prezeau:2003xn}.

The mesonic chiral Lagrangian
for $O_{1,2,3,4,5}^{}$ is given by 
\bea\label{eq:dim9pipi}
\vL^{\rm scalar}_\pi &=&\frac{F_0^4}{4}\left[
\frac{5}{3}g^{\pi\pi}_{1} C_{\rm 1L}^{(9)} L_{21}^\mu L_{21\,\mu}
+\left(g^{\pi\pi}_{2}C_{\rm 2L}^{(9)} + g^{\pi\pi}_{3} C_{\rm 3L}^{(9)}\right) {\rm Tr}\left( U\tau^+U\tau^+\right)\right.\nn\\
&&\left.+
\left(g^{\pi\pi}_{4}C_{\rm 4L}^{(9)}+g_{5}^{\pi\pi}C_{\rm 5L }^{(9)}\right) {\rm Tr}\left( U\tau^+U^\dagger\tau^+\right)
\right]\frac{\bar e_L  C\bar e^T_L}{v^5}+(L\leftrightarrow R)\nn\\
&= &\frac{F_0^2}{2}\left[
\frac{5}{3}g^{\pi\pi}_{1} C^{(9)}_{\rm 1 L}  \,  \partial_\mu \pi^- \partial^\mu \pi^-   
+ \left( g^{\pi\pi}_{4} C^{(9)}_{\rm 4 L}  + g^{\pi\pi}_{5}  C^{(9)}_{\rm 5 L}-g^{\pi\pi}_{2} C^{(9)}_{\rm 2 L}  - g^{\pi\pi}_{3} C^{(9)}_{\rm 3 L}\right)  \pi^- \pi^- 
\right]
\frac{\bar e_L  C\bar e^T_L}{v^5}
\nn 
\\ & & +(L\leftrightarrow R)+\dots\,\,,\eea
where $U=u^2={\rm exp}\left(i \pi\cdot \tau/F_0\right)$ is the matrix of pseudo-Goldstone boson fields, and  $L_{\mu} = iUD_\mu U^\dagger$. 
Assuming NDA,  the LECs  of the non-derivative pion operators are expected to be $g^{\pi\pi}_{2,3,4,5}  = \mathcal O(\Lambda^2_\chi)$,  while $g^{\pi\pi}_{1} = \mathcal O(1)$. 
In Section~\ref{sec:znubb_short},  the LQCD determination of these LECs is discussed. 
The physical amplitudes are scale and scheme independent provided one uses Wilson coefficients $C^{(9)}_i$   
evaluated at the same scale and in the same scheme as used for the $g_i^{\pi \pi}$.

The $\pi N$  terms are only  relevant for the $O_1$ operator and can be written as
\bea\label{eq:dim9PiN}
\vL_{\pi N}^{\rm scalar}& =& g_Ag^{\pi N}_{1}C_{\rm 1L}^{(9)}F_0^2\left[\,\bar N S^\mu u^\dagger\tau^+ u N\,{\rm Tr}\left(u_\mu u^\dagger \tau^+ u \right)\right]\frac{\bar e_L  C\bar e^T_L }{v^5}+(L\leftrightarrow R)\nn\\
&=&\sqrt{2}g_A g^{\pi N}_{1}C_{\rm 1L}^{(9)}F_0 \left[\bar p\, S\cdot (\partial \pi^-)n\right] \, \frac{\bar e_L  C\bar e^T_L}{v^5}+(L\leftrightarrow R)+\dots\,\,,
\eea
where $u_\mu=u^\dagger L_\mu u = i\left[u(\partial_\mu-ir_\mu)u^\dagger -u^\dagger(\partial_\mu-il_\mu)u \right]  $. 
The LEC $g_1^{\pi N}$ is currently unknown, but expected to be $\mathcal O(1)$ by NDA.

In a power counting based on NDA, LNV four-nucleon contact interactions are relevant at LO only for $O_1$,
together with the $\pi \pi$ and $\pi N$ interactions $g^{\pi\pi}_{1}$ and $g^{\pi N}_{1}$.
However, the LNV potential induced by the non-derivative $\pi\pi$ operators in Eq.\ \eqref{eq:dim9pipi}
has the same high-momentum behavior as the neutrino potential mediated by the  Majorana neutrino mass,  $V(\vec q) \sim 1/\vec q^2$ at large $|\vec{q}|$.
In Refs.\ \cite{Cirigliano:2018hja,Cirigliano:2018yza} it has been shown that for these potentials the $nn \rightarrow p p e e$ scattering amplitude  has a logarithmic UV divergence, which must be removed by promoting the $N\!N$ operators 
stemming from  $O_{2,3,4,5}$ to leading order. The relevant  $N\!N$ operators are 
\bea\label{eq:dim9NN}
\vL_{NN}^{\rm scalar} &=& g_{1}^{NN} C_{\rm 1L}^{(9)}  \, (\bar N u^\dagger \tau^+ u N)(\bar N u^\dagger \tau^+ u N)\, \frac{\bar e_L  C\bar e^T_L}{v^5}\nn \\
& & +  \left(   g_{2}^{NN} C_{\rm 2L}^{(9)} +  g_{3}^{NN} C_{\rm 3L}^{(9)}  \right)\, (\bar N u^\dagger \tau^+ u^\dagger N)(\bar N u^\dagger \tau^+ u^\dagger N)\, \frac{\bar e_L  C\bar e^T_L}{v^5}  \nn \\
& & +  \left(  g_{4}^{NN} C_{\rm 4L}^{(9)}  +  g_{5}^{NN} C_{\rm 5L}^{(9)}  \right)\, (\bar N u^\dagger \tau^+ u N)(\bar N u \tau^+ u^\dagger N)\, \frac{\bar e_L  C\bar e^T_L}{v^5}
+(L\leftrightarrow R)
\nn\\
& =& \left(  g_{1}^{NN} C_{\rm 1L}^{(9)} +   g_{2}^{NN}  C_{\rm 2L}^{(9)} +  g_{3}^{NN} C_{\rm 3L}^{(9)}  +
 g_{4}^{NN} C_{\rm 4L}^{(9)}  +   g_{5}^{NN} C_{\rm 5L}^{(9)} \right) \left(\bar pn\right )\,\left(\bar pn\right )\, \frac{\bar e_L  C\bar e^T_L}{v^5}\nn \\ & & +(L\leftrightarrow R)+\dots\,.
\eea
In the Weinberg power counting, the scaling $g_{i}^{NN} \sim \mathcal O(1)$ holds. However, in a properly  renormalized 
$\chi$EFT, the scaling is modified to 
$g_i^{NN}  \sim \mathcal O((4\pi)^2)$  for  
$O_{2,3,4,5}$~\cite{Cirigliano:2018yza}. The renormalization of the scattering amplitude does not require such enhancement for $g_{1}^{NN}$. \footnote{The  $\pi\pi$, $\pi N$, and $N\!N$ Lagrangians for the $O_{1,2,3}'$ operators can be related to the ones for 
$O_{1,2,3}$ by parity considerations,  leading to $\pi\pi$, $\pi N$, and $NN$ vertices of the same form 
as above, with the replacement  $C_{1,2,3}^{(L,R)}\to C_{1,2,3}^{\prime\,(L,R)}$.}

With the above chiral Lagrangians,  the LO $\Delta L=2$ \NLDBD\ potential from the scalar 
dimension-9 operators is given by
\bea\label{scalar}
V_{9}^{(1,2)} &=& -( \tau^{(1) +} \tau^{(2) + }) \,  \frac{2 g_A\sq}{v}    \nonumber \\
& \times &   \!\!\!\! 
\left[ -  
    \left(  \, \boldsigma^{(1)} \cdot \boldsigma^{(2)}-  \, S^{(12)}  \right) \left( \frac{C^{(9)}_{\pi \pi\,\rm L}}{6}   \frac{\vec q^2}{(\vec q^2 + m_\pi^2)^2}  -  \frac{ C^{(9)}_{\pi N\,\rm L}}{3} \frac{\vec q^2}{\vec q^2 + m_\pi^2} \right)
+ \frac{2}{g_A^2} C_{NN\, \rm L}^{(9)}  \right] ~,
\eea
where the combinations $C^{(9)}_{\pi\pi,\, \pi N,\, NN}$ are defined as 
\begin{eqnarray}
C^{(9)}_{\pi \pi\, \rm L} & = &  g^{\pi\pi}_{2} \left(C_{\rm 2L}^{(9)} + C_{\rm 2L}^{(9)\, \prime}\right) +    g^{\pi\pi}_{3}\left(C_{\rm 3L}^{(9)} + C_{\rm 3L}^{(9)\, \prime}\right) 
- g^{\pi\pi}_{4} C^{(9)}_{\rm 4L} -  g^{\pi\pi}_{5} C^{(9)}_{\rm 5L}   \nn \\ 
& & - \frac{5}{3} g^{\pi\pi}_{1} m_\pi^2  \left(C_{\rm 1L}^{(9)} + C_{\rm 1L}^{(9)\, \prime}\right)\,, \nn \\
C^{(9)}_{\pi N\, \rm L} &=& \left(g_{1}^{\pi N} - \frac{5}{6}g^{\pi\pi}_{1}  \right)  \left(C_{\rm 1L}^{(9)} + C_{\rm 1L}^{(9)\, \prime}\right)\,, \nn \\
C_{NN\, \rm L }^{(9)} &=&   g_{1}^{NN}\, \left(C_{\rm 1L}^{(9)} + C_{\rm 1L}^{(9)\, \prime}\right) +   g_{2}^{NN} \, \left(C_{\rm 2L}^{(9)} + C_{\rm 2L}^{(9)\, \prime}\right)  + g_{3}^{NN} \left(C_{\rm 3L}^{(9)} + C_{\rm 3L}^{(9)\prime}\right)   \nn \\
& & +   g_{4}^{NN} C_{\rm 4L}^{(9)} +  g_{5}^{NN} C_{\rm 5L}^{(9)}  \,,
\end{eqnarray}
and similarly for $C_{\{\pi\pi, \, \pi N,\, NN\}\,\rm R }^{(9)}$. 
In the above expressions one has $q^\mu = (q^0,\,\vec q) = (p-p')^\mu$, where $2p$ and $2p'$ are the relative momenta of the ingoing  and outgoing nucleon pairs.  
The potential in Eq.~(\ref{scalar}) 
can be implemented in many-body nuclear calculations to obtain bounds on $C^{(9)}_{\pi\pi,\, \pi N,\, NN}$, 
which, using knowledge of the $g_i^{\pi \pi, \pi N, NN}$, can then be converted into bounds on the 
Wilson coefficients $C^{(9)}_{iL, iR}$, 
and hence on the underlying LNV  model parameters. 
We conclude this discussion by noting 
that the pion contributions to the potential $V_{9}^{(1,2)}$ have  
appeared throughout the \NLDBD\ literature in the context of various 
models, see for example~\cite{Pontecorvo:1968wp,Vergados:1981bm, Faessler:1996ph, Faessler:1998qv,Faessler:2007nz}.

In summary, one finds that  for all scalar operators 
in Eq.~(\ref{eq:Lag}) the $\pi\pi ee$ and $NN$ interactions contribute at the same order (LO) to the two-nucleon  transition operator.  
Moreover, for $O_1$ and $O'_1$, there appears an additional LO contribution from the $\pi N$ interaction. 
Chiral symmetry implies that the contributions from  
$O_1$ to  $V_9^{(1,2)}$  is suppressed by 
$\epsilon_\chi^2$ compared to the contributions 
induced by $O_{2,3,4,5}$.  
The transition operator $V_9^{(1,2)}$ 
induced by the vector operators $O_{6,7,8,9}^\mu$
has the same chiral scaling as the one induced by $O_1$, 
and it is dominated  by the $\pi N$ and $NN$  contributions~\cite{Cirigliano:2018yza}. 
The above considerations imply that 
from the phenomenological point of view 
the most needed LQCD matrix elements are
$\langle \pi^+ |O_{2,3,4,5} | \pi^- \rangle$ 
and $\langle pp |O_{2,3,4,5} |nn \rangle$, 
followed by 
$\langle \pi^+ |O_1 | \pi^- \rangle$, 
$\langle pp |O_1 | nn \rangle$,  
$\langle \pi^+ p |O_1 | n \rangle$,
$\langle pp |O_{6,7,8,9}^\mu | nn \rangle$, and   
$\langle \pi^+ p |O_{6,7,8,9}^\mu | n \rangle$.

\subsection{Lattice QCD  }
\label{sec:LQCD}

The hadronic and nuclear physics inputs needed to study double $\beta$-decay can be calculated from the underlying Standard Model, where the most relevant part is quantum chromodynamics (QCD), using lattice field theory techniques, referred to as {lattice QCD} (LQCD). In this approach,  the relevant information is extracted from various correlation functions that are evaluated from their QCD functional integral representation. As an intermediate stage of these calculations, a Euclidean space-time lattice is used to regulate the divergences of the theory by making the functional integral finite-dimensional. Because the normalized exponential of the Euclidean discretized QCD action has the same form as a Boltzmann distribution, importance sampling Monte-Carlo methods can be used efficiently to stochastically evaluate the requisite integrals. The physical information is then recovered in the limit in which the lattice regulator is removed (the {\it continuum limit})  and the limit in which the space-time volume is taken to infinity (the {\it infinite volume limit}). Many excellent introductions to lattice QCD exist, see for example Refs.~\cite{Rothe:1992nt,Gattringer:2010zz}, and the reader is referred to these works for more complete details. In this brief overview,   aspects of lattice QCD that  impact the discussion of quantities relevant for double $\beta$-decay from LQCD are discussed.

The formulation of LQCD, first proposed by Wilson \cite{Wilson:1974sk},  uses a discrete, space-time geometry which in almost all cases is taken to be  a regular four-dimensional hypercube, $\Lambda_4=\{n_\mu=(n_1,n_2,n_3,n_4) | n_i \in a [0,1,\ldots L_i]\}$, where $a$ is the (dimensionful) lattice spacing and $L_i$ is the extent of the lattice in the $i$th direction. Periodic spatial boundary conditions are typically used on all fields, with periodic temporal boundary conditions on gluon fields and anti-periodic temporal boundary conditions for quarks (thereby implementing finite temperature). In some cases, anisotropy is introduced between the spatial and temporal directions, providing a finer discretization in the temporal direction, and other geometries have been investigated in the past \cite{Celmaster:1982ht}. 

The underlying QCD action must be implemented approximately on this discretized geometry, replacing derivatives by finite differences and implementing the gauge fields in terms of the links between the sites of the lattice. 
For the gauge fields, it is common to use the Wilson action 
\begin{equation}
    S_{\rm Wilson} =\frac{2}{g^2} \sum_{x\in\Lambda_4} \sum_{\mu<\nu} (1-{\rm Re\ Tr}[P_{\mu\nu}(x)])\,,
\end{equation}
where $P_{\mu\nu}(x)$ is the elementary plaquette and corresponds to products of gauge link variables $U_\mu(x)$ around a $1\times1$ elementary cell, $$P_{\mu\nu}(x)={\rm Tr}[U_\mu(x)U_\nu(x+\hat\mu)U_\mu^\dagger(x+\hat \nu) U_\nu^\dagger(x)].$$ The link variables $U_\mu(x)=\exp(iga A_\mu(x))$ are associated with the site $x$ and are parallel transporters to the site $x+\hat\mu$. Expanding this action around the limit $a\to0$ reproduces the continuum QCD action up to ${\cal O}(a^2)$ effects. Variants of this action introduce additional terms that remove higher powers of $a$, providing a closer-to-continuum, improved action.

Naive implementations of lattice fermions have multiple zero modes ($2^d$ of them, where $d$ is the space-time dimension) corresponding to ``doubling'' of the light degrees of freedom. These are avoided with the Wilson quark action \cite{Wilson:1974sk}, the Kogut-Susskind \cite{Kogut:1974ag} quark action, and twisted-mass quark actions \cite{Frezzotti:2003ni}, but these actions explicitly break the chiral symmetry of the massless QCD action. Chiral fermion formulations such as the domain-wall fermion action \cite{Kaplan:1992bt}, which bypasses this issue by introducing an additional space-time dimension, and the overlap fermion action \cite{Narayanan:1994gw} maintain a lattice chiral symmetry. These chiral actions are  more numerically expensive to implement, but  offer advantages in certain  contexts; for example, the additional symmetry can prohibit unwanted operator mixings. As with the gauge actions, the action for fermions can also be improved to reduce discretization artifacts; in this case, there is a unique dimension-five operator to add to the action \cite{Sheikholeslami:1985ij}, $\overline\psi \sigma_{\mu\nu}G^{\mu\nu}\psi$, known as the clover term. The (L)QCD action is bilinear in the fermion fields, $S_{\rm fermion} \sim\int dx \overline\psi {\cal M} \psi$ (where the Dirac operator ${\cal M}$ depends on the choice of action and on the gauge field) and consequently the fermions can be integrated over exactly. For the  action encoded in ${\cal M}$, this results in a fermion determinant ${\rm Det} {\cal M}$ in the gauge field functional integral or equivalently an effective action $S_{\rm eff}= {\rm Tr Ln}[{\cal M}]$.

Given a particular form of the lattice action, LQCD calculations proceed by evaluating the QCD path integrals that define the appropriate correlation functions using importance sampling Monte Carlo based on the distribution defined by that action. 
For an operator ${\cal O}(x_1,x_2,\ldots)$ built from quark and gluon fields, the expectation value is determined by the integral
\begin{equation}
    \langle {\cal O}(x_1,x_2,\ldots)\rangle = \frac{1}{\cal Z}\int {\cal D}U \tilde{\cal O}(x_1,x_2,\ldots) \det[{\cal M}[U]] e^{-S_{\rm gauge}}
\end{equation}
where the partition function is defined as ${\cal Z} = \int {\cal D}U  \det[{\cal M}[U]]e^{-S_{\rm gauge}}$. 
The (multi-local) field operator $\tilde{\cal O}(x_1,x_2,\ldots)$ corresponds to the original operator ${\cal O}$ after the quark fields have been integrated out; this integration results in the ``contraction'' of fermion--anti-fermion pairs in all possible ways, replacing them with quark propagators $S[U]={\cal M}[U]^{-1}$.
To evaluate the integral, Monte-Carlo methods are used; the factor 
${\cal P}[U] = {\cal Z}^{-1}\det[{\cal M}[U]] e^{-S_{\rm gauge}}$ is common to all such integrals and is a probability distribution (non-negative definite and bounded). Sampling the gluon field according to this distribution, the integral above can be approximated as
\begin{equation}
    \langle {\cal O}(x_1,x_2,\ldots)\rangle \approx \frac{1}{N} \sum_{c=1}^N \tilde{\cal O}(x_1,x_2,\ldots)[U_i] 
    + {\cal O}\left(1/\sqrt{N}\right)
\end{equation}
where $\{U_1,\ldots U_N\}$ corresponds to an appropriately distributed set (ensemble) of gauge fields. These requisite configurations are produced with the correct distribution as a  Markov chain Monte Carlo process, with the standard algorithm being Hybrid Monte Carlo \cite{Duane:1987de} (HMC). In the previous millennium, many studies were performed in  ``quenched QCD'' in which the quark determinant above was omitted for computational expediency. Modern calculations do not do this, although the freedom 
of using different value of the quark massfes in the quark determinant (referred to as sea quarks) and the quark propagators (valence quarks) is sometimes used and is referred to as partial quenching.

To perform a lattice calculation, the quark masses and the gauge coupling need to be specified (the  irrelevant operator couplings are also needed if such terms are used to improve the action). To determine these parameters, $N_f+1$ physical quantities must be computed and compared to experiment. While a number of different approaches to the tuning are taken, a  standard method is to use the pion and kaon masses and a quantity that is relatively mass independent, such as the Wilson flow scale $t_0$ \cite{Luscher:2010iy}, for this purpose. 
Having undertaken sets of simulations at a range of different values of the bare parameters, extrapolations to the continuum and infinite volume limits must be performed in order for physical results to be determined. As well as the statistical uncertainties of the simulations, the uncertainties that arise in taking these limits must be carefully investigated and accounted for.  LQCD actions differ from the continuum QCD action by terms of ${\cal O}(a)$ or in some cases ${\cal O}(a^2)$;  many volume effects are controlled by terms ${\cal O}(e^{-m_\pi L})$ where $L$ is the smallest dimension of the lattice geometry and $m_\pi$ is the pion mass, being the lightest hadron. In most cases, LQCD calculations are performed with degenerate up and down quark masses, ignoring the up and down quark mass spitting, and do not include the effects of electromagnetism as these contributions are typically small, although both effects can be (and are) included when necessary. Precision calculations must account for these additional systematic effects.

\subsubsection{Example: the proton mass}

To  further introduce the LQCD method, it is useful to overview calculations of the proton mass which proceed via the evaluation of correlation functions on representative ensembles of gauge configurations. 
The proton mass can be determined from two-point correlation functions (assuming an infinite temporal extent of the lattice geometry for simplicity, and making use of translational invariance):
\begin{eqnarray}
C_{\alpha\beta}(t,{\bf p})= a^3 \sum_{\bf x} e^{-i{\bf p \cdot x}} C_{\alpha\beta}(t,{\bf x})= a^3\sum_{\bf x} e^{-i{\bf p \cdot x}}\langle 0 |\chi_\alpha({\bf x},t) \overline\chi_\beta({\bf 0},0) |0\rangle \,,
\label{eq:c2pt}
\end{eqnarray}
where  ${\bf p}$ is a chosen three-momentum, $x^\mu=(t,{\bf x})$ 
and
\begin{eqnarray}
\chi_\alpha({\bf x},t) = \epsilon^{ijk} u^i_\alpha({\bf x},t) u^j_\gamma({\bf x},t) [C^{-1}\gamma_5]_{\gamma\delta} d^k_\delta({\bf x},t) 
\end{eqnarray}
is an interpolating operator that has the quantum numbers of the proton and $C=\gamma_0\gamma_2$ is the charge conjugation matrix ($C\gamma_0^T C^{-1}=-\gamma_0$). 
After the quark fields are integrated out in the path integral formulation, this correlator is expressed in terms of the gluon field and products of the inverse of the Dirac operator (which depends on the gluon field):
\begin{eqnarray}
 C_{\alpha\beta}(t,{\bf x})
 &=&  - a^3  \sum_{\bf x} e^{-i{\bf p \cdot x}}
 \epsilon^{ijk} \epsilon^{i'j'k'} [C^{-1}\gamma_5]_{\alpha'\alpha''} [\gamma_5 C]_{\beta'\beta''}
 \\ \nonumber && \hspace*{5mm}\times
\left\langle \left[{\cal M}_d^{-1}\right]_{\alpha''\beta'}^{ki'} \left\{ \left[{\cal M}_u^{-1}\right]_{\alpha'\beta''}^{jj'}\left[{\cal M}^{-1}_u\right]_{\alpha\beta}^{ik'} - \left[{\cal M}_u^{-1}\right]_{\alpha\beta''}^{ij'}\left[{\cal M}^{-1}_u\right]_{\alpha'\beta}^{jk'} \right\}  \right\rangle
\end{eqnarray}
where the quark propagator ${\cal M}_f^{-1}={\cal M}^{-1}_f(x,0)$ is the inverse of the Dirac operator for flavor $f$. This correlator can be evaluated stochastically as an average over representative gluon field configurations as discussed above.

By inserting a complete set of states between the source and sink interpolating operators in Eq.~\eqref{eq:c2pt}, it is clear that this two-point correlator has time dependence governed by the energies of states with the quantum numbers of the proton and with three-momentum ${\bf p}$:
\begin{eqnarray}
C_{\alpha\beta}(t,{\bf p})&=& a^3 \sum_{n,\sigma} \frac{e^{-E_n({\bf p}) t}}{2 E_n({\bf p})} \langle 0 | \chi_\alpha | n; {\bf p}, \sigma \rangle  \langle n; {\bf p},\sigma  | \overline\chi_\beta |0\rangle 
\\
&=& a^3 Z({\bf p}) \sum_\sigma u_\alpha(n=0, {\bf p},\sigma) \overline{u}_\beta(n=0, {\bf p},\sigma) \frac{e^{-E_n({\bf p}) t}}{2 E_n({\bf p})} + \ldots
\end{eqnarray}
where $Z({\bf p})$ is an overlap factor and  higher excited states are indicated by the ellipsis but provide only exponentially small contributions in the large time limit. Taking the trace of this correlator with a given Dirac structure, often chosen to be $\Gamma=\frac{1}{2}(1+\gamma_4)$, leads to
\begin{eqnarray}
\Gamma_{\beta\alpha} C_{\alpha\beta}(t,{\bf p}) = \sum_n A_n({\bf p}) e^{-E_n({\bf p})t}, 
\end{eqnarray}
where the $A_n({\bf p})$ are products of overlap factors.
From analyzing the time dependence of correlators determined on a representative  set of gauge configurations, the energy of the   proton state of the specified momentum can be extracted. This can be achieved either using fits to the time dependence at late times or by employing more sophisticated variational approaches based on correlators with sets of different source and sink interpolating operators \cite{Luscher:1990ux,Michael:1985ne,Blossier:2009kd,Detmold:2014hla}.

\subsubsection{Operator renormalization}

An important application of LQCD that is centrally relevant to the topic of this review is in computing the matrix elements of external currents in hadronic and nuclear states. In the continuum, the external currents one might consider are operators such as the axial-vector quark current $\overline\psi\gamma_\mu\gamma_5\psi$,  or four-quark operators $\overline\psi\Gamma\psi \overline\psi\widetilde\Gamma\psi$ (where $\Gamma$ and $\widetilde\Gamma$ are Dirac and flavor structures) arising from integrating out physics above the hadronic scale. In the lattice  theory,  these operators are implemented using the lattice degrees of freedom and differ from the continuum operators by terms ${\cal O}(a)$ (improved lattice operators can be constructed that eliminate the lattice artifacts at a particular order). Even for operators such as the vector and axial-vector current, the lattice operators must be renormalized to connect to the continuum operators.

At a sufficiently high scale (fine lattice spacing), lattice perturbation theory may in principle be used to connect lattice and continuum operators. By performing calculations in lattice perturbation theory with the appropriate lattice action \cite{Capitani:2002mp},  along with the corresponding continuum perturbation theory calculations, matrix elements of lattice operators can be converted to the corresponding continuum operators in a given renormalization scheme (for the case of scale dependent operators). Due to the complexities of lattice perturbation theory, these calculations are typically performed to one loop and introduce matching uncertainties of ${\cal O}(\alpha_s^2(a))$ that are a limiting factor in the precision with which final physical results can be determined.
An alternative is to utilize a non-perturbative  renormalization scheme. Such approaches involve calculating the appropriate vertex function non-perturbatively within the lattice framework itself as an intermediate step before being matched onto perturbative calculations performed 
in the modified minimal subtraction scheme
($\overline{\rm MS}$). A popular choice is the so-called regularisation independent momentum subtraction (RI/MOM) scheme~\cite{Martinelli:1994ty}, in which correlation functions of renormalized operators do not depend on the choice of regulator, up to cutoff effects. The renormalization constants, $Z$, are defined as,
\begin{eqnarray}
O_r(\mu) = \lim_{a\to 0} Z(\mu,a)O_\mathrm{latt}(a) \ .
\end{eqnarray}
In order to suppress nonperturbative effects from chiral symmetry breaking and other infrared effects, as well as truncation error in the conversion ratio from the RI/MOM scheme to the $\overline{\rm MS}$ scheme, the so-called RI/SMOM scheme may be employed~\cite{Sturm:2009kb}. This scheme is an RI/MOM scheme with non-exceptional kinematics, using the renormalization scale $\mu^2 = \left(p_{\mathrm{in}} - p_{\mathrm{out}}\right)^2 = p_\mathrm{in}^2 = p_\mathrm{out}^2$, which is symmetric in the incoming and outgoing momenta ($p_{\mathrm{in}, \mathrm{out}}$, respectively) of the vertex function.

\subsubsection{3-point matrix element calculations}

A common way to extract matrix elements from LQCD calculations is to form a 3-point function, $C_i^{3\mathrm{pt}}({\tau, t};{\bf p},{\bf q})$, 
in which the desired operator is inserted between source and sink operators coupling to the states of interest, which are separated in Euclidean time:  

\begin{equation}\label{eq:C3pt}
C^{3pt}_{\alpha \beta}(\tau,t;{\bf p},{\bf q})=\sum_{{\bf x},{\bf y}}e^{i{\bf p}\cdot{\bf x}}e^{i{\bf q}\cdot{\bf y}}\langle0|\chi_\alpha(0,{\bf 0}) \mathcal{O}(\tau,{\bf y})\chi_\beta(t,{\bf x}) |0\rangle \ ,
\end{equation}
where $\mathcal{O}$ is the operator of interest, and $\chi_{\alpha,(\beta)}(t)$ is an interpolating operator with non-zero overlap onto the desired final (initial) state. For the  purposes of this review,  zero-momentum states and vanishing momentum insertion at the current are sufficient and are chosen henceforth; extraction of matrix elements in states of non-zero momentum and with non-zero momentum transfer at the current are simple extensions.  Inserting two complete sets of states, $|\alpha'\rangle,|\beta'\rangle$ to this expression gives,
\begin{equation}
C^{3pt}_{\alpha \beta}(\tau,t;{\bf 0},{\bf 0})=\sum_{\alpha',\beta'} e^{-E_{\alpha'}\tau}e^{-E_{\beta'}(\tau-t)} \langle \alpha' | \chi_\alpha | 0\rangle \langle 0|\chi_\beta | \beta' \rangle \mathcal{O}_{\alpha'\beta'} \ ,
\end{equation}
where $\mathcal{O}_{\alpha'\beta'}=\langle \alpha'|{\cal O}|\beta'\rangle$ are the matrix elements between eigenstates $|\alpha'\rangle,|\beta'\rangle$ and are the quantities of interest. 
One may extract the desired ground state matrix element, $\mathcal{O}_{00}$, by simultaneously taking the large $\{\tau,\tau-t\}$ limits of this expression
\begin{equation}
C^{3pt}_{\alpha \beta}(\tau,t; {\bf 0},{\bf 0})\stackrel{\tau,\tau-t\to\infty}{\longrightarrow} e^{-E_{\alpha,0}\tau}e^{-E_{\beta,0}(\tau-t)} Z_{\alpha,0} Z_{\beta,0} \mathcal{O}_{00} \ ,
\end{equation}
where $E_{\alpha,0}$ ($Z_{\alpha,0}$) are the ground-state energy (wavefunction overlap) corresponding to interpolating field $\chi_\alpha$. These constants may be determined from a simultaneous fit to the corresponding 2-point correlation functions, or by forming appropriate ratios with the 2- and 3-point functions to eliminate these contributions. Excited state contamination can be shown to be exponentially suppressed in the limit of large $\{t,\tau, |\tau-t|\}$.

\subsubsection{Background field techniques}

An alternative method for extracting hadronic and nuclear matrix elements is by undertaking spectroscopy calculations in the presence of a fixed external field \cite{Chambers:2014qaa,Chambers:2015bka,Chambers:2015kuw,Savage:2016kon,Bouchard:2016heu,Detmold:2006vu,Detmold:2009dx,Detmold:2010ts,Lujan:2014kia,Chang:2015qxa,Lujan:2016ffj,Shanahan:2017bgi,Tiburzi:2017iux,Agadjanov:2016cjc,Agadjanov:2018yxh}. This method was first used to study the proton axial charge and magnetic moment in Refs.~\cite{Fucito:1982ff,Bernard:1982yu,Martinelli:1982cb} and has subsequently been used to extract polarizabilities.
Background fields can be implemented in a number of ways. One approach is to modify the  gluon link field to incorporate an external U(1) field. This results in all-orders couplings of the external field to the quarks, but provided the field is small (such fields must be of quantized strength to be consistent with the periodic lattice geometry), the linear and quadratic responses can be determined. For the background field calculations discussed in Section \ref{sec:tnubb}, an alternative fixed-order approach is used.

In the fixed order approach of Ref.~\cite{Savage:2016kon,Tiburzi:2017iux}, the hadronic and nuclear correlation functions are modified  at the level of the valence quark propagators. Such \emph{compound propagators} in the background field can be written as
\begin{eqnarray}
S_{\{\Lambda_1,\Lambda_2,\ldots \}}(x,y) = S(x,y)
&+& \int \! dz \, S(x,z) \Lambda_1(z)  S(z,y)
\nonumber \\
&+& \int \! dz \int \! dw \, S(x,z) \Lambda_1(z) S(z,w) \Lambda_2(w) S(w,y) + \ldots
\label{eq:bfprop},
\end{eqnarray}
where 
$\Lambda_i(x)$ are space-time-dependent matrices in spinor and flavor space, while $S(x,y)$ is  a matrix in color, spin (and in principle flavor) space. 
Once the background fields $\Lambda_i(z)$ are specified, the  sequential-source technique is used to calculate the second, 
third 
and all subsequent terms in Eq.~(\ref{eq:bfprop}), which are then combined with the first to form the compound propagator (each insertion of a coupling to the field requiring an extra inversion). 
Since couplings to the sea quarks are not included, this approach is only exact for isovector combinations of fields in the isospin-symmetric limit  and, even then, only for  maximally stretched isospin ($I_3=\pm I$) quantities  and thus do not involve operators that couple to the sea quarks. 
At the single-insertion level, this corresponds to isovector quantities such as the isovector axial charges of the proton and triton, and the axial matrix element relevant for the $pp\to d e^+\nu_e$ fusion cross section. 
With two insertions of the background field, either through the third term in Eq.~(\ref{eq:bfprop}) or from single insertions 
on two different propagators, isotensor quantities can be computed correctly. 
To compute more general quantities, the effects of coupling the background fields to the sea quarks need to be included. This can be done either in the generation of dynamical gauge 
configurations~\cite{Chambers:2015bka} or through a reweighing method~\cite{Freeman:2014kka}. 

In order to extract matrix elements of currents that involve zero-momentum insertion, a uniform background field is implemented. 
In the work of Refs.~\cite{Shanahan:2017bgi,Tiburzi:2017iux}, a set of flavor-diagonal background axial-vector fields was used, with operator structure
\begin{eqnarray}
\Lambda^{(u)} = \lambda_u\  \gamma_3 \gamma_5 (1+\tau_3)/2
\qquad {\rm and }\qquad 
\Lambda^{(d)} = \lambda_d\  \gamma_3 \gamma_5  (1-\tau_3)/2,
\label{eq:Lambda}
\end{eqnarray}
where $\lambda_q$ are parameters specifying the strength of the background field. Zero-momentum--projected correlation functions
\begin{eqnarray}
C^{(h)}_{\lambda_u;\lambda_d}(t) 
& = & 
\sum_{\bm x}
\langle 0| \chi_h({\bm x},t) \chi^\dagger_h({\bm 0},0) |0 \rangle_{\lambda_u;\lambda_d}
\label{eq:bfcorr}
\end{eqnarray}
are formed from the compound propagators $S_{\{\Lambda^{(u)}\}}(x,y)$ and $S_{\{\Lambda^{(d)}\}}(x,y)$ that have at most a single insertion of the background field
(indicated by $\langle\ldots\rangle_{\lambda_u;\lambda_d}$). 
Here, $h$ denotes the quantum numbers of the hadronic interpolating operator, $\chi_h$. 
The correlation functions $C_{\lambda_u;\lambda_d}^{(h)}(t)$ are, by construction, polynomials of maximum degree 
$\lambda_u^{N_u}\lambda_d^{N_d}$ in the field strengths, 
where $N_{u(d)}$ is the number of up (down) quarks in the interpolating operator.

\subsubsection{Challenges for nuclear physics}

In principle, either the 3-point or background field methods could be used in conjunction with many-nucleon interpolating operators to directly calculate  \tnubb\ and \znubb\ matrix elements within experimentally relevant nuclei. However, practically speaking, direct LQCD calculations must be limited to few-body systems, currently to $ A\ \lsim 4$. There are several technical reasons for this restriction, as will be discussed below. Central to all of these issues is the use of quark fields as the relevant degrees of freedom. As systems become larger and the relevant energy scales diminish, the use of these microscopic degrees of freedom becomes increasingly inappropriate, manifesting as rapidly increasing computational complexity. Thus, the program outlined in this review, of matching LQCD calculations of small $A$ systems onto effective field theories, to be then utilized within computational many-body techniques, is paradigm.

One of the issues that arises for large systems is simply the number of lattice points that are required to resolve the large range of relevant scales. These scales encompass both the high-energy physics of the quarks and gluons, as well as low-energy excitations, such as those associated with collective motion of the nucleons. Correctly describing these scales requires both very small lattice spacing and large volumes. Another issue is the number of quark propagators that must be produced, as well as the number of  Wick contractions of these propagators that must be computed. The latter na\"ively scales factorially with the number of nucleons, although algorithms have been proposed which can reduce this scaling to power law in some cases ~\cite{Yamazaki:2009ua,Doi:2012xd,Detmold:2012eu,Gunther:2013xj,Detmold:2019fbk}.

A further challenge is the exponentially poor signal-to-noise ratio associated with nucleons and nuclei, decaying roughly as~\cite{Lepage:1989hd,Parisi:1983ae,Beane:2009gs}.
\begin{equation}
    \mathcal{R} \sim \frac{1}{\sqrt{\mathcal{N}}}e^{-A\left( m_N-3/2m_{\pi}\right) t} \ ,
\end{equation}
for large Euclidean time, $t$, and number of configurations, $\mathcal{N}$. The difference in the exponent arises due to the use of quark fields, which can couple both to the  nucleon state in the signal as well as the much lighter pions in the corresponding correlator that determines the variance. This difference is numerically smaller for heavier-than-physical pion masses, which is currently why many calculations are not performed directly at the physical point. For large times where the  ground state dominates, an exponentially large number of configurations is thus necessary to extract the desired signal, increasing as $A$ increases.

\pagebreak

\section{Neutrinoful double-$\beta$ decay}
\label{sec:tnubb}
Neutrinoful double-$\beta$ decay is the rarest SM process whose rate has been measured. As such, this decay provides a crucial test of the SM, and in particular of our understanding of weak interactions in nuclei.
Achieving controlled predictions of $2\nu\beta\beta$ decay rates from the SM is, however, challenging; the nuclei which undergo this decay are too large for the application of LQCD or {\it ab initio} methods, and there is considerable model-dependence inherent in the more phenomenological many-body methods which can be applied, leading to significant model uncertainties in current best theory calculations of these rates. A promising approach to improving the reliability of these predictions is to couple LQCD and {\it ab initio} methods, as outlined in this review.
First progress has been made towards this goal; the second-order weak $\beta\beta$-decay matrix element of the two-nucleon system was recently computed from LQCD for the first time~\cite{Shanahan:2017bgi,Tiburzi:2017iux}. With sufficiently precise and systematically-controlled calculations of few-body \tnubb\ decay matrix elements, the free parameters of few- and many-body methods, including those based on EFTs, can be constrained from LQCD, effectively anchoring phenomenological approaches in the SM. It can be expected that this approach will reduce the model-dependence implicit in many-body calculations of double-beta decay rates, enabling reliable predictions for these rates with systematically-improvable uncertainties.

The rate of the neutrinoful double-$\beta$ decay resulting in the nuclear transition $A_i\rightarrow A_f$ (and also the corresponding neutrinoless double-$\beta$ decay in a light Majorana-neutrino scenario), is dictated by second-order weak interactions. Since the long-distance contribution from the 
Fermi (vector) piece is suppressed by isospin symmetry, the dominant contribution arises from the Gamow-Teller (axial-vector) piece of the weak current.
Precisely, neglecting lepton-mass effects, the inverse half-life of the neutrinoful double-$\beta$ decay $[T^{2\nu}_{1/2}]^{-1}$, can be expressed as~\cite{Engel:2016xgb}
\begin{eqnarray}
[T^{2\nu}_{1/2}]^{-1}=
 G_{2\nu}(E_i-E_f,Z_i) | M_{GT}^{2\nu}|^2,
\label{eq:MGT1}
\end{eqnarray}
where the matrix element $M_{GT}^{2\nu}$ is defined from the time-ordered product of two axial currents by
\begin{eqnarray}
M_{GT}^{2\nu}&=& 
6 \times \frac{1}{2}
\int{d^4x} \, {d^4y} 
\, 
\langle A_f |  T \left[ J^+_{3}(x) J^+_{3}(y)  \right] | A_i \rangle = 6\sum_{{\frak n}}\frac{\langle A_f |  \tilde{J}_3^+ |{\frak n}\rangle\langle {\frak n} |  \tilde{J}_3^+ |A_i \rangle}{E_{\frak n}-(E_{i}+E_{f})/2},
\label{eq:MGT}
\end{eqnarray}
where $Z_i$ is the proton number of the initial nuclear state $A_i$, $E_{i,f}$ are the energies of the initial and final states, and $G_{2\nu}(\Delta E,Z_i)$ is a known phase-space factor~\cite{Kotila:2012zza,Stoica:2013lka}. The spatial component of the $\Delta I_3=1$ zero-momentum axial current in the 3-direction is expressed as 
\begin{equation}
    \tilde{J}_3^a \equiv \tilde{J}_3^a(\bm{0},t=0)  = \int d^3 \bm{x} {J}_3^a(\bm{x},t=0), \ \ \ \ {\rm with} \ \ \ \ {J}_3^a(x)= \overline{q}(x) \frac{\gamma_3\gamma_5}{2} \tau^a q(x),
\end{equation}
where $\tau$ denotes a Pauli matrix in isospin space, and $\tau^+ = \frac{1}{\sqrt{2}} \left(\tau^1 + i\; \tau^2 \right)$. A complete set of zero-momentum states is indexed by ${\frak n}$, and the factor of $6$ in $M_{GT}^{2\nu}$ is a consequence of rotational symmetry (as $M_{GT}^{2\nu}$ is written using the third spatial component of the axial currents) and the convention for the normalization of the currents.

The second-order Gamow-Teller transition matrix element $M_{GT}^{2\nu}$, as well as individual contributions to this matrix element, can in principle be determined for various nuclear transitions from LQCD calculations, providing refined inputs for nuclear many-body calculations of double-beta decay rates. In Refs.~\cite{Shanahan:2017bgi,Tiburzi:2017iux}, the first LQCD calculation of $M_{GT}^{2\nu}$ was undertaken, for the $nn\to pp$ transition. While this is not an allowed transition in nature because the dineutron is not bound, the corresponding matrix element itself is well-defined, calculable, and related to the two-body sub-process of double-$\beta$ decays of larger nuclei. This calculation was performed without the inclusion of electromagnetism, at a single lattice spacing and volume, and at the SU(3) flavor-symmetric point with degenerate up, down and strange quark masses corresponding to a larger-than-physical pion mass of $m_\pi \sim 806$ MeV. While all of these caveats are possibly important, the key qualitative result of that work was to reveal the potential significance of an operator that contributes to the $\beta\beta$-decay of nuclei, but not to single-$\beta$ decays, namely the isotensor axial polarizability, $\beta_A^{(2)}$, of the $\si$ two-nucleon system. This polarizability is defined from $M_{GT}^{2\nu}$ by subtracting the term corresponding to an intermediate deuteron state, i.e., the `Born' term as in forward Compton scattering:
\begin{equation}
 \frac{1}{6}M_{GT}^{2\nu} = \beta_A^{(2)} - \frac{ |\langle pp|\tilde{J}_3^+|d\rangle|^2} {E_{pp}-E_d}. 
\label{eq:axial-polz}
\end{equation}
Since terms of this form have not been included in phenomenological analyses of double-$\beta$ decay, its significance in the numerical calculation of Refs.~\cite{Shanahan:2017bgi,Tiburzi:2017iux} implies that theoretical predictions of double-$\beta$ decay rates with fully quantified uncertainties will require constraints on the isotensor axial polarizabilities of nuclei.
In Refs.~\cite{Shanahan:2017bgi,Tiburzi:2017iux} it was also explicitly demonstrated how LQCD results can provide input to many-body methods to constrain second-order electroweak properties of nuclear systems, by constraining the leading $\Delta I = 2$ low-energy constant of pionless EFT (see Section \ref{sec:pionless}) from the LQCD two-nucleon transition matrix element. The remainder of this section will review that calculation, with a particular focus on the difficulties, especially related to the bi-local nature of weak processes, which must be overcome in order to undertake such calculations with controlled uncertainties at the physical quark masses, and for larger nuclear systems. 

\subsection{Lattice QCD calculations}
\label{sec:tnubblqcd}

An efficient way to determine the matrix elements relevant to double-$\beta$ decay processes in LQCD calculations is via the background field technique discussed in Sec.~\ref{sec:LQCD}. 
From the isospin structure of the operator inducing the $nn \rightarrow pp$ transition, it is clear that no self-contractions of the quark fields in the axial-current operators, no contractions of quark fields between the two axial-current operators, and no double insertions of axial-current operators on a single quark line, contribute to the matrix element. The matrix element of interest can thus be constructed from correlators formed from propagators computed in a background-field corresponding to a single axial-current insertion.
For an axial current $J_3^a(x)$, and hadron $h$, these background-field correlators have the form 
\begin{eqnarray}
C^{(h)}_{\lambda_a}(t) 
&=&
\sum_{\bf x}
\langle 0| \chi_{h}({\bf x},t) \chi^\dagger_{h}(0) |0 \rangle  
+ \lambda_a
\sum_{{\bf x},{\bf y}}\sum_{t_1=0}^t
\langle 0| \chi_{h}({\bf x},t) J_3^{(a)} ({\bf y},t_1)   \chi^\dagger_{h}(0) |0 \rangle 
\nonumber \\ 
&&+ \frac{\lambda_a^2}{2}
\sum_{{\bf x},{\bf y},{\bf z}}\sum_{t_{1,2}=0}^t 
\langle 0| \chi_{h}({\bf x},t) J_3^{(a)} ({\bf y},t_1)  
J_3^{(a)} ({\bf z},t_2)   \chi^\dagger_{h}(0) |0 \rangle +  \cO(\lambda_a^3), 
\label{eq:quad1}
\end{eqnarray}
where $\chi_h^{(\dagger)}$ defines an interpolating operator with the quantum numbers of $h$. The second-order term in the field strength, i.e., the piece proportional to $\lambda_a^2$, can be extracted from fits to calculations of the background-field correlators at multiple values of $\lambda_a$. While this construction of the background-field correlator involves sums over all possible insertion times $t_{1,2}$ of the two axial currents, which is sufficient for a determination of the matrix element of the $nn\rightarrow pp$ transition in Refs.~\cite{Shanahan:2017bgi,Tiburzi:2017iux}, background-field constructions with the current insertions restricted to smaller temporal regions provide additional constraints which allow the Euclidean time-dependence of the correlators to be further decomposed. This will likely be necessary in extensions of this approach to calculations of the double-$\beta$ decay transitions of larger nuclei, at values of the quark masses corresponding to lighter pion masses.  

\begin{figure}
	\centering
	\includegraphics[width=0.35\columnwidth]{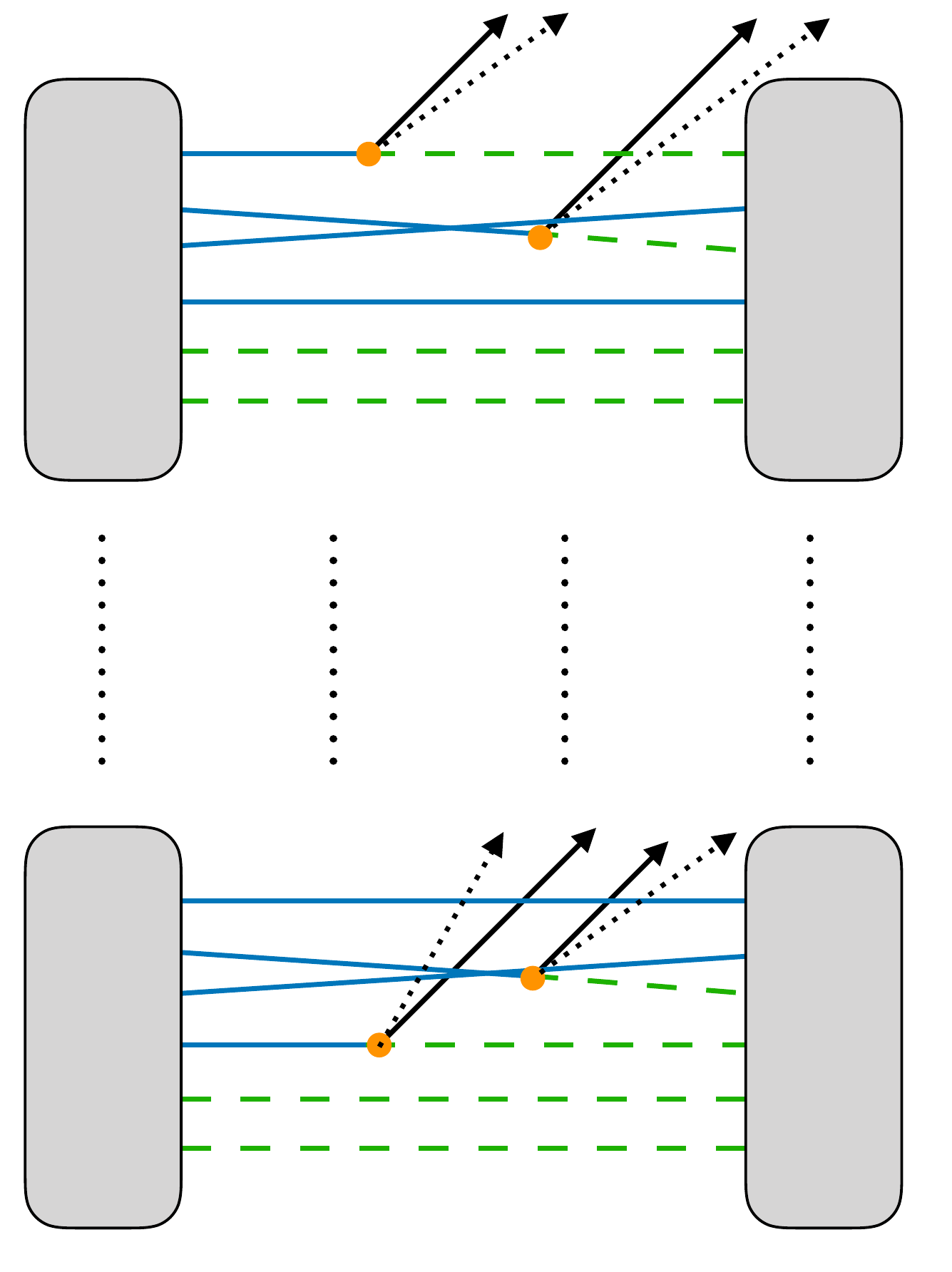}
	\caption{Example contraction for the $nn\to p p e^- e^-$ transition corresponding to Eq.~(\ref{eq:Ct}). The solid blue and dashed green lines represent down and up quark propagators respectively, while the solid orange circles represent the $\Delta I=1$ weak interaction vertices. The dotted and solid black arrows represent the  neutrino and electron final states, respectively.}
	\label{fig:nn2ppeenunucontraction}
\end{figure}

For a calculation with the background field inserted at all times, the key correlator which is extracted from a LQCD calculation is 
\begin{eqnarray}\label{eq:Ct}
C(t)
&=&\sum_{{\bf x},{\bf y},{\bf z}}\sum_{t_{1,2}=0}^t 
\langle 0| \chi_{pp}({\bf x},t) T\left[J_3^{+} ({\bf y},t_1)  
J_3^{+} ({\bf z},t_2) \right]   \chi^\dagger_{nn}(0) |0 \rangle.
\end{eqnarray}
This can be determined either using a $\tau^+$ background field insertion, or alternatively using only flavor-diagonal fields via the isospin relation (detailed in Ref.~\cite{Tiburzi:2017iux}):
\begin{eqnarray}
\langle p p | J^+_{3}(x) J^+_{3}(y) | n n \rangle 
=
\langle n p | J_3^{(u)}(x) J_3^{(u)}(y) | n p \rangle 
-\frac{1}{2} \langle n n | J_3^{(u)}(x) J_3^{(u)}(y) + J_3^{(d)}(x)  J_3^{(d)}(y) | n n \rangle
\label{eq:recipe}.
\quad
\end{eqnarray}
An example of the quark contractions which contribute to this correlation function is displayed in Fig.~\ref{fig:nn2ppeenunucontraction}.
Inserting complete sets of states, the correlation function $C(t)$ of Eq.~\eqref{eq:Ct} can be expanded as
\begin{eqnarray} \label{eq:corr}
C(t) & =& 
\frac{2V}{a^2} 
\sum_{\frak{n},\frak{m}, \frak{l}'} 
\langle 0| \chi_{pp} | \frak{n}\rangle \langle \frak{m}| \chi_{nn}^\dagger | 0\rangle
e^{- E_\frak{n} t} 
\frac{
\langle \frak{n}| \tilde{J}_3^{+} | \frak{l}' \rangle 
\langle \frak{l}' | \tilde{J}_3^{+}  | \frak{m} \rangle
}{E_{\frak{l}'}- E_\frak{m}} 
\left(
\frac{
e^{- \left(E_{\frak{l}'} - E_\frak{n}\right) t} -1
}
{E_{\frak{l}'} - E_\frak{n}}
+
\frac{
e^{\left(E_\frak{n}-E_\frak{m}\right)t}-1
}
{E_\frak{n}-E_\frak{m}}\right),
\end{eqnarray}
where zero-momentum energy eigenstates with the quantum numbers of the $pp$, $nn$ and deuteron systems are expressed as $|\frak{n} \rangle$, $|\frak{m}\rangle$ and $|\frak{l}' \rangle$ respectively, and $E_{\frak{l}'}=E_{nn}+\delta_{\frak{l}'}$ and $E_\frak{n}=E_{nn}+\delta_\frak{n}$ are the energies of the $\frak{l}'$th and $\frak{n}$th excited states in the $\siii$ and $\si$ channels. 
From the Euclidean time-dependence of this correlation function, the short-distance isotensor axial polarizability, defined in Eq.~\eqref{eq:axial-polz}, can be extracted.
The long-distance deuteron pole contribution, however, which is dictated by the single-current matrix element $\langle pp| \tilde{J}_3^{+}  |d\rangle$, can be determined most precisely from the linear contributions to the background field correlation functions defined in Eq.~\eqref{eq:quad1}.  

Forming a ratio of the background field correlation function, given in Eq.~\eqref{eq:corr}, to the zero-field two-point function of the $nn$ system ($C_0^{(nn)}$ in the notation of Eq.~\eqref{eq:quad1}), and subtracting the term with a deuteron intermediate state, defines
\begin{eqnarray}\nonumber
a^2 \hat\cR(t) & = &  \frac{a^2C(t)}{2 C^{(nn)}_{0}(t)} - \frac{|\langle pp| \tilde{J}_3^{+}  |d\rangle|^2 }{\Delta} 
\left[ \frac{e^{\Delta t}-1}{\Delta} -t \right] \\
\label{eq:rnnppsub}
& = & 
 t \sum_{{\frak l}' \ne d}
{
	\langle pp| \tilde{J}_3^{+}  |{\frak l}' \rangle \langle {\frak l}' | \tilde{J}_3^{+}| nn \rangle
	\over E_{{\frak l}'}-E_{nn}}
+ c  + d\ e^{\Delta t} + \cO(e^{-\hat\delta t}) ,
\end{eqnarray}
where $\Delta = E_{nn}-E_{d}$, and $\tilde{\delta}\sim \delta_\frak{m},\delta_{\frak{n}^\prime}$ denotes a generic gap between eigen-energies of two-nucleon systems. In the final expression, the terms $c$ and $d$ collect $t$-independent factors involving overlap factors, energy gaps, and ground and excited-state transition amplitudes. In this expansion, $\Delta$ is assumed to be small relative to the inverse of the time separation between the source and the sink, and the gaps between eigen-energies are assumed to be large, i.e., $\tilde{\delta} \gg\Delta$. While these assumptions are valid for the analysis of Refs.~\cite{Shanahan:2017bgi,Tiburzi:2017iux}, in the generic case the simplifications applied in Eq.~\eqref{eq:rnnppsub} can not be used. In the limits of physical quark masses and large simulation volumes in particular, $\tilde{\delta}\rightarrow 0$ and $\Delta\rightarrow 2.22$ MeV, so that contributions involving the transition matrix elements of excited states can no longer be neglected. In this scenario, alternative strategies involving insertions of the background field over ranges of timeslices that are  separated from the source and sink, will need to be pursued~\cite{Christ:2012se}.

\begin{figure}
    \centering
    \includegraphics{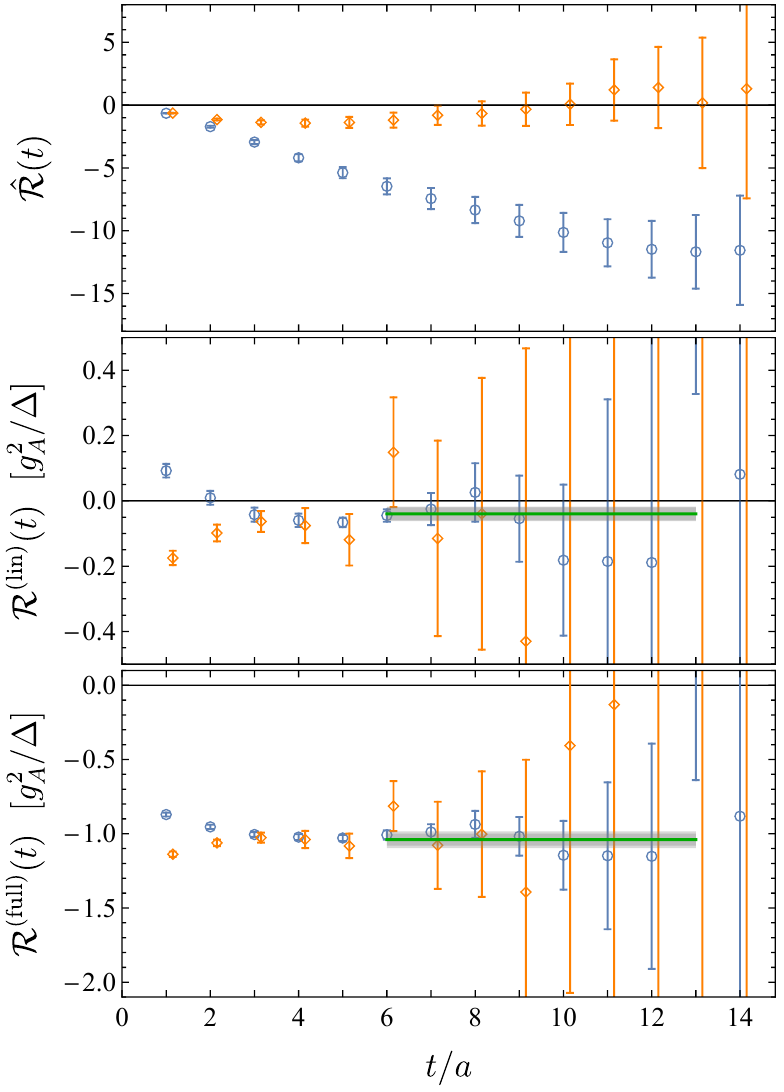}
    \caption{$\hat\cR(t)$, $\cR^{\text{(lin)}}(t)$, and $\cR^{\text{(full)}}(t)$ (Eqs.~\eqref{eq:rnnppsub}--\eqref{eq:Rfull}) determined in the calculation of Refs.~\cite{Shanahan:2017bgi,Tiburzi:2017iux}. In each panel, the blue circles and orange diamonds indicate results obtained from correlation functions constructed with different sink smearing prescriptions. Note that $\hat{\mathcal{R}}$ is expected to differ between the two sets of results, as it includes the terms in Eq.~\eqref{eq:rnnppsub} with coefficients $c$ and $d$ which contain ground-state and excited-state overlap factors and are thus smearing-dependent. (From Ref.~\cite{Tiburzi:2017iux}) }
    \label{fig:RhatRlinRfull}
\end{figure}

The coefficient of the term linear in $t$ in Eq.~\eqref{eq:rnnppsub}, which determines the isotensor axial polarizability $\beta_A^{(2)}$ defined in Eq.~\eqref{eq:axial-polz}, can be extracted from the large-time limit of
\begin{eqnarray}
\cR^{\text{(lin)}}(t)= 
{(e^{ a \Delta} +1) \hat\cR(t+a) - \hat\cR(t+2 a) - e^{a \Delta} \hat\cR(t) \over 
 e^{a \Delta} -1 } ~\stackrel{t \to \infty }{\longrightarrow}~ \frac{1}{aZ_A^2}\frac{\beta_A^{(2)}}{6}.
\nonumber \\
\label{eq:Rlin}
\end{eqnarray}
Here, the isotensor axial polarizability is renormalized by the square of the axial current renormalization constant, $Z_A^2$; this is correct up to lattice-spacing suppressed artefacts arising from radiatively-generated four-quark operators. Such corrections necessarily occur  in the background field approach, which includes contributions where both insertions of the axial current are localized around the same space-time point.
In the analysis of Ref.~\cite{Tiburzi:2017iux}, it is concluded that this mixing results in sub-percent effects, which are neglected in that analysis. 
Finally, the complete bare Gamow-Teller matrix element is defined by the combination of $\cR^{\text{(lin)}}(t)$, defined in Eq.~\eqref{eq:Rlin}, with the deuteron-pole contribution:
\begin{eqnarray}
\cR^{\text{(full)}}(t)= \cR^{\text{(lin)}}(t) - \frac{|\langle pp| \tilde{J}_3^{+}  |d\rangle|^2 }{a \Delta}  \stackrel{t\to\infty}{\longrightarrow} {M^{2\nu}_{GT}\over 6 \, a Z_A^2} . \ \ \
\label{eq:Rfull}
\end{eqnarray}
Numerical results for $\hat\cR(t)$, $\cR^{\text{(lin)}}(t)$, and $\cR^{\text{(full)}}(t)$, determined in the calculation of Refs.~\cite{Shanahan:2017bgi,Tiburzi:2017iux}, are shown in Fig.~\ref{fig:RhatRlinRfull}. Two different analyses were undertaken, with correlation functions constructed with different smearing prescriptions; the results from both analyses are consistent, with the isotensor axial polarizability resolved from zero to two standard deviations.

\subsection{Phenomenological consequences}
\label{sec:tnubbpheno}

\begin{figure}[!t]
        \includegraphics[width=0.95\columnwidth]{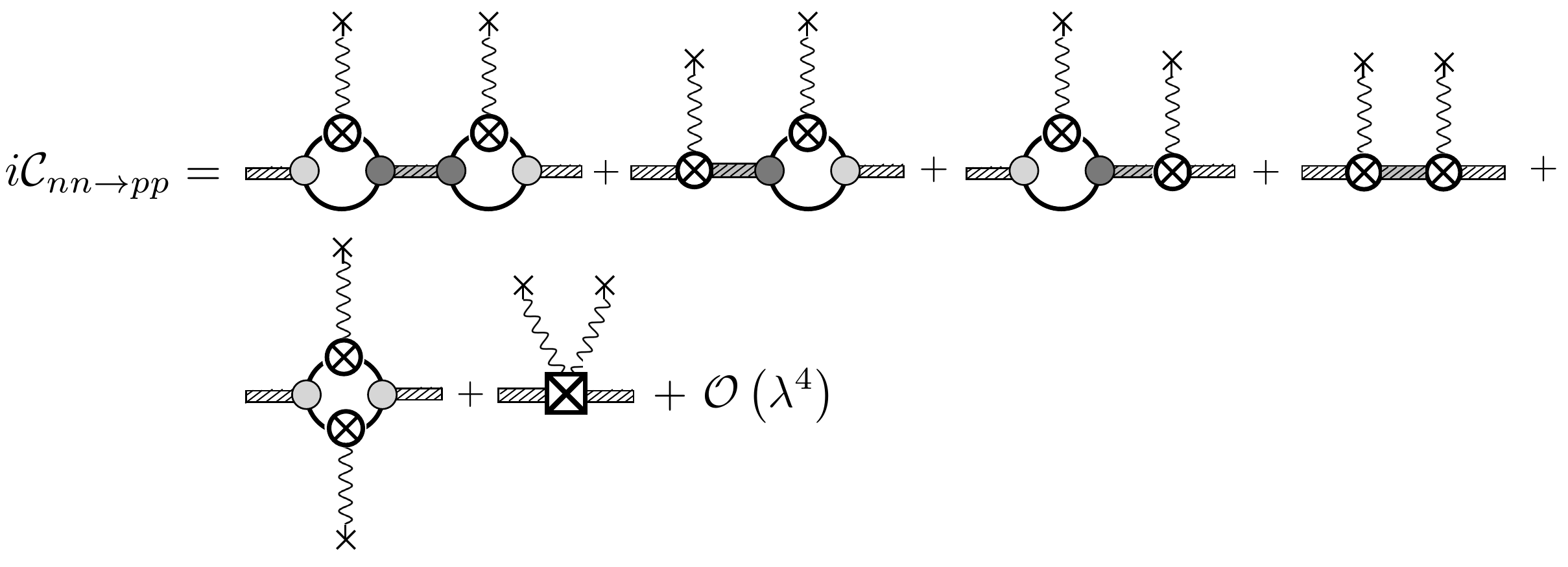}
        \caption{
        Diagrammatic representation of the $nn \to pp$ correlation function in the pionless dibaryon EFT formulation. The light (dark) gray circles denote the isotriplet (isosinglet) strong dibaryon coupling to two nucleons, when inserted on the nucleon line the crossed circle represents the singly-weak single-nucleon coupling to the background field, and when inserted on the dibaryon line it denotes the singly weak dibaryon coupling. The crossed square represents the doubly weak dibaryon coupling to the background field. The thick dashed light (dark) gray lines denote the fully-dressed isotriplet (isosinglet) dibaryon propagator, and the thin black lines represent nucleon propagators. (From Ref.~\cite{Tiburzi:2017iux})
        }
        \label{fig:twoax}
\end{figure}

As discussed in previous sections, the Gamow-Teller matrix element for the $nn\rightarrow pp$ transition determined in lattice QCD calculations can be used to constrain counterterms in effective field theory descriptions of the same system. 
Then, by matching to few-body methods, the potentially key contribution from the axial polarizability can be included in calculations of the decay rates of larger nuclei than are accessible to LQCD; this approach was explored for spectroscopy in Ref.~\cite{Barnea:2013uqa,Contessi:2017rww}. For LQCD calculations undertaken with the heavy pion mass of Refs.~\cite{Shanahan:2017bgi,Tiburzi:2017iux}, it is natural to consider pionless EFT descriptions, as discussed in Sec.~\ref{sec:pionless}. At lighter quark masses approaching the physical point, pionful EFTs as discussed in Sec.~\ref{sec:EFTforDBD} will likely be required.
The general approach to matching LQCD results for double-$\beta$ decay transitions to EFT is to equate correlation functions constructed in the two formalisms, where couplings to the background fields are included in the effective Lagrangian of the EFT. To study the $nn \to pp$ matrix element, then, the correlation function matrix in the $\{ nn, np({^3}S_1),pp \}$ EFT channel space can be constructed (since the axial background field changes both spin and isospin, there is no coupling of the $np(^1S_0)$ state):
\begin{eqnarray}
\mathcal{C}_{NN \to NN} \equiv
\left(
\begin{array}{ccc}
\cC_{nn \to nn} & \cC_{nn \to np({^3S_1})} & \cC_{nn \to pp} \\
\cC_{np({^3S_1}) \to nn} & \cC_{np({^3S_1}) \to np({^3S_1}) } & \cC_{np({^3S_1}) \to pp} \\
\cC_{pp \to nn} & \cC_{pp \to np({^3S_1})} & \cC_{pp \to pp} \\
\end{array}
\right).
\label{eq:Cmat}
\end{eqnarray}
This matrix is constructed as an expansion in terms of LECs, including couplings to the background axial field. In Refs.~\cite{Shanahan:2017bgi,Tiburzi:2017iux}, a dibaryon formulation of pionless EFT~\cite{Beane:2000fi,Phillips:1999hh} was used; matching to the LQCD calculation then enabled the coefficient of a short-distance, two-nucleon, second-order axial-current operator in the that formalism to be determined. In that approach, the $nn\rightarrow pp$ transition, expanded to second order in the background axial field, can be expressed as shown in Fig.~\ref{fig:twoax}. The correlation function can then be expressed in a cubic spatial volume with periodic boundary conditions, Fourier-transformed in time, and Wick-rotated to Euclidean space by $x_0\rightarrow it$. 
Taking the ratio of the $nn \to pp$ transition correlator to two times the zero-field two-point function, as done for the LQCD correlation function in Eq.~\eqref{eq:rnnppsub}, and taking the second derivative with respect to the background-field strength to extract the terms linear in time which determine the second-order finite-volume matrix element, yields an  expression that can be matched precisely to the LQCD result.
The key matrix element can be expressed as
\begin{eqnarray}
M_{nn \to pp}
& = &
-\frac{|g_A(1+S)+\mathbb{L}_{1,A}|^2}{\Delta}
\ +\
\frac{Mg_A^2}{4 \gamma_s^{2}}
-\mathbb{H}_{2,S}.
\label{eq:M-nnpp-EFTb}
\end{eqnarray}
The first term corresponds to the deuteron pole, while the remaining terms are short-distance contributions; the quantities $\mathbb{L}_{1,A}$ and $\mathbb{H}_{2,S}$
denote the correlated two-nucleon axial contribution to the phenomenological quenching of $g_A$ and the correlated two-nucleon two-axial coupling contribution, respectively. 
In terms of the parameters in Sec.~\ref{sec:pionless}, 
\begin{equation}
\mathbb{L}_{1,A}=\frac{Z_s Z_t
	\sqrt{\gamma_t \gamma_s}}{2 M}\tilde{l}_{1,A}, 
\qquad 
\mathbb{H}_{2,S}=\frac{\gamma_sZ_s^2}{2M}(\tilde{h}_{2,S}-\frac{M^2r_s}{2\gamma_s^2}g_A^2),
\end{equation}
where $\gamma_{s,t}=\sqrt{M(2M-E_{s,t})})$ and $Z_{s,t}=1/(1-\gamma_{s,t} r_{s,t})$.
These can be constrained using the LQCD matrix elements for the $nn\rightarrow pp$ transition, in addition to the values of the proton axial charge, and the binding momenta and effective ranges.
In Eq.~\eqref{eq:M-nnpp-EFTb}, $S$ is an $SU(4)$ Wigner symmetry-breaking factor, $\gamma_{s}=\sqrt{MB_{s}}$ (where $B_{s}=2M-E^{(0)}_{s}$ is the binding energy), and the tower of shape parameters has been ignored. From the results of Refs.~\cite{Shanahan:2017bgi,Tiburzi:2017iux}, the correlated two-nucleon two-axial coupling contribution can be constrained to be $\mathbb{H}_{2,S}=4.7(1.3)(1.8)$ {fm}. While the deuteron-pole term dominates (the full matrix element and the Born term differ by approximately $5\%$), the contribution from $\mathbb{H}_{2,S}$ is of the same order of magnitude as the term proportional to $g_A^2$, and is thus non-negligible. While this result suggests that the two-nucleon two-axial coupling should not be neglected in analyses of double-$\beta$ decay, the numerical result is valid only at the heavier-than-physical quark masses used in the LQCD calculation. Connecting directly to phenomenology will require LQCD calculations at, or close to (with chiral extrapolation), the physical masses. 

For future calculations via this approach to reduce the model-dependence implicit in many-body calculations of double-beta decay rates, significant progress is still required. In particular, repeating the LQCD calculation of Refs.~\cite{Shanahan:2017bgi,Tiburzi:2017iux} at quark masses corresponding to the physical point presents several challenges. Firstly, the hierarchy of mass splittings changes from that which was exploited in that work to isolate the matrix element of interest. As discussed above, this could be addressed by separating the source and sink from the background field region, and considering calculations with several different regions. Moreover, the initial and final states will no longer be  bound, complicating the relationship between the finite-volume bi-local matrix elements and the infinite-volume transition amplitudes; a generalization of the formalism presented in Ref.~\cite{Christ:2015pwa} will be required. One might also consider exploring extensions to investigate the many-body effects directly by calculations of the $nnp\rightarrow ppp$ or $nnn\rightarrow npp$ transitions. This requires I=$\frac{3}{2}$ states such as $ppp$ to be constructed in the LQCD calculation, which will need non-local source structures, not studied in previous work, to be constructed. 
Nevertheless, the path to achieving such calculations is reasonably clear, despite the above challenges, and LQCD calculations with physical quark masses have the potential to provide critical input to many-body calculations of double-$\beta$ rates that cannot be obtained through any other known method.
\pagebreak

\section{Neutrinoless double-$\beta$ decay}
\label{sec:znubb}

As discussed in the introduction and in Section~\ref{sec:EFTforDBD}, 
there are a number of potential BSM scenarios that result in a double beta decay process in which there are no neutrinos in the final state.
The two important classes of new physics are a) light Majorana neutrinos and b) new short distance lepton number violating processes at scales beyond the electroweak scale. To understand the effects of either scenario in physical states requires knowledge of  nuclear matrix elements. 
Experimental searches for this process make use of large nuclei which are at present too difficult to study in LQCD. 
However, as discussed  in Section~\ref{sec:EFTforDBD}, effective field theory 
allows one to obtain critical non-perturbative input for the many-body transition operators 
by performing LQCD calculations of simpler systems.
To this end, recent work has focused on calculations in pionic systems where the numerical complexities of nuclei in LQCD are absent and the techniques necessary to study \znubb\ can be developed.

\subsection{Short distance contributions in pion matrix elements}
\label{sec:znubb_short}

As discussed  in Section~\ref{sec:LNV9},  if there is BSM physics contributing to \znubb\ at scales above the electroweak scale, then the effects manifest at lower scales as local composite operators whose contributions arise from integrating out the new physics.
Generically, a given high-scale physics scenario will produce multiple different operators at low energies, 
listed in Eq.~\eqref{eq:Lag}, 
of which the most phenomenologically relevant are the five  four-quark scalar operators, $O_{1,2,3,4,5}$. 
In Section~\ref{sec:LNV9}, the chiral EFT realization of these operators is reviewed in detail,  
in various power counting schemes.  The EFT analysis implies  that in order to construct the leading-order 
transition operators one needs to determine $\langle \pi^+ |O_{1,2,3,4,5} | \pi^- \rangle$ 
and $\langle pp |O_{1,2,3,4,5} |nn \rangle$. 
So far the LQCD community has focused on the pionic matrix elements, as the two-nucleon matrix elements suffer 
from technical complications. In addition to the generic issues related to LQCD calculations involving nucleons (Sec.~\ref{sec:LQCD}), difficulties arise from the fact that 
%
%
%
(i) the quark line contractions for two-nucleon \znubb\ calculations are more involved, resulting in the need for either highly improved position-space operators for the nucleons or stochastic methods for projecting onto definite momenta; (ii) the connection with infinite volume physics for two-body systems requires the calculation of scattering phase shifts coupled with a sophisticated finite-volume formalism~\cite{Luscher:1986pf,Briceno:2015tza}.

While the $\pi^- \to \pi^+ e^- e^-$ in vacuum is itself unphysical due to kinematic considerations, this transition can occur within an off-shell pion exchanged inside a nucleus. Computing the pure QCD portion of this transition at unphysical kinematics 
gives access to the relevant EFT LEC's, 
namely the $g_i^{\pi \pi}$ introduced in Eq.~\eqref{eq:dim9pipi}. 

\begin{figure*}
\centering
\includegraphics[width=0.42\textwidth]{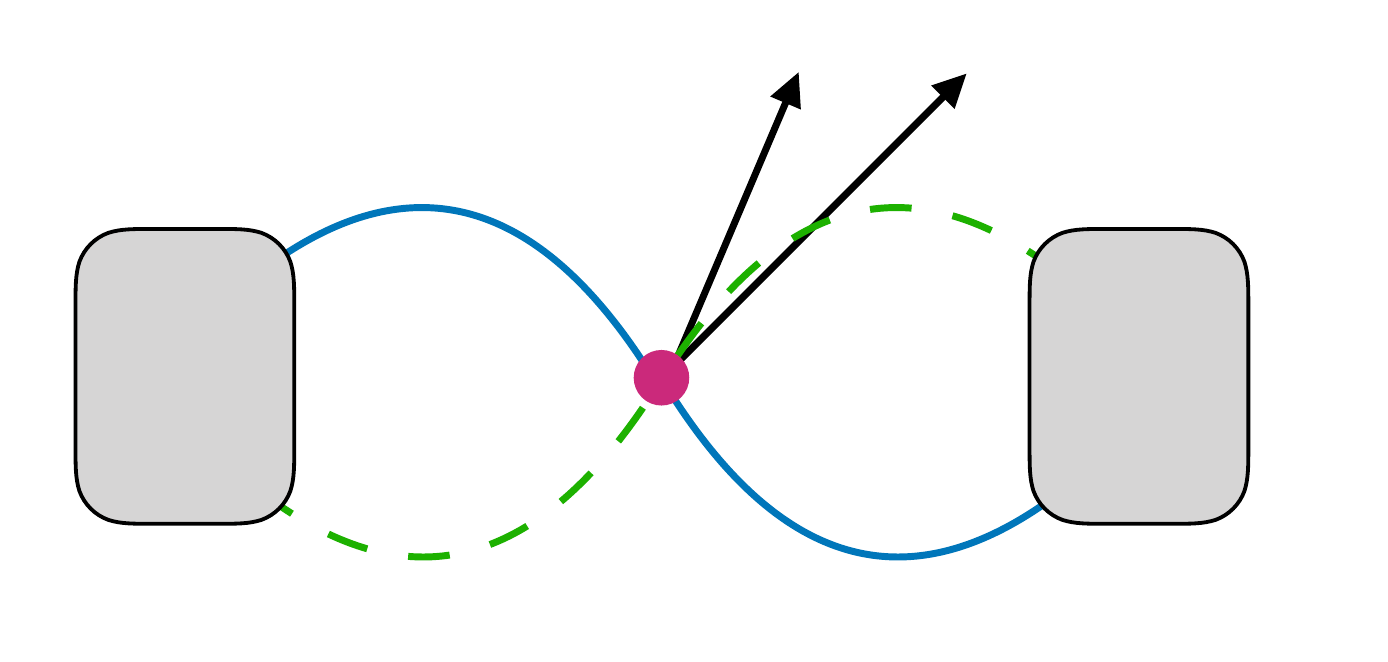}
 \caption{\label{fig:shortdist}Contractions for the $\pi^- \to \pi^+ e^- e^-$ transition induced by short distance four-quark operators. The solid blue and dashed green lines represent down and up quark propagators respectively, and the dark circle represents the $\Delta I=2$ operator. The solid black lines represent the electron final states.}
\end{figure*}

The most straightforward method for performing these calculations is to use a traditional 3-point function, $C_i^{3\mathrm{pt}}(t,T-t)$ (see Sec.~\ref{sec:LQCD}), in which a four-quark operator is inserted between source and sink operators for the pion which are sufficiently separated in Euclidean time: 
\begin{equation}\label{eq:C3pt_znubb}
C^{3pt}_i(t_i,t_f)=\sum_{\alpha}\sum_{\mathbf{x},\mathbf{y}}e^{-E_\alpha T}\langle\alpha|\Pi^{+}(t_f,\mathbf{x}) {O}_i(0,\mathbf{0})\Pi^+(t_i,\mathbf{y}) |\alpha\rangle
\end{equation}
where $\Pi^+(t_f,\mathbf{x})=\bar{d}\gamma^5u$ and $\Pi^+(t_i,\mathbf{y})=\Pi^{-\dagger}(t_i,\mathbf{y})$ are annihilation and creation operators with the quantum numbers of a $\pi^+$ and a $\pi^-$  respectively. The corresponding quark contraction is shown in Fig.~\ref{fig:shortdist}. Note that here the operator is chosen to be held at a fixed space-time point while the times at which the source and sink pion interpolating operators are inserted are free to vary.

The contractions for these 3-point functions can be easily performed by creating a single pion ``block" \cite{Detmold:2010au} 
\begin{equation}
\Pi_{a,\alpha,b,\beta} = \sum_{c,\gamma}\sum_{\mathbf{x}} \left[ S_{d}\left(\mathbf{x},t;\mathbf{0},0\right) \gamma_5 \right]_{b,\beta,c,\gamma}\left[ S^{\dagger}_{u}\left(\mathbf{x},t;\mathbf{0},0\right) \gamma_5 \right]_{a,\alpha,c,\gamma} \ ,   
\end{equation}
from $d$ and $\bar{u}$ propagators, $S_d(x;y)$ and $S^{\dagger}_u(x;y)$, respectively. The pion block has indices contracted at time $t=0$, and open spin and color indices at the other time. These open indices are then contracted by the  operator at $t=0$ and a single space-time point. This same pion block is time reversed, utilizing the periodic boundary conditions, such that the sink pion propagates backward toward the operator insertion (see Fig.~\ref{fig:shortdist}). The spatial indices at source and sink are summed over in order to project onto zero momentum. This setup, requiring all quark propagators to be contracted by the operator, is similar in spirit to calculations of $K^0$-~\cite{Aoki:2010pe,Durr:2011ap,Boyle:2012qb,Bertone:2012cu,Bae:2013tca,Bae:2014sja,Carrasco:2015pra,Jang:2015sla,Garron:2016mva}, $D^0$-~\cite{Carrasco:2015pra,Bazavov:2017weg} and $B^0_{(s)}$-meson mixing~\cite{Gamiz:2009ku,Carrasco:2013zta,Aoki:2014nga,Bazavov:2016nty} or $N\bar{N}$ oscillations~\cite{Buchoff:2012bm,Rinaldi:2018osy}. 

\begin{figure*}
\centering
\includegraphics[width=0.54\textwidth]{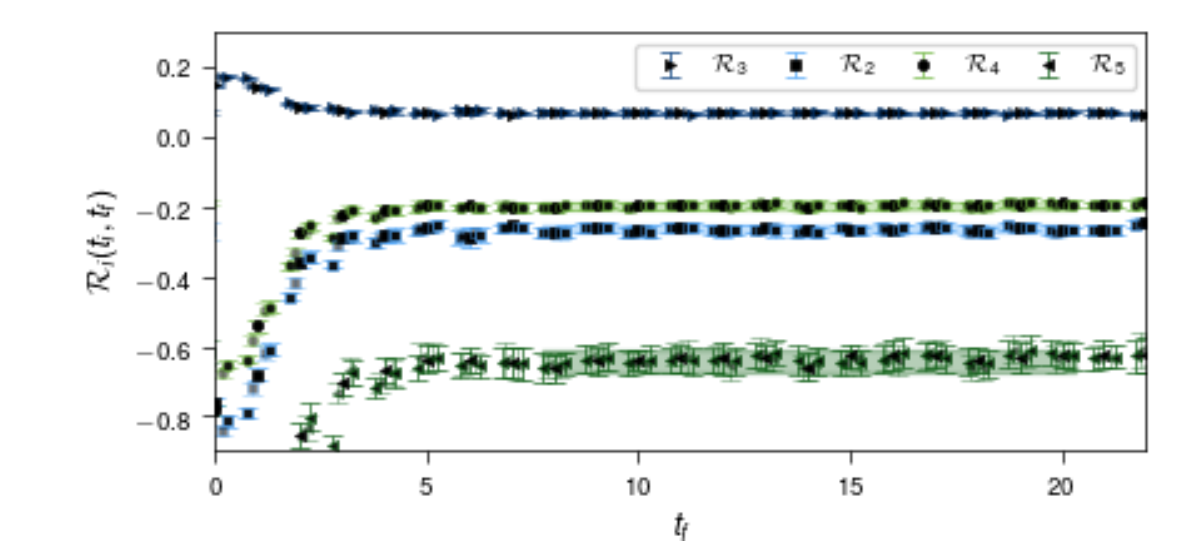}
\includegraphics[width=0.40\textwidth]{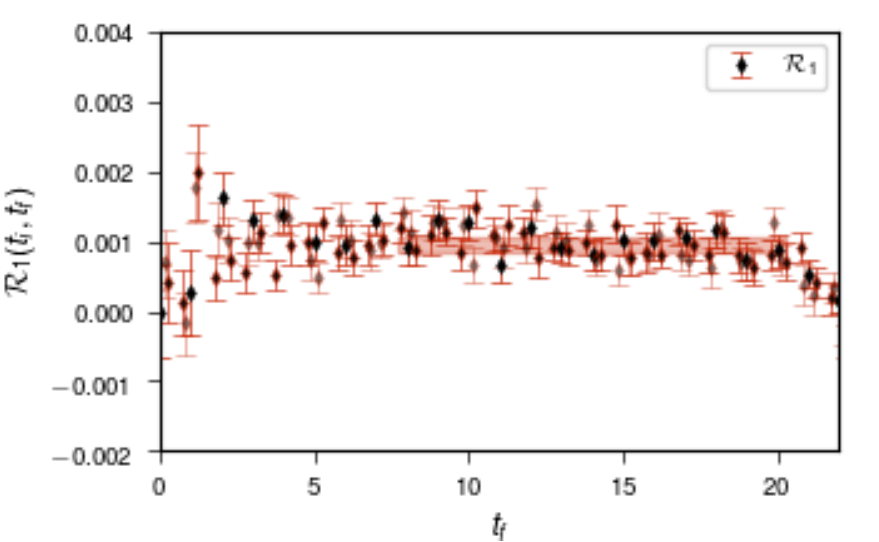}
 \caption{\label{fig:0nubbshortsigs}
 An example of the $t_i$ and $t_f$ dependence of the ratio correlation functions, $R_i(t_i, t_f )$ for the five relevant \znubb\ operators on a near physical pion mass ensemble with $a \approx 0.12$ fm. The filled black symbols correspond to the diagonal components $|t_i| = t_f$. The neighboring points with open symbols correspond to, from left to right, $|t_i| = t_f +[-2,-1,1,2]$. The horizontal bands are the ground state contributions to $R_i$ extracted from single-state fits.  (data and fits from Ref.~\cite{Nicholson:2018mwc}, converted to the basis of Eq.~\ref{LagSca})}
\end{figure*}

From $C_i^{3\mathrm{pt}}(t,T-t)$, Eq.~(\ref{eq:C3pt_znubb}), ratios $\mathcal{R}_i(t) $ with the pion correlation function, $C_{\pi}(t)$, can be formed and related to the pion matrix element of operator ${\cal O}_i$  as:
\begin{align}\label{eq:ratio}
\mathcal{R}_i(t) \equiv C_i^{3\mathrm{pt}}(t,T-t)/\left(C_{\pi}(t)C_{\pi}(T-t)\right)\underset{t,T-t\to\infty}{\longrightarrow}\frac{a^4\langle\pi|{O}_{i}|\pi\rangle}{(a^2Z_0^\pi)^2}+\mathcal{R}_\textrm{e.s.}(t) \ ,
\end{align}
where $Z_0^\pi$ gives the overlap of the pion operator onto the pion ground state and may be extracted from the pion two-point correlation function $C_{\pi}(t)$, analogous to Eq.~(\ref{eq:c2pt}). Residual effects from excited state contamination in $\mathcal{R}_{e.s.}(t)$ can be shown to fall of exponentially with the time separations $\{t,|T-t|\}$. Forming this ratio, which removes the need to extract the pion masses in a separate calculation, has the added benefit of canceling the contributions from the first thermal state in the pion correlation functions. In Fig.~\ref{fig:0nubbshortsigs}, an example of this ratio calculated in Ref.~\cite{Nicholson:2018mwc} is reproduced, showing the clear ground-state plateaus obtained with this method. 

Once matrix elements have been extracted on various  ensembles of gauge fields, extrapolations to the continuum, physical pion mass, and infinite volume limits must be performed.
To determine the pion mass dependence, it is straightforward to include the operators in $\chi$PT and to derive the virtual pion corrections which arise at next-to-leading order in the chiral expansions,
\begin{align}\label{eq:Oi_chipt}
\frac{O_1}{\e_{\pi}^2} = \frac{b_1 \Lambda_{\chi}^4}{(4\pi)^2} \bigg[
		1 -\e_\pi^2
		\left(
			3\ln(\e_\pi^2)
			+1
			- c_1
		\right)
		\bigg]\, ,
		\cr
O_{4,5} = \frac{b_{4,5}\Lambda_{\chi}^4}{(4\pi)^2} \bigg[
		1 +\e_\pi^2
		\left( \phantom{3}
			\ln(\e_\pi^2)
			-1
			+ c_{4,5}
		\right)
		\bigg]\, ,
\cr
O_{2,3} = \frac{b_{2,3} \Lambda_{\chi}^4}{(4\pi)^2} \bigg[
		1 +\e_\pi^2
		\left( \phantom{3}
			\ln(\e_\pi^2)
			-1
			+ c_{2,3}
		\right)
		\bigg]\, ,
\end{align}
where $\Lambda_\chi = 4\pi F_\pi$, and $\e_\pi = \frac{m_\pi}{\Lambda_\chi}$ is the small expansion parameter. The dimensionless LEC's, $b_i$ and $c_i$ may be constrained from fits to the data, with the $b_i$ determining the $g^{\pi\pi}_i$ in 
Eq.~\eqref{eq:dim9pipi}. 
These expressions can be generalized to incorporate finite lattice spacing corrections arising from a particular lattice action~\cite{Sharpe:1998xm} and finite volume corrections arising from virtual pions becoming sensitive to the finite periodic volume~\cite{Gasser:1987zq}.

Finally, the bare quark operators must be renormalized and evolved to the appropriate scale for matching onto a given BSM model. While the combined currents that enter the chiral Lagrangian are necessarily ``color blind", gluon interactions will intermingle the colors amongst the quarks in a given four-quark operator. In particular, as the scale is run between the electroweak and QCD scales, some of the operators defined in Eq.~\eqref{LagSca} will mix under renormalization. In particular, $O_2$ will mix with $O_3$, while $O_4$ mixes with $O_5$. Therefore, one must compute the renormalization of the full matrix of operators including the off-diagonal mixing which will become non-zero as the scale is varied. 

One group~\cite{Nicholson:2018mwc} has calculated the dominant $\pi^- \to \pi^+$ matrix elements arising from short-range operators relevant for experimental searches for \znubb, performing a full extrapolation to the physical point. The calculation was performed on  Highly-Improved Staggered Quark (HISQ) ensembles produced by the MILC collaboration~\cite{Bazavov:2012xda}, and includes ensembles with pion masses ranging from $ 130 \lsim m_{\pi} \lsim 310\ $MeV, lattice spacings $0.09 \lsim a \lsim 0.15\ $fm, and several volumes corresponding to $m_{\pi} L \sim 3.2 - 5.4$, allowing for all systematics to be controlled. 
A mixed action approach is taken by solving for M\"obius Domain Wall Fermion (MDWF) propagators on these gauge field configurations, after applying  gradient flow smearing  to the ensembles to reduce noise stemming from  UV fluctuations~\cite{Luscher:2010iy}. 
While more costly to produce, the MDWF valence quark action has improved chiral symmetry properties, resulting in smaller discretization errors beginning at $\mathcal{O}(a^2)$. The effects of the mixed action are incorporated into the extrapolation formulae using a partially quenched version of $\chi$PT~\cite{Berkowitz:2017opd}.  

\begin{figure*}
\centering
\includegraphics[width=0.44\textwidth]{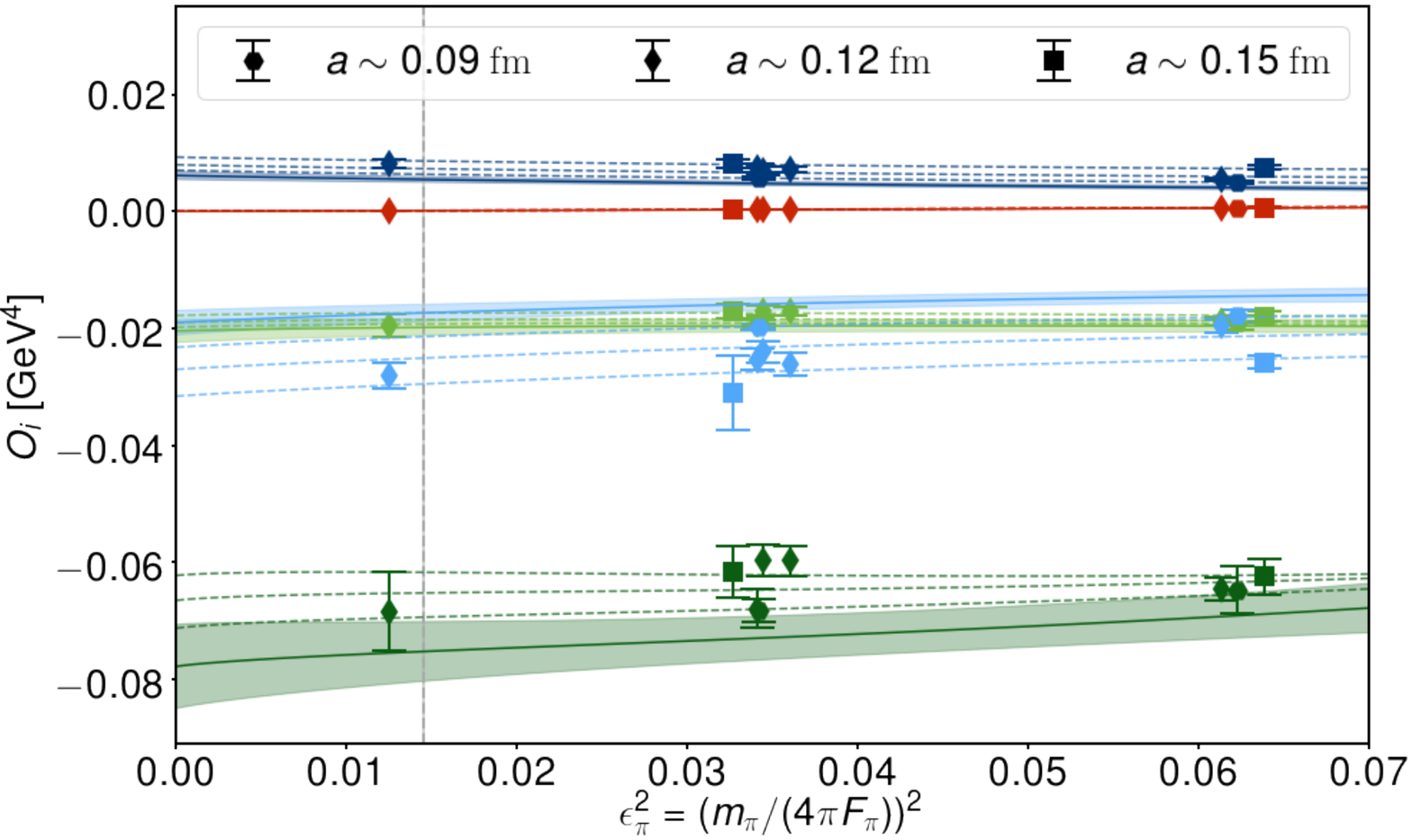}
\includegraphics[width=0.43\textwidth]{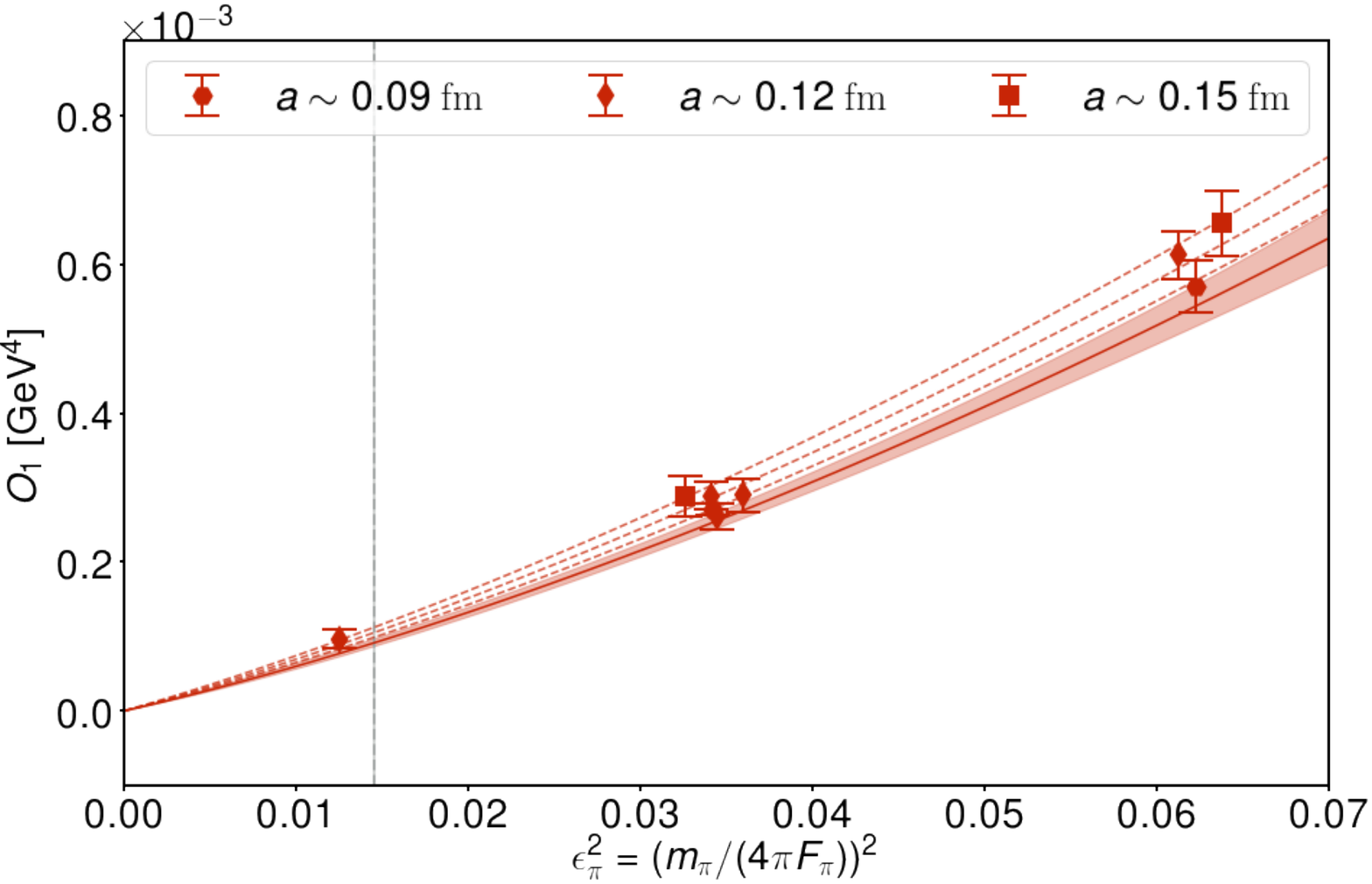}
 \caption{\label{fig:0nubbshortextraps} 
 Interpolations/extrapolations of the pion matrix elements. Light and dark blue correspond to $O_{2,3}$, respectively, light and dark green to $O_{4,5}$, and red to $O_1$. The fit bands are
constructed with $\Lambda_{\chi}$ held fixed while changing $\epsilon_{\pi}$, so the
corresponding LQCD results are adjusted by $(F_{\pi}/F_\pi^{\mathrm{latt}})^4$ (where $F_\pi^{\mathrm{latt}}$ is the value of the pion decay constant determined on a given ensemble)
for each lattice ensemble in order to be consistent with this interpolation. The bands represent the 68\% confidence interval of the continuum, infinite volume extrapolated values of the matrix elements. The vertical gray band highlights the physical pion mass point. (data and fits from Ref.~\cite{Nicholson:2018mwc}, converted to the basis of Eq.~\ref{LagSca})}
\end{figure*}

The relevant matrix of renormalization constants is computed non-perturbatively following the Rome-Southampton method~\cite{Martinelli:1994ty} with a non-exceptional kinematics-symmetric point~\cite{Sturm:2009kb}, using the RI/SMOM $(\gamma_\mu,\gamma_\mu)$-scheme~\cite{Boyle:2017skn}.
Momentum sources are implemented to achieve a high statistical precision~\cite{Gockeler:1998ye}, and non-perturbative step-scaling techniques~\cite{Arthur:2010ht,Arthur:2011cn} are used to run the Z-factors to a common scale (reported at $\mu = 3\ $GeV in the original publication).
The mixed-action setup provides an additional benefit because the renormalization pattern is the same as in the continuum
(to a very good approximation) and does not require the spurious subtraction of operators of different chirality. Examples of the chrial and continuum extrapolations are  shown in  Fig.~\ref{fig:0nubbshortextraps}.

At the scale  $\mu=2$ GeV in the $\overline{\rm MS}$ scheme, Ref.~\cite{Nicholson:2018mwc} finds:
 $g^{\pi\pi}_{1} =  0.36 \pm 0.02$, 
  $g^{\pi\pi}_{2}=   2.0  \pm 0.2 $  GeV$^2$, 
 $g^{\pi\pi}_{3} =  -(0.62 \pm 0.06)$   GeV$^2$,   
  $g^{\pi\pi}_{4}= -(1.9  \pm 0.2)$   GeV$^2$,    
 $g^{\pi\pi}_{5} =   -(8.0  \pm 0.6)$    GeV$^2$.  
 Note that the scaling of these LECs agrees with the NDA estimate.   
Prior to the direct LQCD calculation, 
these LECs were also  extracted in Refs.~\cite{Savage:1998yh,Cirigliano:2017ymo} using 
$SU(3)$ relations between $\pi^+$-$\pi^-$ and the kaon processes $K$-$\overline{K}$ and $K \rightarrow \pi \pi$  matrix elements, for which LQCD results are also available~\cite{Carrasco:2015pra,Bertone:2012cu,Jang:2015sla,Boyle:2012qb,Garron:2016mva,Blum:2012uk,Blum:2015ywa}. 
Ref.~\cite{Cirigliano:2017ymo} found
$g^{\pi\pi}_{1} = 0.38 \pm 0.08$, $g^{\pi\pi}_{2} = 2.9 \pm 0.6$ GeV$^2$, $g^{\pi\pi}_{3} = -( 1.0 \pm 0.3)$ GeV$^2$, $g^{\pi\pi}_{4} = -(2.5 \pm 1.3)$ GeV$^2$, $g^{\pi\pi}_{5} = -(11 \pm 4)$ GeV$^2$. 
This indirect LQCD extraction is in good agreement with the more precise results of Ref.~\cite{Nicholson:2018mwc}.

\subsection{Long distance contributions in pion matrix elements}
\label{sec:znubb_long}

A light Majorana neutrino  propagates over ``long distances'' that are resolvable at the QCD scale, and therefore the non-locality of the second order weak-process must be incorporated into the evaluation of the  QCD matrix elements. This leads to more complicated calculations than those discussed in Section \ref{sec:znubb_short}. In contrast to the \tnubb\ process discussed in Section \ref{sec:tnubb}, the momentum carried by the neutrino propagator must be integrated over, also leading to additional complications.

The \znubb\ process between an initial state $i$ and final state $fe^-e^-$ is induced at second order in  perturbation theory, with two insertions of the $\Delta I=1$ weak Hamiltonian in Eq.~(\ref{eq:intro.1.5}) leading to the bi-local matrix element 
\begin{equation}
  \int d^{4} x \, d^{4} y \left\langle f e e \right\vert T \left\{ \mathcal{H}_{W}(x) \mathcal{H}_{W}(y) \right\} \left\vert i \right\rangle = 4 m_{\beta \beta} G_{F}^{2} V_{ud}^{2} \int d^{4} x \, d^{4} y \, H_{\alpha \beta}(x,y) L_{\alpha \beta}(x,y),
  \label{eqn:dbd_me}
\end{equation}
where the leptonic tensor is given by
\begin{equation}
  L_{\alpha \beta} \equiv \overline{e}_{L}(p_{1}) \gamma_{\alpha} S_{\nu}(x,y) \gamma_{\beta} e_{L}^{C}(p_{2}) e^{- i p_{1} \cdot x} e^{- i p_{2} \cdot y}
  \label{eqn:leptonic_tensor}
\end{equation}
and the hadronic tensor is given by
\begin{equation}
  H_{\alpha \beta} \equiv \left\langle f \right\vert T \left\{ J_{\alpha L:}(x) J_{\beta L}(y) \right\} \left\vert i \right\rangle\,,
\label{eqn:hadronic_tensor}
\end{equation}
and $J_{\mu L}(x)=\overline{u}_{L}(x) \gamma_{\mu} d_{L}(x)$.
The neutrino propagator is given by
\begin{eqnarray}
S_\nu(x,y) = S_\nu(x-y) = \int \frac{d^4q}{(2\pi)^4} \frac{ e^{iq\cdot (x-y)}}{q^2}\,.
\end{eqnarray}
%
%
The convolution with the leptonic tensor and resulting integration mean that evaluations of the hadronic tensor are required for all sets of space-time points. The leptonic tensor results in an integration kernel that has significant support for momenta of up to  ${\cal O}(100)$ MeV. 
An additional complication is that LQCD calculations are performed in finite volume Euclidean space-time, while the physical matrix elements are required in Minkowski space. The  analytically continued information can be extracted but requires careful treatment, particularly when the initial and/or final state involve multiple hadronic states.

\subsubsection{The transition $\pi^-\to \pi^+ e^- e^-$}

Two groups \cite{Detmold:2018zan,Tuo:2019bue} have pursued calculations of the \znubb\ matrix element that induces a transition between an initial $\pi^-$ state and a final $\pi^+e^-e^-$ state. While this transition is not physical due to the electron masses and degeneracy of the $\pi^\pm$, it can be studied at unphysical kinematics. This matrix element is equivalent to the charge exchange zero-momentum scattering process $\pi^- e^+ \to \pi^+ e^-$. From the point of view of LQCD, it is the simplest \znubb\ process that can be investigated as it does not suffer from an exponential signal-to-noise problem at increasing Euclidean time, and the initial and final states contain only single hadrons. 

This process is a low energy process and $\chi$PT provides a prediction for it. At next-to-leading order, this is determined by a single low energy constant {LEC}~\cite{Cirigliano:2017tvr}
\begin{equation}
\label{eq:ChPT2}
\begin{split}
    \frac{\mathcal{A}(\pi^-\to\pi^+ee)}{\mathcal{A}^{LO}}=1+\frac{m_\pi^2}{(4\pi
    F_\pi)^2}\left(3\log\frac{\mu^2}{m_\pi^2}+6+\frac{5}{6}
    g_\nu^{\pi\pi}(\mu)\right)\,,
\end{split}
\end{equation}
where ${\cal A}^{LO} = -8 G_F^2 |V_{ud}|^2m_{\beta\beta} \bar{u}_L (p_{e1}) C u_L^T (p_{e2}) \times F_\pi^2 $ is the leading order prediction (corresponding to the vacuum saturation approximation \cite{Tuo:2019bue}). Here, the LEC $g_\nu^{\pi\pi}(\mu)$ from Eq.~(\ref{eq:CTN2LO}) depends on the renormalization scale in such a way that the amplitude is renormalization-scale independent.

\begin{figure*}[t]
\subfigure[]{
      \raisebox{0mm}{\includegraphics[width=0.44\textwidth]{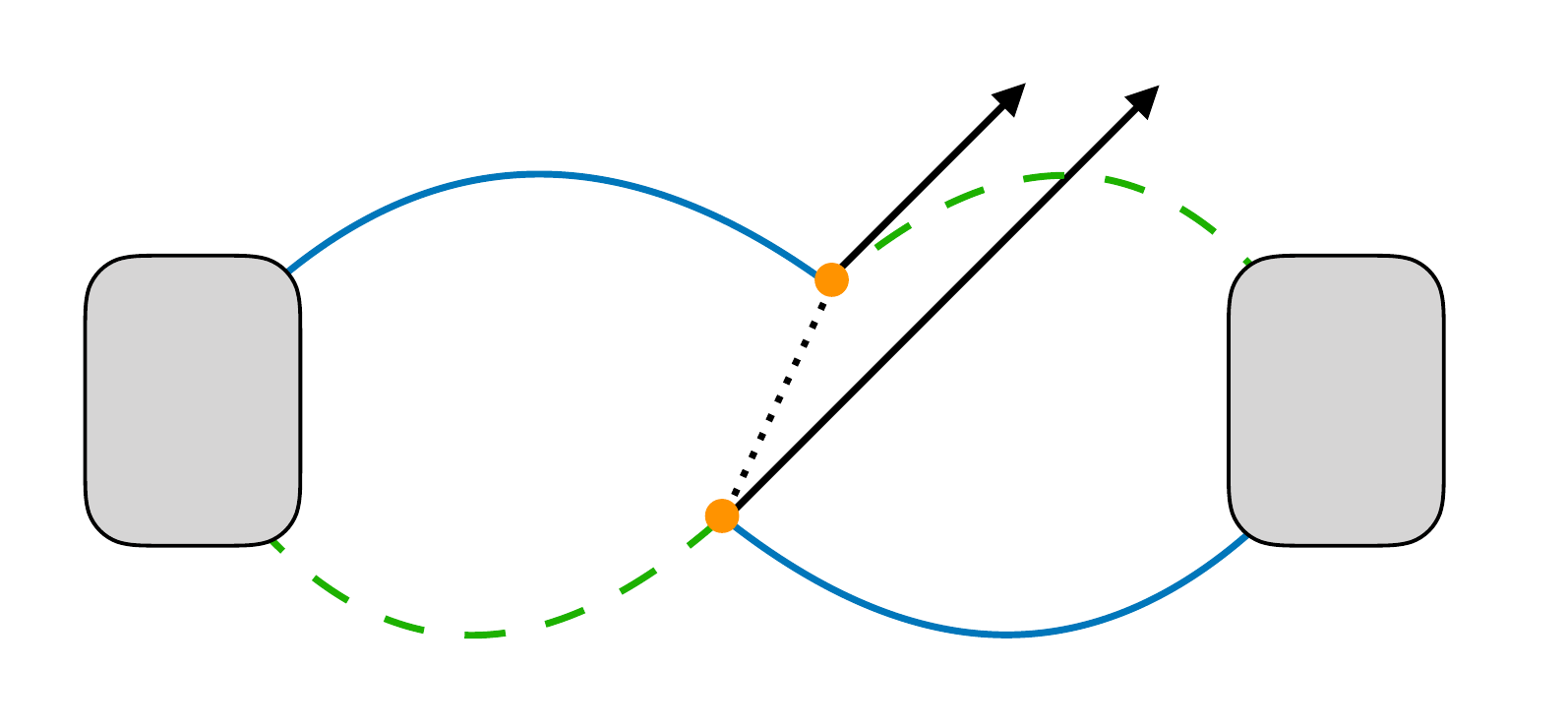}}
       }\qquad \qquad
   \subfigure[]{
       \centering
       \includegraphics[width=0.44\textwidth]{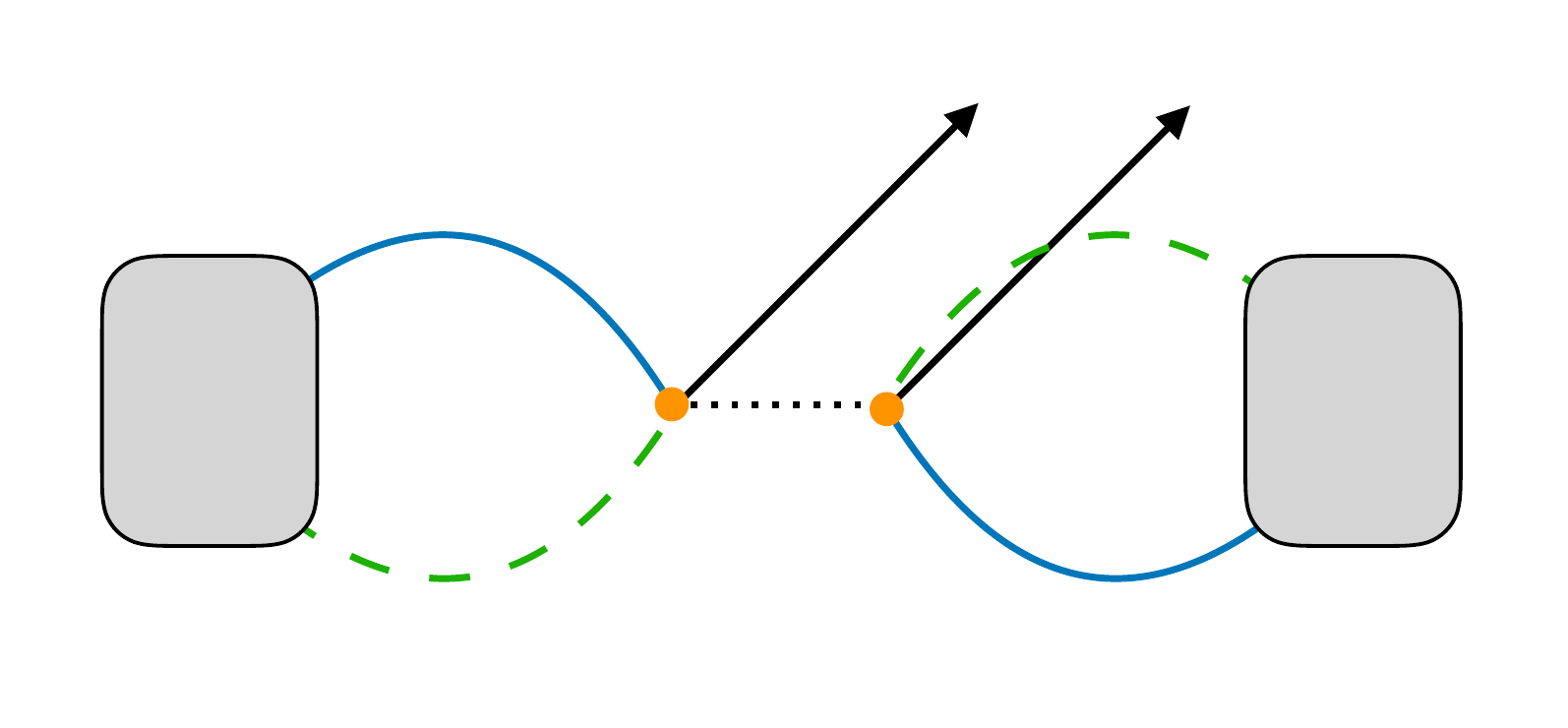}
       }   
 \caption{\label{fig:pi2piee}Contractions for the $\pi^- \to \pi^+ e^- e^-$ transition in Eqs. (\ref{eq:pitopia}) and (\ref{eq:pitopib}). The solid blue and dashed green lines represent down and up quark propagators respectively and the circles represent the $\Delta I=1$ weak vertices. The dotted and solid black lines represent the Majorana neutrino propagator and electron final state respectively.}
\end{figure*}

\begin{figure*}
\centering
\includegraphics[width=0.47\textwidth]{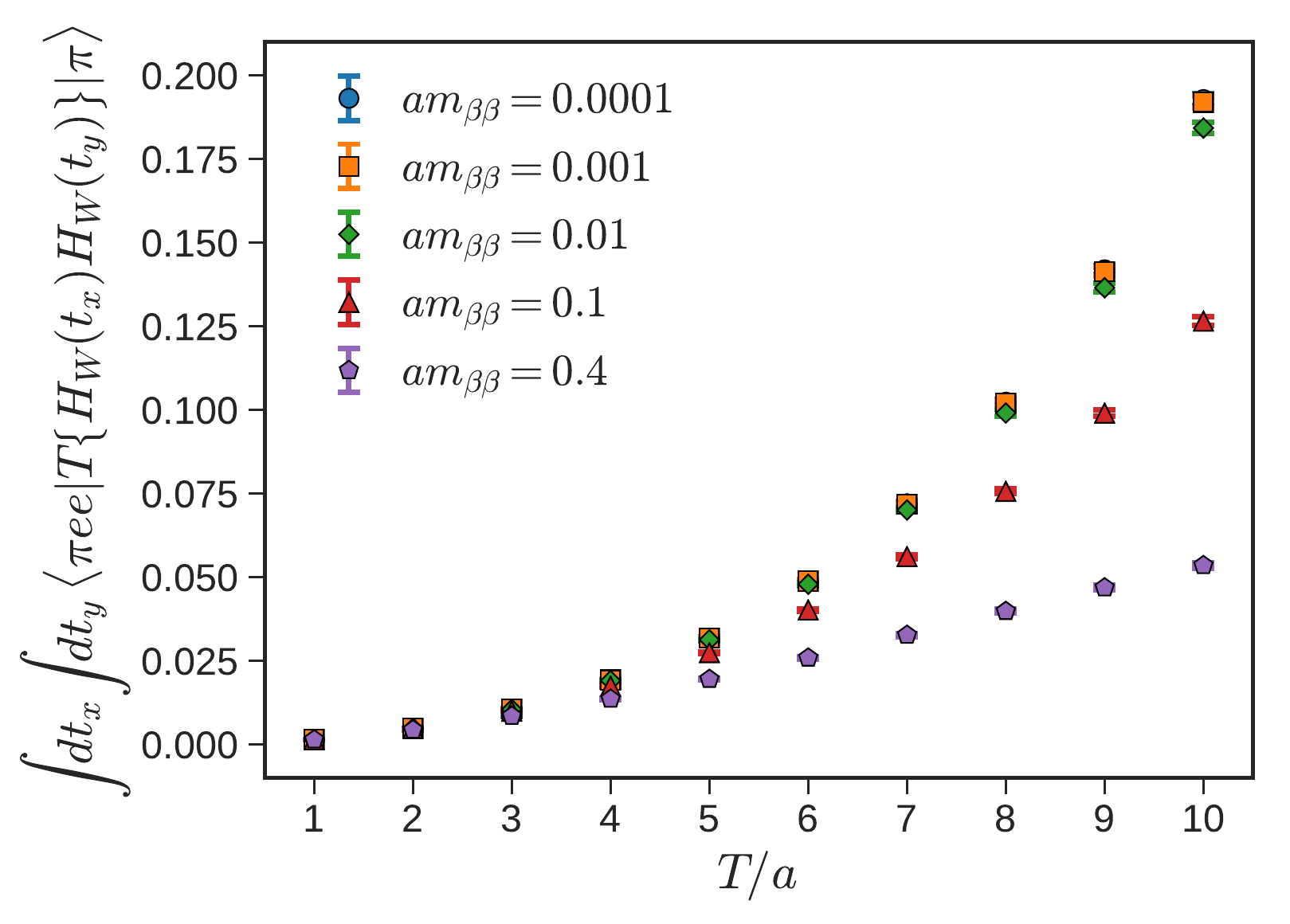}
\qquad
\includegraphics[width=0.47\textwidth]{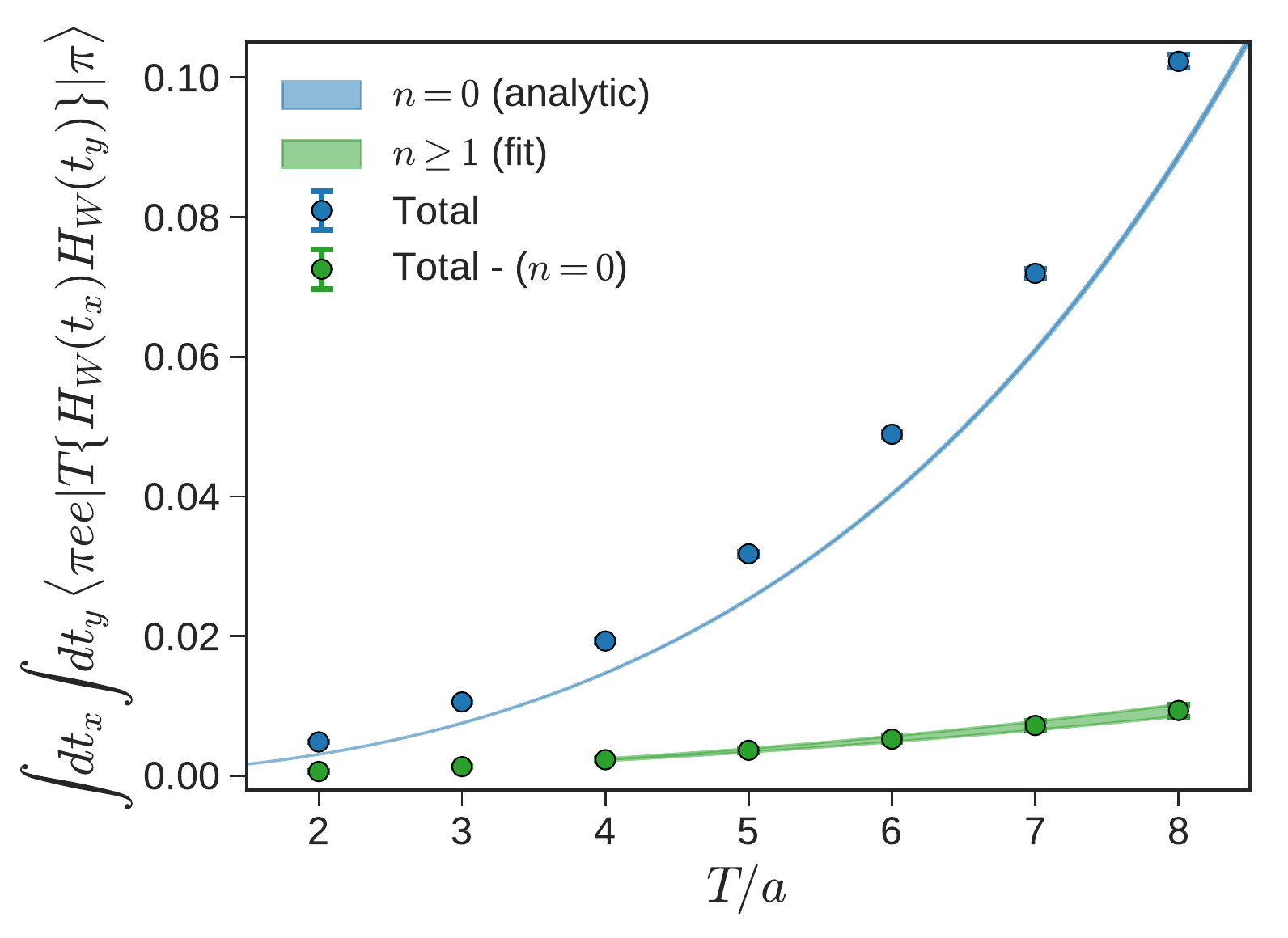}
 \caption{\label{fig:znubblongmurphy1} 
 The  integrated transition amplitude for various different neutrino masses (left) and decomposed into the various terms contributing in Eq.~(\ref{eq:largeT}). (From Ref.~\cite{Murphy:2018tdc})}
\end{figure*}

In Ref.~\cite{Detmold:2018zan}, a preliminary study of this process is presented using domain wall fermions at a pion mass of $m_\pi\sim 420$ MeV with further developments underway \cite{MurphyDetmold2019_inprogress}. The calculation builds on the pioneering studies  of rare kaon decays  by the RBC collaboration \cite{Bai:2017fkh,Bai:2018hqu} and the NPLQCD collaboration study \cite{Shanahan:2017bgi,Tiburzi:2017iux} of the \tnubb\ process discussed in the previous Section.  Specifying to the particular initial and final states in Eq.~(\ref{eqn:dbd_me}), the hadronic matrix elements of interest can be determined from the correlation function
 \begin{equation}
 \begin{split}
 C_{\mu\nu}^{\pi\to\pi}(t_+,x,y,t_-)=\langle0 |T\left\{\chi_{\pi^+}(t_+)J_{\nu L}(x)J_{\mu L}(y)\chi^\dagger_{\pi^-}(t_-)\right\}|0\rangle
 \end{split}
 \label{eq:pipi4pt}
 \end{equation} 
with $\chi_{\pi^+}$ and $\chi^\dagger_{\pi^-}$ being interpolating operators for zero-momentum pion states. 
The correlation function above is constructed from two different types of Wick contractions which are shown in  Fig.~\ref{fig:pi2piee}. The forms of the quark line components of these contractions are 
\begin{equation}
\label{eq:pitopia}
   C^{\pi\to\pi}_{(a)} = \mathrm{Tr} \left[ S_{u}^{\dagger}(t_{+} \rightarrow x)  S_{d}(t_{+} \rightarrow y) \gamma_{\mu} \left( 1 - \gamma_{5} \right)  S_{u}^{\dagger}(t_{-} \rightarrow y) S_{d}(t_{-} \rightarrow x) \gamma_{\nu} \left( 1 - \gamma_{5} \right) \right]
\end{equation}
and  
\begin{equation}
\label{eq:pitopib}
  C^{\pi\to\pi}_{(b)} = \mathrm{Tr} \left[ S_{u}^{\dagger}(t_{-} \rightarrow x) \gamma_{\mu} \left( 1 - \gamma_{5} \right) S_{d}(t_{-} \rightarrow x) \right] \cdot \mathrm{Tr} \left[ S_{u}^{\dagger}(t_{+} \rightarrow y) \gamma_{\nu} \left( 1 - \gamma_{5} \right) S_{d}(t_{+} \rightarrow y) \right]\,,
\end{equation}
respectively. Here, a simple local pion interpolator, $\chi_{\pi^+}=\overline{u}\gamma_5 d$, has been used and terms with $\mu\leftrightarrow\nu$ and $x\leftrightarrow y$ are also required.
Inserting a complete set of states between the two currents in Eq.~(\ref{eq:pipi4pt}) as in Refs.~\cite{Bai:2014cva,Bai:2018hqu}, it is clear that the correlation function has the asymptotic time dependence
\begin{eqnarray}
{\cal C}^{\pi\to\pi}(t;T) &\equiv&
  \sum_{{\bf x},{\bf y}} \sum_{t_x=0}^{T}\sum_{t_y=0}^{T} \frac{ L^{\mu\nu}(x,y)C_{\mu\nu}^{\pi \rightarrow \pi}(t_+,x,y,t_-) }{C_\pi(t)}  \nonumber \\
  &\propto& 
  \sum_{n}  \frac{\left\langle \pi e e \right\vert \mathcal{H}_{W} \left\vert n \right\rangle \left\langle n \right\vert \mathcal{H}_{W} \left\vert \pi \right\rangle}{E_n(E_{n} - m_{\pi})} \left[ T + \frac{e^{-(E_{n}-m_{\pi}) T} - 1}{E_{n} - m_{\pi}} \right]
  \label{eq:largeT}
\end{eqnarray}
for pions at rest, where $T$ is the size of the temporal integration window for the weak current insertions and $t = \vert t_{+} - t_{-} \vert$ is the $\pi^{-} - \pi^{+}$ source-sink separation. In deriving this formula, it is assumed that the current insertions are held sufficiently far from the pion source and sink 
($t_- \ll 0 \ll T \ll t_+$) 
so that the couplings to excited states before and after the integration window may be safely neglected. The states contributing to the sum are: $\left|e\overline\nu_e\right\rangle$, $\left|\pi e\overline\nu_e\right\rangle$, $\left|n=2 \right\rangle$, $\ldots$, with energies $E\sim m_e<m_\pi$, $E\sim m_\pi$ and $E > m_\pi$. For the lowest-energy state, the terms in the square brackets in Eq.~(\ref{eq:largeT}) are growing exponentially with $T$ and the matrix element is the square of the pion decay constant; for the second state, $\left|\pi e\overline\nu_e\right\rangle$, the terms in the square brackets behave approximately quadratically; for  the remaining $n\ge 2$ terms, the large $T$ behaviour of Eq.~\eqref{eq:largeT} is linear. Combining these pieces, the matrix element governing \znubb\
\begin{equation}
  M^{0 \nu} = \sum_{n} \frac{\left\langle \pi e e \right\vert \mathcal{H}_{W} \left\vert n \right\rangle \left\langle n \right\vert \mathcal{H}_{W} \left\vert \pi \right\rangle}{E_n(E_{n} - m_{\pi})}
  \label{eqn:spectral_me}
\end{equation}
can be determined. The procedure by which this can be achieved is illustrated in Fig.~\ref{fig:znubblongmurphy1} ad discussed in detail in Ref.~\cite{Murphy:2018tdc}.

\begin{figure*}
\centering
\includegraphics[width=0.62\textwidth]{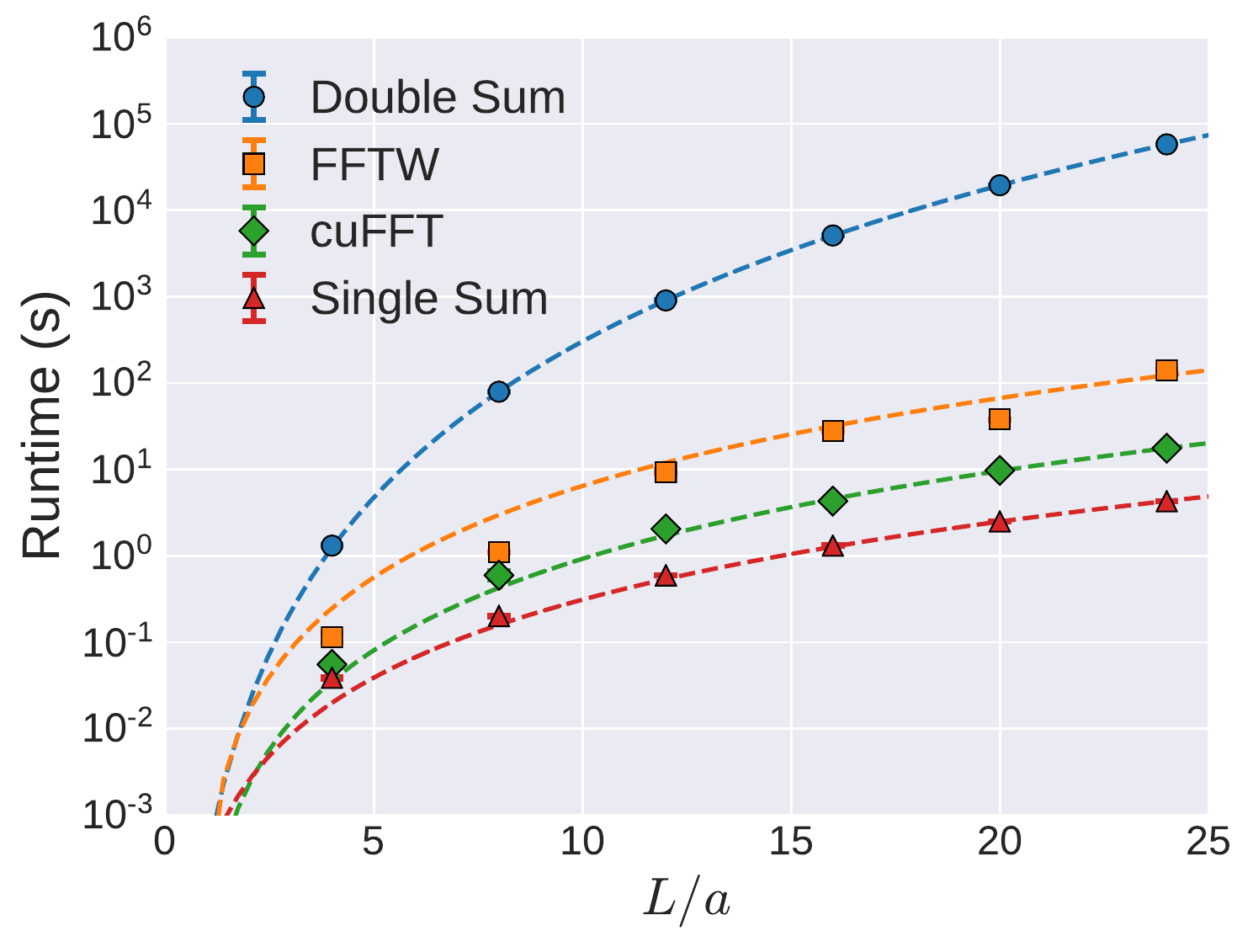}
 \caption{\label{fig:znubblongmurphy2} The runtime of the various implementations of the double summation over the neutrino creation and annihilation points. The results correspond to the naive double summation, a CPU implementation (FFTW) of the FFT,  a GPU implementation of the FFT and, simply setting one of the ends of the neutrino propagator to be fixed (single sum), albeit this last method has significantly larger variance. Further details on the comparison are given in Ref.~\cite{Murphy:2018tdc}. 
 	(From Ref.~\cite{Murphy:2018tdc})}
\end{figure*}

The  calculation of the double sum over the spatial volume in Eq.~(\ref{eq:largeT}) is naively numerically prohibitive for all but the smallest volumes.
Fortunately, the translational invariance of the neutrino propagator can be exploited to reduce this to $\mathcal{O}(V \log V)$ using the convolution theorem
\begin{equation}
  \int d^{3} x \, d^{3} y \, f_{\alpha}(x) L_{\alpha \beta}( x - y ) g_{\beta}(y) = \int d^{3} x \, f_{\alpha}(x) \left[ \mathcal{F}^{-1} \left\{ \mathcal{F}(L_{\alpha \beta}) \cdot \mathcal{F}(g_{\beta}) \right\} \right] (x)
  \label{eqn:convolution_theorem}
\end{equation}
and the  Fourier transform, $\mathcal{F}(\ldots)$. 
As discussed in Ref.~\cite{Murphy:2018tdc}, this convolution can be implemented using the fast Fourier transform (FFT). The second type of contraction, $C^{\pi\to\pi}_{(b)}$, in Fig.~\ref{fig:pi2piee}, requires the FFT to be performed for the various spin-color components separately. The speed-up achieved using FFT is significant and is shown in Fig.~\ref{fig:znubblongmurphy2}.

\begin{figure*}
	\centering
	\includegraphics[width=0.62\textwidth]{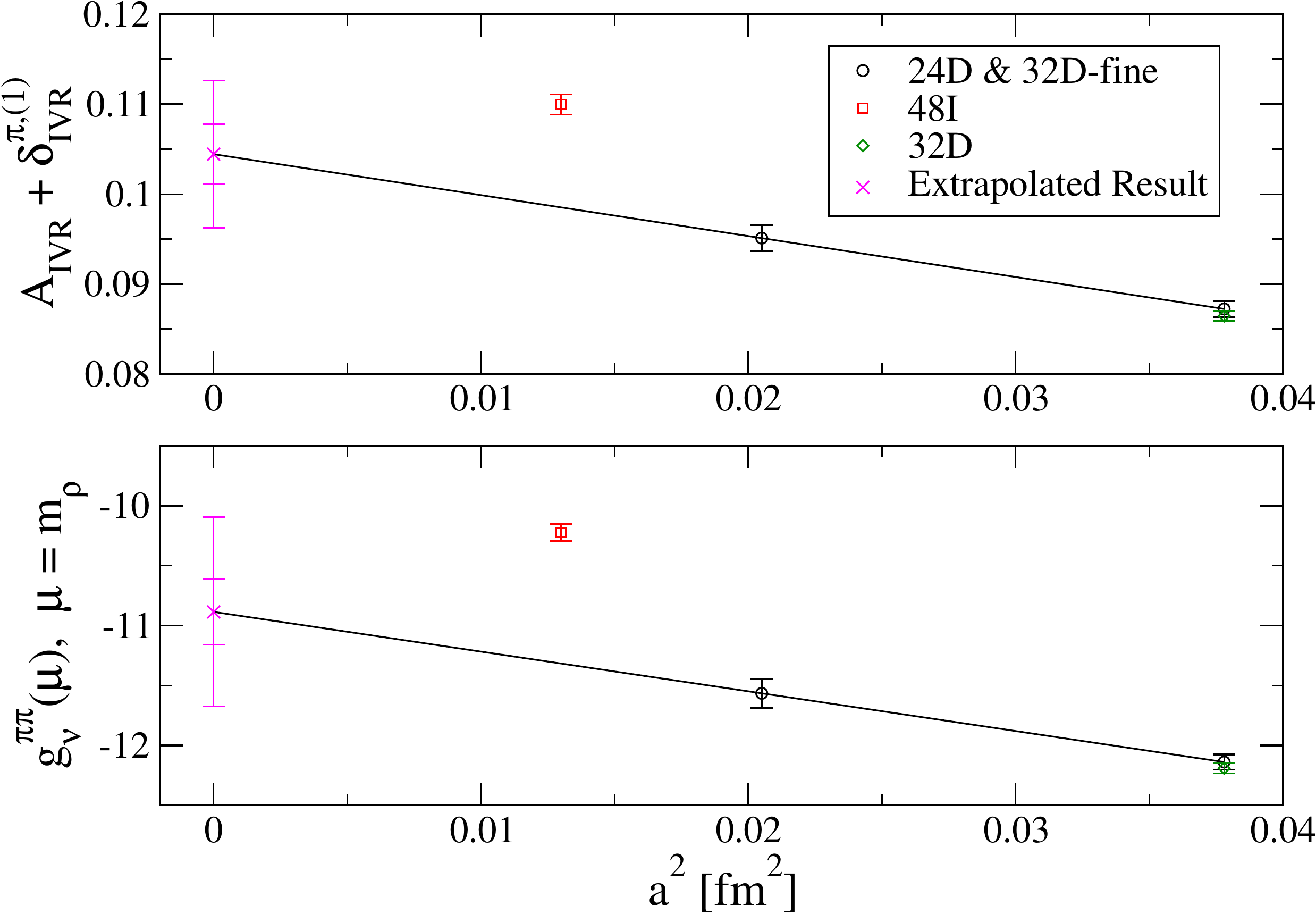}
	\caption{\label{fig:znubblongtou}Continuum limit of the infinite volume reconstructed amplitude for $\pi^-\to\pi^+ e^- e^-$ \cite{Tuo:2019bue}. The notation 24D, 32D, 32D-fine, and 48I refers to the different ensembles used in Ref.~\cite{Tuo:2019bue}. In each case the extrapolated result is obtained as a linear fit vs $a^2$ ignoring the 48I point which corresponds to a different discretization. The lower panel shows the extraction of the \xpt \ LEC $g_\nu^{\pi\pi}(\mu=m_\rho)$. (From Ref.~\cite{Tuo:2019bue})}
\end{figure*}

In Ref.~\cite{Tuo:2019bue}, Tuo \textit{et al.} have presented a further, more complete study of this $\pi^-\to\pi^+ e^- e^-$  process also using domain wall fermion ensembles. This calculation differs in technical details from that of Ref.~\cite{Murphy:2018tdc}  and also contains results at the physical values of the quark masses for the first time.  This calculation uses sophisticated statistical methods and also implement the infinite volume reconstruction technique \cite{Feng:2018qpx} to ameliorate the finite volume effects of the almost massless neutrino propagator. In addition, the calculation proceeds by dividing the full correlator by the completely disconnected version of contraction $C^{\pi\to\pi}_{(b)}$ in Fig.~\ref{fig:pi2piee} (performing separate averages  over gluon field configurations of the two traces) which corresponds to subtracting the leading order \xpt\ result.
Noting the similarity of the $\pi^-\to\pi^+ e^- e^-$ hadronic matrix element to the electromagnetic self energy of the pion (computed in Feynman gauge), Tuo \textit{et al.} used the infinite volume reconstruction method \cite{Feng:2018qpx} 
which in this case removes all power-law suppressed finite-volume effects caused by the (almost) massless neutrino propagator. The infinite volume reconstruction method works due to the fact that at large enough time separation, the matrix elements are dominated by single pion intermediate states contribution (significant effort is put into assessing the systematic effects of this and the remaining exponentially suppressed finite-volume effects).
Four different DWF ensembles at quark masses very close to their physical values were used with volumes ranging from $m_\pi L=3.3$--4.5. Different lattice spacings and actions were also used, enabling an estimate of the continuum limit. Figure \ref{fig:znubblongtou} shows the continuum limit extrapolation.  To achieve clean signals, all mode averaging and low-mode deflation were used and wall sources for the quark  fields were placed on every timeslice.
The final result in Tuo \textit{et al.} is  the amplitude
${\cal A} = 0.1045(34)(50)_L(55)_a$ where the uncertainties are from statistics, finite volume and continuum extrapolation respectively. This allows a determination of the $\chi$PT low energy constant 
\begin{equation}
\left.g_{\nu}^{\pi \pi}(\mu)\right|_{\mu=m_{\rho}}=-10.89(28)(33)_{L}(66)_{a}~, 
\end{equation}
more precise and in  reasonable agreement with the large-$N_C$ estimate $ g_{\nu}^{\pi \pi}(\mu)|_{\mu=m_{\rho}}=-7.6 $~\cite{Cirigliano:2017tvr,Ananthanarayan:2004qk}. 

\subsubsection{The transition $\pi^-\pi^-\to  e^- e^-$}

Second order weak interactions also generate the crossed channel transition $\pi^-\pi^-\to  e^- e^-$. This
transition, while kinematically allowed, is not accessible experimentally but provides a similarly useful theoretical arena as the $\pi^-\to \pi^+ e^- e^-$ transition. In Ref.~\cite{Feng:2018pdq}, Feng \textit{et al.} have investigated the corresponding transition amplitude in the light neutrino exchange scenario. The calculations were performed using domain-wall fermion ensembles with quark masses corresponding to  pion masses $m_\pi=420$ and 140 MeV. While exploratory, this calculation demonstrated the feasibility of the methods used.

\begin{figure*}
\centering
\subfigure[]{
      \raisebox{0mm}{\includegraphics[width=0.38\textwidth]{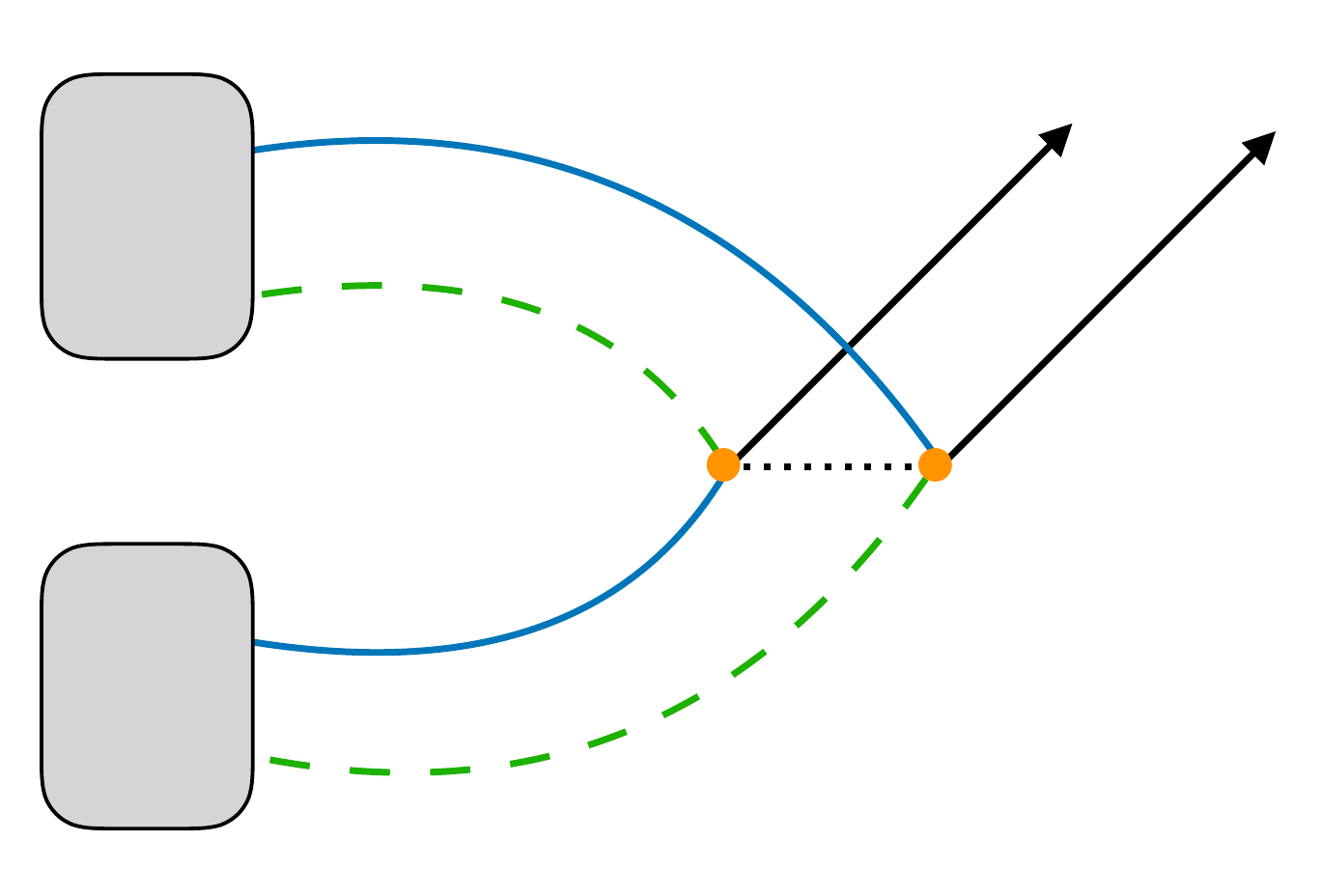}}
       }\qquad \qquad
   \subfigure[]{
       \centering
       \includegraphics[width=0.38\textwidth]{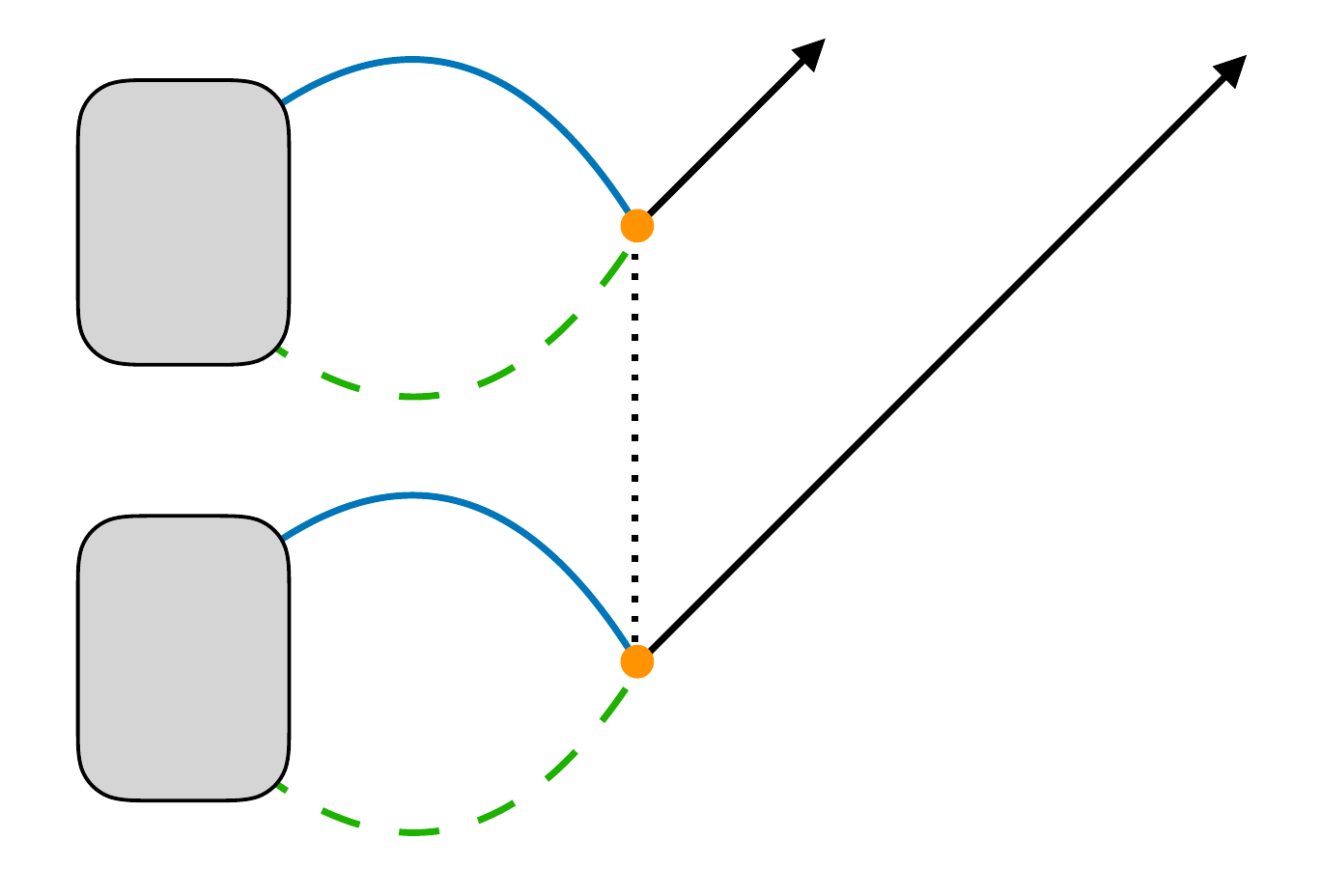}
       }   
 \caption{\label{fig:pipi2ee}Contractions for the $\pi^-\pi^- \to e^- e^-$ transition. The solid blue and dashed green lines represent down and up quark propagators respectively, and the circles represent the $\Delta I=1$ weak vertices. The dotted and solid black lines represent the Majorana neutrino propagator and electron final state respectively. }
\end{figure*}

\begin{figure*}
\centering
\includegraphics[width=0.62\textwidth]{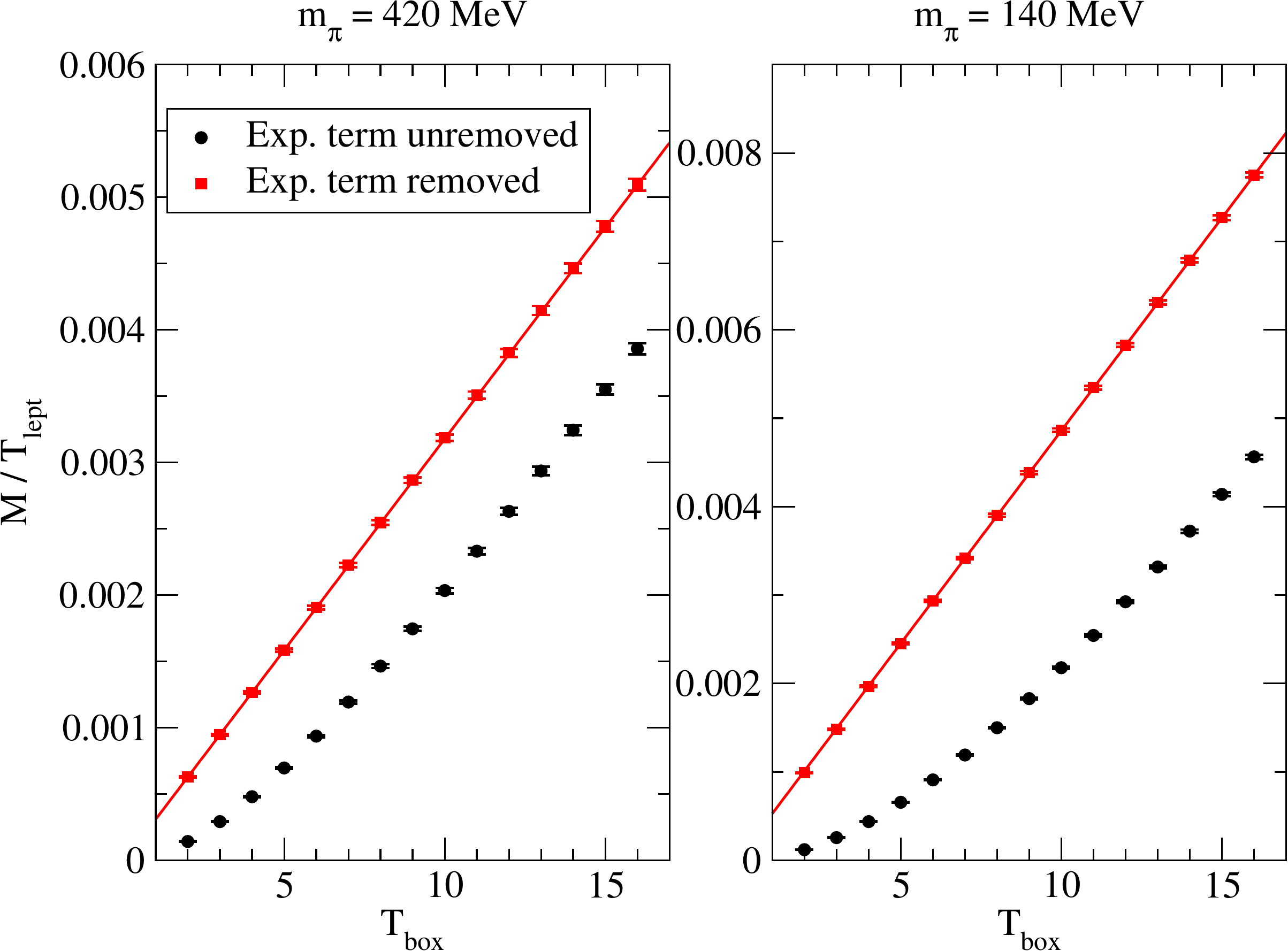}
 \caption{\label{fig:znubblongfeng} The integrated $\pi^-\pi^-\to  e^- e^-$ matrix element $M$ modulo the leptonic contribution as a function of the period of time over which the two weak currents are integrated, $T_{\rm box}$. The red and black circles show the integrated matrix element with and without the exponential term for the ground intermediate state subtracted. (From Ref.~\cite{Feng:2018pdq})}
\end{figure*}

As with the $\pi^-\to\pi^+ e^- e^-$ calculations above, there are two types of contractions involved in constructing the LQCD correlators from which the transition amplitudes can be extracted; these are shown in Fig.~\ref{fig:pipi2ee}.
A key difficulty in this calculation is that the initial state in the correlation functions is a  multi-particle state $\left| \pi^- \pi^- \right\rangle$ and is a finite volume state that must be converted to the desired infinite volume state using the Lellouch-L\"uscher factor \cite{Lellouch:2000pv,Lin:2001ek}, requiring knowledge of the $\pi\pi$-scattering phase shifts. Figure \ref{fig:znubblongfeng} shows the 
integrated matrix element extracted in this calculation for the two different quark masses.
The transition amplitudes that were obtained are $24\%$ and $9\%$ smaller than the predication from leading order
chiral perturbation theory at the two different quark masses. The results provide lattice QCD constraints on the NLO  $\chi$PT counterterms. As discussed in Ref.~\cite{Tuo:2019bue}, these  are compatible with those obtained from the $\pi^-\to \pi^+ e^- e^-$ transition amplitude.
Follow-on calculations in which all the systematic uncertainties are  fully controlled will be possible with larger-scale computational resources.

\subsection{Two-nucleon $nn\to pp e^-e^-$ and other matrix elements}
\label{sec:znubb_nn}

\begin{figure*}
\subfigure[]{
      \raisebox{0mm}{\includegraphics[width=0.35\textwidth]{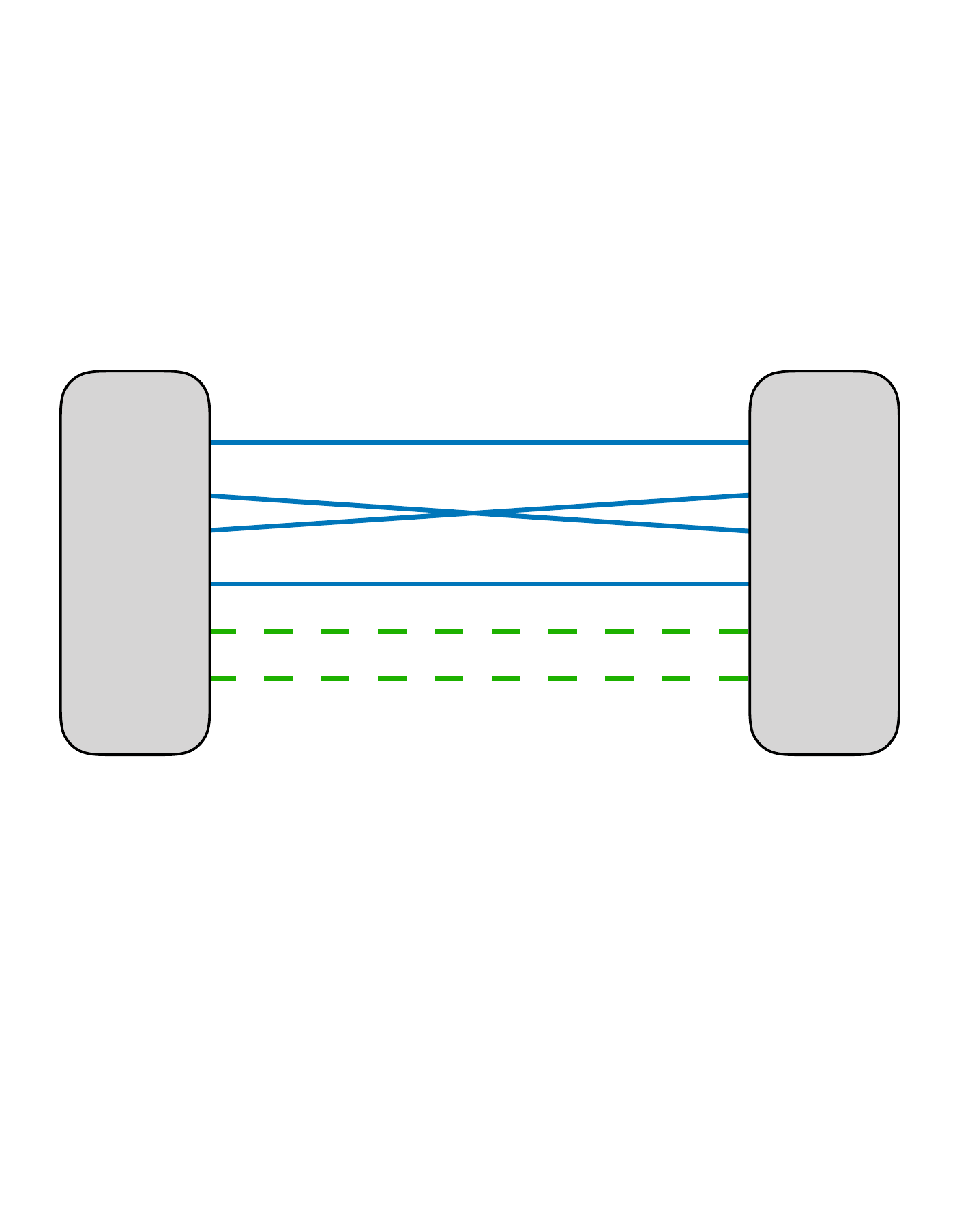}}
       \label{fig:ZRvsBz}
       }\qquad  \raisebox{45mm}{$\Longrightarrow$} \qquad
   \subfigure[]{
       \centering
       \includegraphics[width=0.35\textwidth]{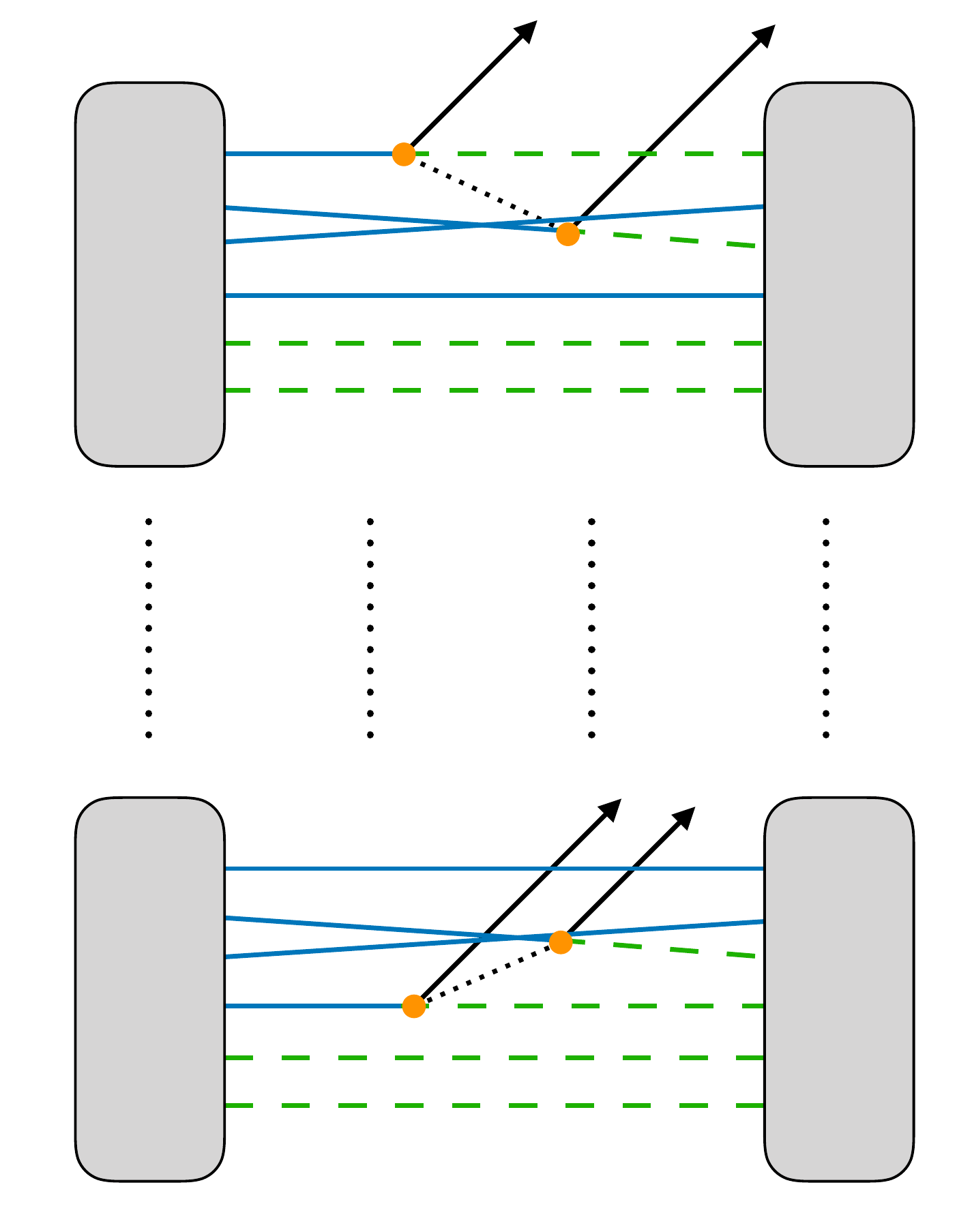}
       \label{fig:ZIvsBz}
       }   
 \caption{\label{fig:my_label} Example contraction for (a) $nn$ two-point correlation functions and (b) the corresponding set of contractions for the long-distance \znubb\ $nn\to p p e^- e^-$ transition. The solid blue and dashed green lines represent down and up quark propagators respectively, and the circles represent the $\Delta I=1$ weak vertices. The dotted and solid black lines represent the Majorana neutrino propagator and electron final state respectively.}
\end{figure*}

The pion transition calculations discussed above have served to investigate technical aspects of the methodology of both short and long-distance contributions to \znubb, and also constrain pionic contributions within nuclear decays. However, the next stage of development is to move to the more directly phenomenologically relevant two-nucleon process $nn\to pp e^-e^-$, as has been done for \tnubb\ decay. 
There are some differences in the computational details for the two-nucleon processes from those of the pion. There is no disconnected type contribution, but the complexity of the connected contractions is significantly higher. Considerable effort has been put into efficient contraction methods for spectroscopy of nuclear systems \cite{Yamazaki:2009ua,Doi:2012xd,Detmold:2012eu,Gunther:2013xj,Detmold:2019fbk} and this can be extended to the contractions relevant for the \znubb\ and \tnubb\ decays of nuclear systems. For each contraction necessary for spectroscopy of a nuclear system, there are $C^{n_d}_2$ contractions for the case of the long distance $\beta\beta$ decay transition where $n_d$ is the number of down quarks in the initial state interpolating field. This is shown in Fig.~\ref{fig:my_label} for the long-distance case (the short distance contributions scale similarly). 

Beyond the two-nucleon transition, the next relevant systems to consider are ${}^4{\rm H} \to {}^4{\rm Li}\ e^- e^-$, ${}^6{\rm H} \to {}^6{\rm Li}\ e^- e^-$ and ${}^6{\rm He} \to {}^6{\rm Be}\ e^- e^-$. Calculations of these transitions are considerably more complicated than the two-nucleon calculations, requiring significantly higher statistics, new types of interpolating operators, and sophisticated approaches to the much larger computational complexity of the quark contractions. These calculations would serve as additional benchmarks for nuclear many-body methods, in particular allowing determinations of LECs in EFT-based many-body approaches and providing an assessment of the convergence of the EFT for larger nuclear systems. 
It is also possible to consider \znubb\ matrix elements in any other hadronic or nuclear states with $I\ge1$. As an example, the $\Sigma^-\to \Sigma^+ e^- e^-$ transition is a simpler 
transition to study numerically which involves the same type of contractions as are 
necessary for the two nucleon transitions.

\pagebreak

\section{Outlook and prospects}
\label{sec:outlook}

While first principles studies of \znubb\ and \tnubb\ decays  are in their early stages, the last few years have seen the application of lattice QCD techniques to calculations relevant for hadronic and nuclear inputs needed for a concrete predictions for these processes. This review highlights recent calculations of the  \tnubb\ transition $nn\to pp e^- e^- \bar\nu_e \bar\nu_e$, and on various pionic \znubb\ transitions induced through either short-distance higher-dimensional operators originating from high scales, or long-distance contributions from light Majorana neutrinos.  A central aspect of this topic is the connection between these LQCD calculations and the effective field theories needed to make connection to phenomenology. 

Important progress in both the LQCD calculations and the EFT matching can be anticipated in the coming years. The scope of LQCD calculations will increase; already there are efforts underway to pursue the \znubb\ transition $nn\to ppe^-e^-$, and one might anticipate the extension of these calculations to larger atomic systems with $A=4,6$. The control of systematic uncertainties for two- and higher-nucleon systems will also improve dramatically with multiple lattice spacings and volumes being used; for the more challenging nuclear calculations an understanding of the quark mass dependence of these amplitudes will also emerge.
In the context of effective field theory, a major improvement will come from performing finite volume EFT calculations of these processes, first for pionic amplitudes and subsequently for the nuclear systems. Such an advance will enable more precise matching to the (necessarily finite volume) LQCD calculations.

As this area matures, LQCD and EFT will have an important impact on $\beta\beta$ phenomenology:
\begin{itemize}
    \item Two-nucleon amplitudes for both short- and long-range mechanisms are essential to determine the 
LO transition operators to be used in nuclear many-body calculations. 
Currently, ignorance of the LO LECs induces an order-of-magnitude uncertainty in the predicted decay rates  even before considering nuclear structure uncertainties. 
Therefore, the calculation of two-nucleon amplitudes 
in LQCD and their extrapolation to the continuum and physical quark masses are of highest priority;

\item  LQCD calculations in systems with A=4 and A=6 
will  help in quantifying uncertainties in the 
nuclear many-body methods. 
In particular, even if one knew the LECs by matching to 
two-body amplitudes,  one expects additional 
uncertainty in nuclear calculations from residual scheme 
dependence and regulator dependence of the results. 
Benchmarking against LQCD may help constrain the 
size of these uncertainties.  
Transitions that can be used for benchmarking purposes include:
$^6{\rm He} \to  \ ^6{\rm Be}\  e^- e^- $  ($\Delta I=0$)
(nuclear structure results exist from several groups), 
$^6 {\rm H} \to \ ^6{\rm Li}\  e^- e^- $  ($\Delta I=2$) 
(more difficult for nuclear structure methods 
because $^6{\rm H}$ is unstable) and possibly
$^4{\rm H} \to \ ^4{\rm Li}\  e^- e^-$ ($\Delta I=0$) and
$4n \to \ ^4{\rm He}\  e^- e^-$.

\end{itemize}

In summary, LQCD coupled to EFT 
has the potential to greatly impact double beta decay phenomenology, 
by dramatically reducing current uncertainties arising  
from matching of quark to hadronic degrees of freedom,  and further 
benchmarking nuclear structure methods in $A \leq 6$ systems.  
While this program requires very challenging calculations,  significant progress is expected 
in the next five to ten years, 
commensurate with the time-scale of next generation 
`ton-scale' experimental searches for \znubb.

\section*{Acknowledgements}

We thank Xu Feng and Luchang Jin for helpful comments and discussions.
WD and PS were supported in part by the U.S.~Department of Energy, Office of Science, Office of Nuclear Physics under grant Contract Number DE-SC0011090. WD was also supported within the framework of the TMD Topical Collaboration of the U.S.~Department of Energy, Office of Science, Office of Nuclear Physics, and  by the SciDAC4 award DE-SC0018121. PES was also supported by the National Science Foundation under CAREER Award 1841699. VC and AN were supported in part by the DOE topical collaboration on “Nuclear Theory for Double-Beta Decay and Fundamental Symmetries”. VC was supported in part by the US DOE Office of
Nuclear Physics under award number DE-AC52-06NA25396. 

\pagebreak

\bibliographystyle{kp}
\bibliography{dbd}
\end{document}